\documentclass[12pt]{article}
\usepackage[top=30truemm,bottom=30truemm,left=25truemm,right=25truemm]{geometry}
\usepackage{amsfonts,amsmath,amssymb,epsf}
\usepackage{amsmath,braket}
\usepackage{here}
\usepackage{physics}
\usepackage{graphicx,hyperref,color}
\usepackage[dvipsnames]{xcolor}
\usepackage{cite}
\usepackage{url}
\usepackage{comment}
\usepackage{appendix}
\usepackage{chngcntr}
\usepackage{etoolbox}
\usepackage{lipsum}

\AtBeginEnvironment{subappendices}{%
\addcontentsline{toc}{section}{Appendices}
\counterwithin{figure}{subsection}
\counterwithin{table}{subsection}
}
\hypersetup{
    unicode=false,          
    pdftoolbar=true,        
    pdfmenubar=true,        
    linktocpage=true,       
    pdffitwindow=false,     
    pdfstartview={FitH},    
    pdfnewwindow=true,      
    colorlinks=true,       
    linkcolor=blue,          
    citecolor=blue,        
    filecolor=blue,      
    urlcolor=blue           
}

\numberwithin{equation}{section}									

\newcommand{\de}{\partial}
\newcommand{\be}{\begin{equation}}
\newcommand{\ba}{\begin{eqnarray}}
\newcommand{\ea}{\end{eqnarray}}
\newcommand{\ee}{\end{equation}}

\newcommand{\s}{\sqrt}

\newcommand{\ti}{\tilde}

\newcommand{\ddd}{\cdot\cdot\cdot}
\newcommand{\no}{\nonumber \\}

\newcommand{\lb}{\rangle}
\newcommand{\bea}{\begin{eqnarray}}
\newcommand{\eea}{\end{eqnarray}}
\newcommand{\bes}{\begin{equation*}}
\newcommand{\beas}{\begin{eqnarray*}}
\newcommand{\eeas}{\end{eqnarray*}}
\newcommand{\bas}{\begin{array*}}
\newcommand{\eas}{\end{array*}}
\newcommand{\ees}{\end{equation*}}
\newcommand{\nn}{\nonumber}
\newcommand{\p}{\partial}
\newcommand{\ep}{\epsilon}

\newcommand{\Arcsin}[1]{\mathrm{Arcsin}{#1}}

\newcommand{\mL}{\mathcal{L}}
		
\newcommand{\Arg}[1]{\mathrm{Arg}\qty(#1)}

\let\a=\alpha \let\b=\beta   \let\e=\epsilon     \let\l=\lambda \let\m=\mu 
 \let\p=\phi \let\r=\rho 
\let\t=\tau \let\th=\theta     \let\z=\zeta
 \let\D=\Delta \let\G=\Gamma        
\def\nn{\nonumber}
\def\inf{\infty}

\DeclareMathOperator{\arctanh}{arctanh}

\newcommand{\EllipticF}[2]{F\left( \left.#1  \right| #2\right)}
\newcommand{\EllipticK}[1]{K\left(#1\right)}
\newcommand{\EllipticPi}[3]{\Pi\left(\left.#1;#2 \right| #3\right)}
\newcommand{\EllipticPiC}[2]{\Pi\left(#1| #2\right)}
\newcommand{\greone}{\mathrm{I}}
\newcommand{\gretwo}{\mathrm{I}\hspace{-1.2pt}\mathrm{I}}
\newcommand{\grethree}{\mathrm{I}\hspace{-1.2pt}\mathrm{I}\hspace{-1.2pt}\mathrm{I}}
\usepackage{tikz}
\usetikzlibrary{arrows,arrows.meta,intersections, calc,positioning,decorations.pathreplacing,decorations.pathmorphing,shapes}
\usetikzlibrary{patterns}
\usetikzlibrary{decorations.markings}
\usetikzlibrary{knots}

\usepackage{subcaption}

\begin{document}

\begin{titlepage}
\thispagestyle{empty}

\vspace*{-2cm}
\begin{flushright}
YITP-23-148
\\
\end{flushright}

\bigskip

\begin{center}
\noindent{\bf \large {Entanglement Phase Transition in Holographic Pseudo Entropy}}\\
\vspace{1.2cm}

Hiroki Kanda$^a$, Taishi Kawamoto$^a$, Yu-ki Suzuki$^a$, 
Tadashi Takayanagi$^{a,b}$, \\
Kenya Tasuki$^a$ and Zixia Wei$^{c,d}$
\vspace{1cm}\\

{\it $^a$Center for Gravitational Physics and Quantum Information,\\
Yukawa Institute for Theoretical Physics, Kyoto University, \\
Kitashirakawa Oiwakecho, Sakyo-ku, Kyoto 606-8502, Japan}\\
\vspace{1.5mm}
{\it $^b$Inamori Research Institute for Science,\\
620 Suiginya-cho, Shimogyo-ku,
Kyoto 600-8411, Japan}\\
\vspace{1.5mm}
{\it $^{c}$ Center for the Fundamental Laws of Nature,\\ Harvard University, Cambridge, MA 02138, USA}\\
\vspace{1.5mm}
{\it $^{d}$ Society of Fellows, Harvard University, Cambridge, MA 02138, USA}
\bigskip \bigskip
\vskip 1em
\end{center}

\begin{abstract}
In this paper, we present holographic descriptions of entanglement phase transition using AdS/BCFT. First, we analytically calculate the holographic pseudo entropy in the AdS/BCFT model with a brane localized scalar field and show the entanglement phase transition behavior where the time evolution of entropy changes from the linear growth to the trivial one via a critical logarithmic evolution. In this model, the imaginary valued scalar field localized on the brane controls the phase transition, which is analogous to the amount of projections in the measurement induced phase transition. Next, we study the AdS/BCFT model with a brane localized gauge field, where the phase transition looks different in that there is no logarithmically evolving critical point. Finally, we discuss a bulk analog of the above model by considering a double Wick rotation of the Janus solution. We compute the holographic pseudo entropy in this model and show that the entropy grows logarithmically.

\end{abstract}

\end{titlepage}

\newpage

\tableofcontents

\newpage

\section{Introduction}

Entanglement entropy is a very useful probe of quantum phases in many-body systems and field theories\cite{Bombelli:1986rw,Srednicki:1993im,Casini:2009sr,Calabrese:2009qy}. For example, it plays a role of order parameter in quantum phase transitions \cite{Vidal:2002rm,Latorre:2003kg} and it can detect topological phases \cite{Levin:2006zz,Kitaev:2005dm}. At the critical points, it measures the degrees of freedom of conformal field theories (CFTs),  namely central charges \cite{Holzhey:1994we,Calabrese:2004eu}. It also offers a nice entropic quantity of thermalization for non-equilibrium quantum processes, such as quantum quenches \cite{Calabrese:2005in,Calabrese:2007mtj}. 

As a more recent development, entanglement entropy has been employed to probe a novel dynamical quantum phase transition, so called measurement induced phase transition or entanglement phase transitions \cite{Skinner:2018tjl,Li:2018mcv}. This phase transition occurs in monitored quantum circuits, where a quantum many-body system is time evolved by a chaotic or random Hamiltonian and subsequently by projection measurements. Under the unitary time evolution of the Hamiltonian, the entanglement entropy grows linearly. However, in the presence of projection measurements, the entanglement growth is suppressed. 
Due to the competition between these two effects, an intriguing phase transition occurs. In (1+1)-dimensional quantum spin systems,
the time evolution of entanglement entropy changes as 
\ba
&& (i)\ \  p<p_*:\ \ \   S_A\propto t,\no
&& (ii)\ \ p=p_*:\ \ \   S_A\propto \log t,\label{EPTa}\\
&& (iii)\ \ p>p_*:\ \ \  S_A=\mbox{finite},  \nonumber  
\ea
where $p$ is the rate of projection measurement and $p_*$ denotes the phase transition point. A similar phase transition was found for quantum systems with dissipating dynamics in \cite{Kawabata:2022biv}.

It is intriguing to see if such entanglement phase transitions exist in broader quantum systems such as interacting quantum field theories, including those in higher dimensions. For this purpose a powerful method is the holographic calculation of entanglement entropy \cite{Ryu:2006bv,Ryu:2006ef,Hubeny:2007xt} based on the AdS/CFT \cite{Maldacena:1997re}. Even though there have been interesting progresses on holographic descriptions of measurement induced phase transition \cite{Milekhin:2022bzx, Antonini:2022lmg,Antonini:2022sfm,Goto:2022fec,Antonini:2023aza}, they are either discussed in (0+1)-dimensional models \cite{Milekhin:2022bzx, Antonini:2022lmg} or by realizing a limited number of projection measurements \cite{Milekhin:2022bzx, Antonini:2022lmg,Antonini:2022sfm,Goto:2022fec,Antonini:2023aza}, and hence the entanglement phase transition with the behavior (\ref{EPTa}) has not been realized so far. 

Recently, a different approach to the entanglement phase transition in a two dimensional holographic CFT was suggested in 
the paper \cite{Kanda:2023zse}. Instead of considering a holographic description of projection measurements, a certain dissipative effect is taken into account as an imaginary valued scalar field on the end-of-the-world brane (EOW brane) in the AdS/BCFT formulation \cite{Takayanagi:2011zk,Fujita:2011fp}. This can be regarded as a modification of the gravity dual of quantum quenches \cite{Hartman:2013qma}. The resulting gravity dual describes a time-dependent transition matrix \cite{Nakata:2020luh} in CFT viewpoints, instead of a regular quantum state. Therefore, strictly speaking, the geodesic length computes the holographic pseudo entropy \cite{Nakata:2020luh}.  The pseudo entropy 
is a natural generalization of entanglement entropy so that it depends not only on the initial state (a given quantum state) but also on the final state (a final state projection), which has been successfully applied to detect quantum phase transitions \cite{Mollabashi:2020yie,Mollabashi:2021xsd,Nishioka:2021cxe} in quantum many-body systems. Refer also to e.g. \cite{Goto:2021kln,Miyaji:2021lcq,Murciano:2021dga,Mukherjee:2022jac,Ishiyama:2022odv,Guo:2022jzs,Akal:2021dqt,Akal:2022qei,Guo:2022sfl,Doi:2022iyj,Li:2022tsv,Doi:2023zaf,Chu:2023zah,He:2023ubi,Narayan:2022afv,He:2023eap,Narayan:2023ebn,Jiang:2023loq,Kawamoto:2023nki,Chen:2023eic,Chen:2023gnh,Chandra:2023rhx,Carignano:2023xbz,Parzygnat:2023avh,Guo:2023aio,Omidi:2023env,Narayan:2023zen} for further developments. In \cite{Kanda:2023zse}, it was conjectured that the time evolution of pseudo entropy shows the same behavior as the entanglement transition (\ref{EPTa}), where the 
phases for $p<p_*$, $p=p_*$ and $p>p_*$ correspond to the three different geometries: AdS black hole, Poincar\'e AdS and the thermal AdS, respectively. 

The purpose of this paper is to explore this connection more quantitatively and generalize to other setups. First we will fully work out the behavior of holographic pseudo entropy and show the following quantitative behavior in a holographic model constructed by introducing a massless free scalar on the end-of-the-world brane in
AdS$_3$/BCFT$_2$:
\ba
&&(i)\ \  \Delta \phi<\Delta\phi_*:\ \ \   \Delta S_A\simeq \frac{c}{6a} t,\no
&& (ii)\ \  \Delta \phi=\Delta\phi_*:\ \ \   \Delta S_A\simeq \frac{c}{3}\log t, \label{EPTb} \\
&& (iii)\ \  \Delta \phi>\Delta\phi_*:\ \ \  \Delta S_A=\mbox{finite},   \nonumber
\ea
where $\Delta \phi$ is the amount of scalar field perturbation which controls the dissipative effect in Lorentzian time evolution and $\Delta S_A$ is the pseudo entropy with the ground state entanglement entropy subtracted. The constant $a$ increases as $\Delta\phi$ gets larger and becomes divergent at $\Delta\phi \rightarrow \Delta\phi_*$, and $c$ is the central charge of the corresponding BCFT$_2$. 

One advantage of our approach is that it is straightforward to be extended to higher dimensional cases and we will show that the intermediate logarithmic behavior is not present in higher dimensions. Moreover, We will discuss a generalization of our setup with a gauge field on the EOW brane. We will also explore a purely bulk model which mimics a general measurement by employing the Janus solution.

A brief summary of the contents in this paper is as follows: 

In section \ref{sec:PEfromAdSBCFT}, we explore the model proposed in \cite{Kanda:2023zse}, where an EOW brane with a brane-localized massless free scalar field was considered in the AdS$_3$/BCFT$_2$. Especially we work out the global structures of its Lorentzian time evolution obtained by a Wick rotation.  We compute explicitly the growth of pseudo entropy including its higher dimensional generalization and determine the coefficient of both its linear and logarithmic growth as presented in (\ref{EPTb}). We will also analyze some brane profiles when the brane-localized scalar field has a nontrivial potential. Besides, We find a higher dimensional result for the growth at the critical point. In this calculation the global extension beyond the Poincar\'e horizon turns out to be important. We also present a simple model of dissipating time evolution which shows the logarithmic growth and which is universal for any two dimensional CFTs. 

In section \ref{sec:gauge_field}, we analyze another setup of AdS/BCFT with a gauge field on the EOW brane. We give solutions of EOW branes with a localized gauge field in Poincar\'e AdS$_{d+1}$ setup. Interestingly, for $d>2$ we find that the solution with the localized gauge flux in Poincar\'e AdS$_{d+1}$ look identical to that with the localized scalar in  Poincar\'e AdS$_{d}$. For $d=2$, we find that the EOW brane solution with the gauge flux is not available in Poincar\'e AdS$_3$. Instead, 
for the three dimensional gravity dual, the EOW brane solution with the flux can be obtained for BTZ and thermal AdS geometry. We examine the phase transition between the two phases. Note that this section is written in an independent manner of the above sections to be self-contained.

In section \ref{sec:Janus}, which is different from the above two sections involving the EOW brane, we consider the model with a bulk scalar field in an asymptotically AdS spacetime. One motivation for this section is to provide the soft wall model for the EOW brane model, similar to the argument in the gluon condensation\cite{Csaki:2006ji,Kim:2007qk}. To this end, we introduce the spacelike brane in the CFT, by taking the double Wick rotation of Janus solution, described by the deformation caused by the line defect whose tangential direction is in the spacelike direction. We discuss the dual Janus solution for both the Euclidean and Lorentzian time cases. Especially in the Lorentzian model, the dual spacetime exhibits highly exotic behavior. We compute the pseudo entropy using the HRT formula for these spacetimes. We find that the analytic formula for the entanglement pseudo entropy, defined with the subsystem parallel to the defect, shows logarithmically decreasing behavior around the defect.

In section \ref{sec:conclusions}, we summarize our conclusions and discuss future problems. 

Some math formulas used in section \ref{sec:Janus} will be summarized in appendix \ref{app:tech_details}. An alternative interpretation of the setup analyzed in section \ref{sec:Janus} can be found in appendix \ref{app:int_Janus}. Besides, while all the analyses appearing in the main text are performed holographically on the gravity side, we present an analogous CFT analysis in \ref{app:CFT_analysis} where the pseudo R\'enyi entropy is computed perturbatively. 

\section{Entanglement phase transition from AdS/BCFT}\label{sec:PEfromAdSBCFT}

As proposed in \cite{Kanda:2023zse}, an asymptotically AdS$_3$ background with an end-of-the-world brane (EOW brane) where a massless free scalar field is localized, provides an interesting holographic setup of entanglement phase transition. In this section, we would like to analyze this setup by focusing on the behavior of its holographic pseudo entropy and would like to quantitatively confirm that it shows the expected behavior of entanglement phase transition (\ref{EPTa}) with explicit coefficients as given in (\ref{EPTb}). 
We would like to mention that a brane-localized scalar field was also considered in earlier works on the holographic Kondo effect \cite{Erdmenger:2013dpa,Erdmenger:2014xya,Erdmenger:2015xpq,Erdmenger:2015spo}. A scalar field in AdS/BCFT was also analyzed in a probe limit by the paper \cite{Suzuki:2022tan}. There is another interesting approach to AdS/BCFT with multiple boundary conditions and brane localized matter,
by introducing a corner defect on the EOW brane \cite{Miyaji:2022dna,Biswas:2022xfw,Kusuki:2022ozk}.

\subsection{End-of-the-world brane with a localized scalar in AdS/BCFT}
\begin{figure}[h]
    \centering
    \includegraphics[width=8cm]{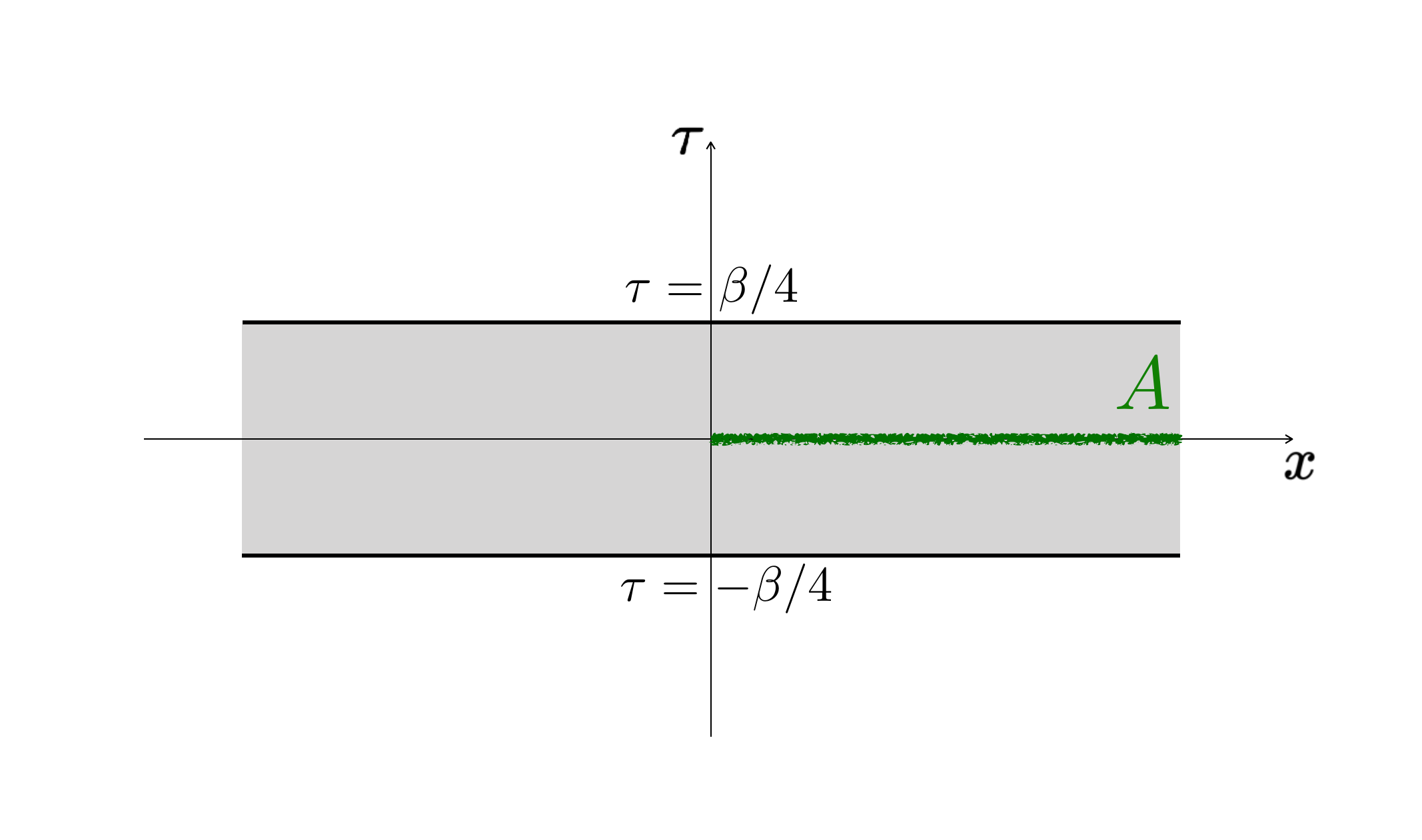}
    \includegraphics[width=8cm]{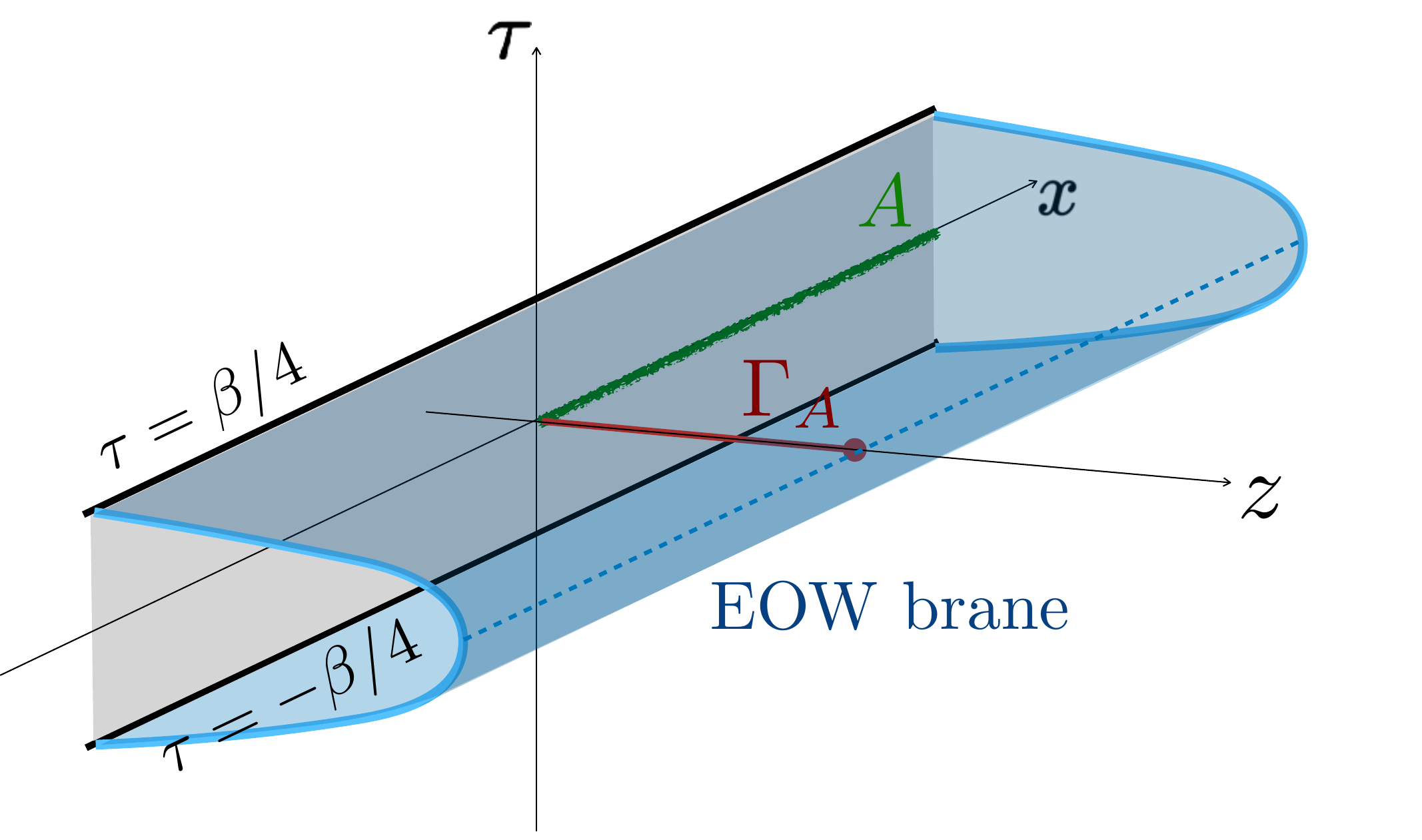}
    \caption{A sketch of the AdS/BCFT model. The left figure shows a BCFT defined on a strip (light grey region) with the width $\frac{\beta}{2}$. The right figure shows the dual gravity setup which is given by the AdS$_3$ region surrounded by the strip and the EOW brane (blue surface). The red curve $\Gamma_A$ is the geodesic whose length calculates the holographic pseudo entanglement entropy for the subregion $A$ shaded in green.
    }
    \label{fig:AdSBCFT}
\end{figure}

Before we present our calculations of holographic entropy, we would like to start with a brief review of the model introduced in 
\cite{Kanda:2023zse}. The AdS/BCFT is an extension of AdS/CFT to the cases where the CFT is defined on a manifold $\Sigma$ with boundaries \cite{Takayanagi:2011zk,Fujita:2011fp,Karch:2000gx}. 
It argues that the CFT on $\Sigma$ is dual to an asymptotically AdS region which is surrounded by the $\Sigma$ at the asymptotic boundary and by an EOW brane $Q$. The EOW brane is a dynamical brane beyond which no spacetime exists, and it satisfies a Neumann boundary condition in the metric fluctuations. Refer to figure \ref{fig:AdSBCFT} for a sketch.

In the model introduced in \cite{Kanda:2023zse}, one ingredient was added to the minimal model of AdS/BCFT explained above in three dimensions. It is a scalar field $\phi$ localized on the EOW brane. From the viewpoint of boundary conformal field theory (BCFT), this scalar field corresponds to an external field which couples to a boundary scalar operator. The Neumann boundary condition on $Q$ in the presence of the scalar field (with the potential energy $V(\phi)$) reads 
\begin{equation}\label{eq:the-general-eom-of-the-EOW-brane-in-Loretzian}
    K_{ab} -h_{ab} K=T_{ab}=-h_{ab}\left( \partial _{c} \phi \partial ^{c} \phi +V( \phi )\right) +2\partial _{a} \phi \partial _{b} \phi ,
\end{equation}
where the indices represent the coordinates on the EOW brane. Below, unless otherwise noted, we will set $V(\phi) = 0$, i.e. we will consider a massless free scalar field. 


In particular, we consider a BCFT defined on a two dimensional strip whose coordinate is described by $(\tau,x)$ in the Euclidean signature. The width of this strip is $-\frac{\beta}{4}\leq \tau\leq \frac{\beta}{4}$, while the coordinate $x$ is non-compact. In the absence of the localized scalar field, this is identical to the setup in \cite{Takayanagi:2011zk,Fujita:2011fp} and \cite{Hartman:2013qma} and the bulk geometry is found to be the BTZ black hole with the temperature $1/\beta$. We write the metric of BTZ black hole as follows:
\begin{equation}
    ds^2=\frac{dz^{2}}{h( z) z^{2}} -\frac{h( z) dt^{2}}{z^{2}} +\frac{dx^{2}}{z^{2}},\quad h(z)=1-\frac{z^2}{a^2},  \label{BTZm}
\end{equation}
where $T_{BTZ}=\frac{1}{2\pi a}$ is the temperature of the BTZ black hole. 
We have $\beta=2\pi a$ when there is no brane-localized scalar field. 
The EOW brane connects the two boundaries of the strip in the bulk as in figure \ref{fig:AdSBCFT}. 

\begin{figure}[h]
    \centering
    \includegraphics[width=10cm]{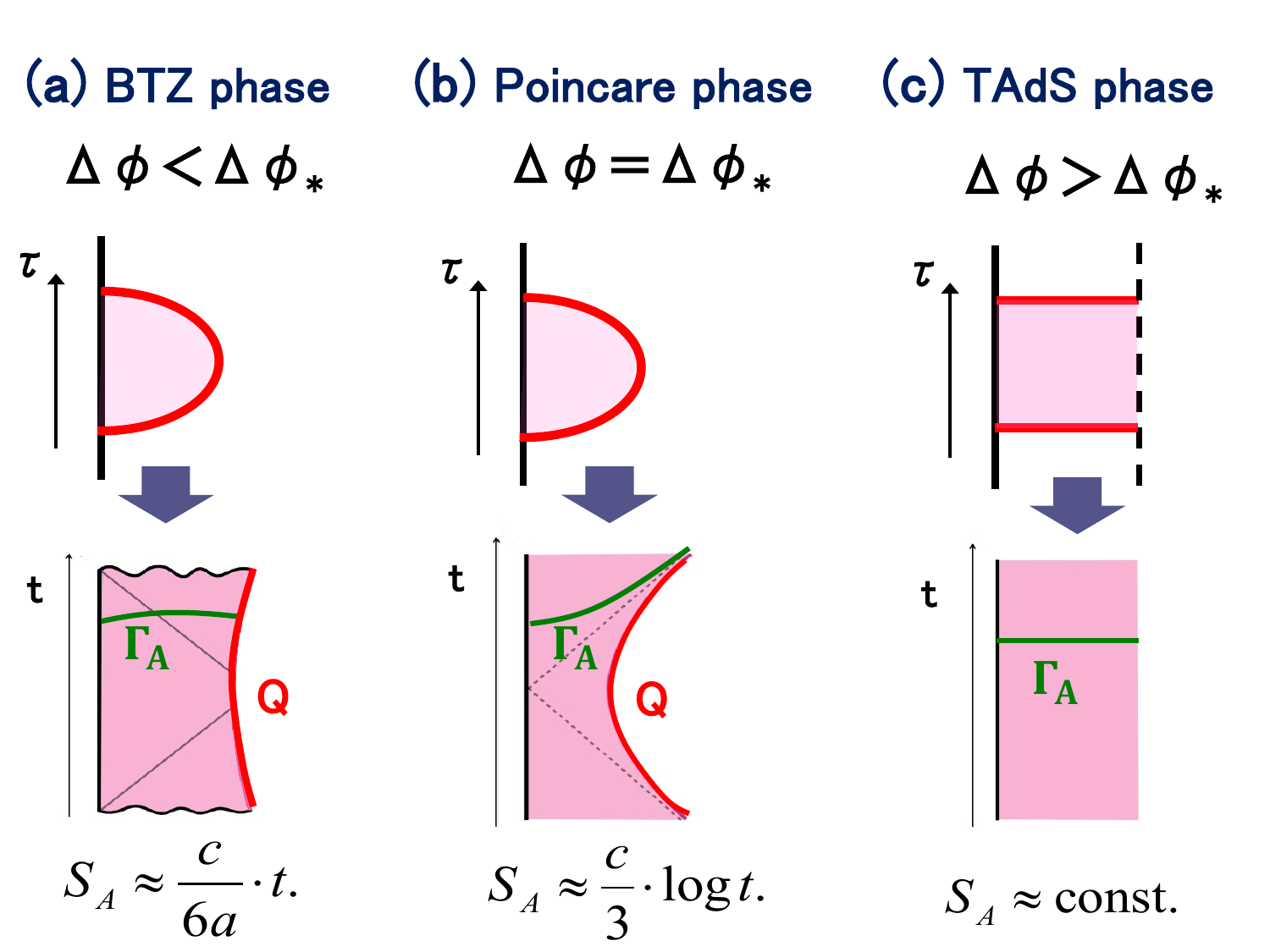}
    \caption{A sketch of gravity duals in our AdS/BCFT model with a brane localized scalar. Depending on the value of the shift of scalar field $\Delta \phi$, there are three phases (a) BTZ phase, (b) Poincar\'e phase and (c) TAdS phase. The upper three pictures describe the Euclidean geometries, while the lower three are the corresponding Lorentzian time evolution obtained from analytic continuation.
    }
    \label{fig:phases}
\end{figure}

We then introduce the localized scalar field on $Q$ such that it takes the values $-\frac{\Delta\phi}{2}$ and  $\frac{\Delta\phi}{2}$ on each of the two boundaries of the strip. We further analytically continue at $\tau=0$ to the Lorentzian time as $\tau=it$. This realizes the following transition matrix in a CFT: 
\begin{equation}
    \mathcal{T}=e^{-iHt}e^{-\b H/4}\ket{B(\D \p/2)}\bra{B(-\D\p/2)}e^{-\b H/4}e^{iHt},
\end{equation}
where $|B(\cdots)\lb$ are boundary states (or Cardy states) \cite{Cardy:2004hm} labeled by different boundary values of $\phi$.
The gravity dual of this transition matrix can be found by solving (\ref{eq:the-general-eom-of-the-EOW-brane-in-Loretzian}) together with the bulk Einstein equation. The non-trivial profile of the scalar field gives the backreaction, which makes the inverse temperature $2\pi a$ deviate from $\beta$, where $\beta$ is also twice the width of strip. As the scalar field shift $\Delta \phi$ increases, with $\beta$ fixed, $a$ grows and gets divergent at the critical point $\Delta \phi=\Delta \phi_*$. Its explicit value is given by $\Delta\phi_*=2K[-1]\simeq 2.6$.
The fact that $a=\infty$ means that the dual geometry now looks like the Poincar\'e AdS$_3$. For larger values $\Delta \phi>\Delta \phi_*$, the EOW brane is no longer connected and the dual geometry becomes the thermal AdS$_3$ where the EOW brane consists of two disconnected surfaces, which disappear in the Lorentzian counterpart obtained by performing analytic continuation. These three phases are sketched in figure \ref{fig:phases}. 

The quantity we focus on below is the pseudo entropy defined by 
\begin{equation}
   S_A=-\mbox{Tr}[{\mathcal{T}}_A\log {\mathcal{T}}_A],
\end{equation}
where ${\mathcal{T}}_A$ is the reduced transition matrix obtained by tracing out the complement $\bar{A}$ of the subsystem $A$ as
\begin{equation}
    {\mathcal{T}}_A=\mbox{Tr}_{\bar{A}}[{\mathcal{T}}].
\end{equation}
Even though ${\mathcal{T}}_A$ is not Hermitian in general, it turns out that $S_A$ is real and non-negative for the examples we study in this paper. Since it is obvious that in the final phase $\Delta \phi>\Delta \phi_*$, the time evolution of the bulk geometry and hence the time evolution of $S_A$ is trivial, we will focus on the BTZ phase $\Delta \phi<\Delta \phi_*$ and critical phase $\Delta \phi=\Delta \phi_*$ below. As a result, we will get the behavior shown in \eqref{EPTb}

Below we will calculate the holographic counterpart of $S_A$ from the bulk geometry. In the AdS/CFT correspondence, the pseudo entropy $S_A$ for a subsystem $A$ can be computed from the area of the extremal surface $\Gamma_A$ which surrounds $A$ and takes the minimal value as \cite{Ryu:2006bv,Ryu:2006ef,Hubeny:2007xt}
\be
S_A=\frac{\mbox{Area}(\Gamma_A)}{4G_N}\label{RTF}.
\ee
In the AdS/BCFT, the holographic pseudo entropy can be computed similarly except that $\Gamma_A$ can also end on the EOW brane $Q$ \cite{Takayanagi:2011zk,Fujita:2011fp}. Since our analysis in this paper always assumes that the subsystem $A$ covers a half of the total space, $\Gamma_A$ always ends on $Q$.
In this case, we need to extremalize by changing the end points of $\Gamma_A$ on $Q$.

Before we proceed, we would like to point out an interesting possibility which may give another interpretation of our results. In our example, the pseudo entropy is always real and non-negative, though the pseudo entropy in general can take complex values, as we have already mentioned. Moreover, it is also natural that the R\'enyi extensions of pseudo entropy (pseudo R\'enyi entropy) also take real and non-negative values in our setup as the replicated geometries are expected to be described by certain real valued metrics, though it would be difficult to construct them explicitly. If all pseudo R\'enyi entropies are real and non-negative, then there is a chance that the eigenvalues of the reduced transition matrix are also real and non-negative.
Such a matrix is called pseudo Hermitian and can be diagonalized by a similarity transformation $\mathcal{T}_A=P\cdot \Lambda\cdot P^{-1}$, where $\Lambda$ is a diagonal matrix with real and non-negative eigenvalues. This implies, in this case, that the calculation of pseudo entropy for the transition matrix can be regarded as that of entanglement entropy for certain quantum states. It is an interesting future problem to examine this clearly and relate our results to entanglement entropy of definite quantum states in some quantum systems, though we will not pursue this in the present paper.

\subsection{Complete analysis of holographic pseudo entropy in BTZ phase\texorpdfstring{: $\Delta \phi<\Delta \phi_*$}{}}

Let us consider the BTZ black hole phase and compute $S_A$.
Since the geodesic extends beyond the horizon in the coordinate patch (\ref{BTZm}), we need to work with the Kruskal coordinates
\begin{equation}
  ds^{2} =\frac{-4dudv+( 1-uv)^{2} dx^{2} /a^{2}}{( 1+uv)^{2}},
\end{equation}
where $u$ and $v$ are related to $z$ and $t$ outside of the horizon as 
\begin{equation}\label{eq:coord_change_from_poi_to_kru}
    u=-\sqrt{\frac{a-z}{a+z}} e^{-t/a},\quad v=\sqrt{\frac{a-z}{a+z}} e^{t/a}.
\end{equation}
In this coordinate, the horizon locates at $uv=0$. The center and the asymptotic boundary of the spacetime correspond to $uv=1$ and $uv=-1$ respectively. Note that since we do not compactify $x$, the points on $uv=1$ are not singular.

Now, the trajectory of the EOW brane and the scalar field on the EOW brane can be described as $v=v(u)$ and $\p=\p(u)$, assuming the translational invariance in $x$ direction. After some algebra, we obtain
\begin{align}
    V(\phi)&=-\frac{(1-u^{2}v^{2})v^{\prime\prime}+2(3-uv)(v-uv^\prime)v^\prime}{8(1-uv)v^{\prime3/2}},\\
    \phi^{\prime2} &=\frac{( 1-uv) v^{\prime\prime} -2( v-uv^\prime ) v^\prime }{2\left( 1-u^{2} v^{2}\right)\sqrt{v^\prime }},
\end{align}
where primes denote the derivative with respect to $u$. As we have already mentioned, we consider the case where $V(\phi)=0$. This implies
\begin{equation}\label{eq:eow_of_btz_v_less}
    v^{\prime\prime} =-\frac{2v^\prime ( 3-uv)( v-uv^\prime )}{1-( uv)^{2}}.
\end{equation}
We impose
\begin{equation}\label{eq:init_cond_of_btz_v_less}
    v(u_0)=-u_0,\quad v^\prime(u_0)=1, \quad u_{0} :=-\sqrt{\frac{1-\zeta_{0}}{1+\zeta_{0}}},\quad \zeta_0:=\frac{z_0}{a}<1,
\end{equation}
as the initial conditions. Besides, $\zeta_0$ and $a$ are fixed by $\D \p$ and $\beta$ via
\begin{equation}
    \Delta \phi =2\left( 1-\zeta _{0}^{2}\right)^{1/4} K\left( -\left( 1-\zeta _{0}^{2}\right)\right),
\end{equation}
\begin{equation}
    \frac{\beta }{a} =\frac{4}{\zeta _{0}\sqrt{\left( 1-\zeta _{0}^{2}\right)\left( 2-\zeta _{0}^{2}\right)}}\left( \Pi \left( -\frac{\zeta _{0}^{2}}{1-\zeta _{0}^{2}} \middle|\frac{1-\zeta _{0}^{2}}{2-\zeta _{0}^{2}}\right) -\left( 1-\zeta _{0}^{2}\right) K\left(\frac{1-\zeta _{0}^{2}}{2-\zeta _{0}^{2}}\right)\right).
\end{equation}
The conditions \eqref{eq:init_cond_of_btz_v_less} are equivalent to $z(t=0)=z_0$ and $\dot{z}(t=0)=0$ in the Poincar\'e coordinate. Using the Poincar\'e coordinate $z$ as the parameter of the curve, we can solve this equation as follows:
\begin{equation}
    u=-\frac{1-z/a}{A( z/a)} e^{-B( z/a)} ,\quad v=\frac{A( z/a)}{1+z/a} e^{B( z/a)},
\end{equation}
\begin{equation}
    A( \zeta ) :=\frac{\sqrt{\left( 1-\zeta _{0}^{2}\right)\left( \zeta _{0}^{2} +\zeta ^{2} -\zeta _{0}^{2} \zeta ^{2}\right)} +\zeta \sqrt{\zeta ^{2} -\zeta _{0}^{2}}}{\sqrt{\zeta ^{2} +\zeta _{0}^{2} -\zeta _{0}^{4}}},
\end{equation}
\begin{equation}
    B( \zeta ) :=\frac{\zeta _{0}}{\sqrt{\left( 1-\zeta _{0}^{2}\right)\left( 2-\zeta _{0}^{2}\right)}}\left[ F\left(\cos^{-1}\frac{\zeta _{0}}{\zeta } \middle|\frac{1}{2-\zeta _{0}^{2}}\right) -\Pi \left(\frac{1-\zeta _{0}^{2}}{2-\zeta _{0}^{2}} ;\cos^{-1}\frac{\zeta _{0}}{\zeta } \middle|\frac{1}{2-\zeta _{0}^{2}}\right)\right],
\end{equation}
where $F(\phi|m)$ and $\Pi(n;\phi|m)$ are the incomplete elliptic integrals of the first kind and of the third kind, respectively. Here, $z$ runs from $z_0$ to $\infty$ as a parameter.

Consider the extended region by embedding BTZ metric into the global AdS:
\begin{equation}
    ds^{2} =\frac{1}{\cos^{2} \rho }\left( d\rho ^{2}-d\tau ^{2} +\sin^{2} \rho d\theta ^{2}\right),
\end{equation}
where the AdS radius is normalized to unity. Explicitly, embedding is accomplished by
\begin{align}
    u&=\frac{-\sin \rho \cos \theta +\sin \tau }{\cos \rho +\sqrt{\cos^{2} \tau -\sin^{2} \rho \sin^{2} \theta }}\nonumber\\
    v&=\frac{\sin \rho \cos \theta +\sin \tau }{\cos \rho +\sqrt{\cos^{2} \tau -\sin^{2} \rho \sin^{2} \theta }}\nonumber\\
    x &=a\sinh^{-1}\left(\frac{\sin \rho \sin \theta }{\sqrt{\cos^{2} \tau -\sin^{2} \rho \sin^{2} \theta }}\right).\label{eq:coord_change_from_glo_to_kru}
\end{align}
The original BTZ region is a subregion of the global AdS, which is
\begin{equation}
    \left\{( \rho,\tau ,\theta ) |\cos^{2} \tau  >\sin^{2} \rho \sin^{2} \theta ,\tau \in [ -\pi/2 ,\pi/2 ] ,\rho \in [ 0,\pi /2) ,\theta \in [ 0,2\pi )\right\}.
\end{equation}

\begin{figure}[h]
    \centering
    \includegraphics[width=0.25\linewidth]{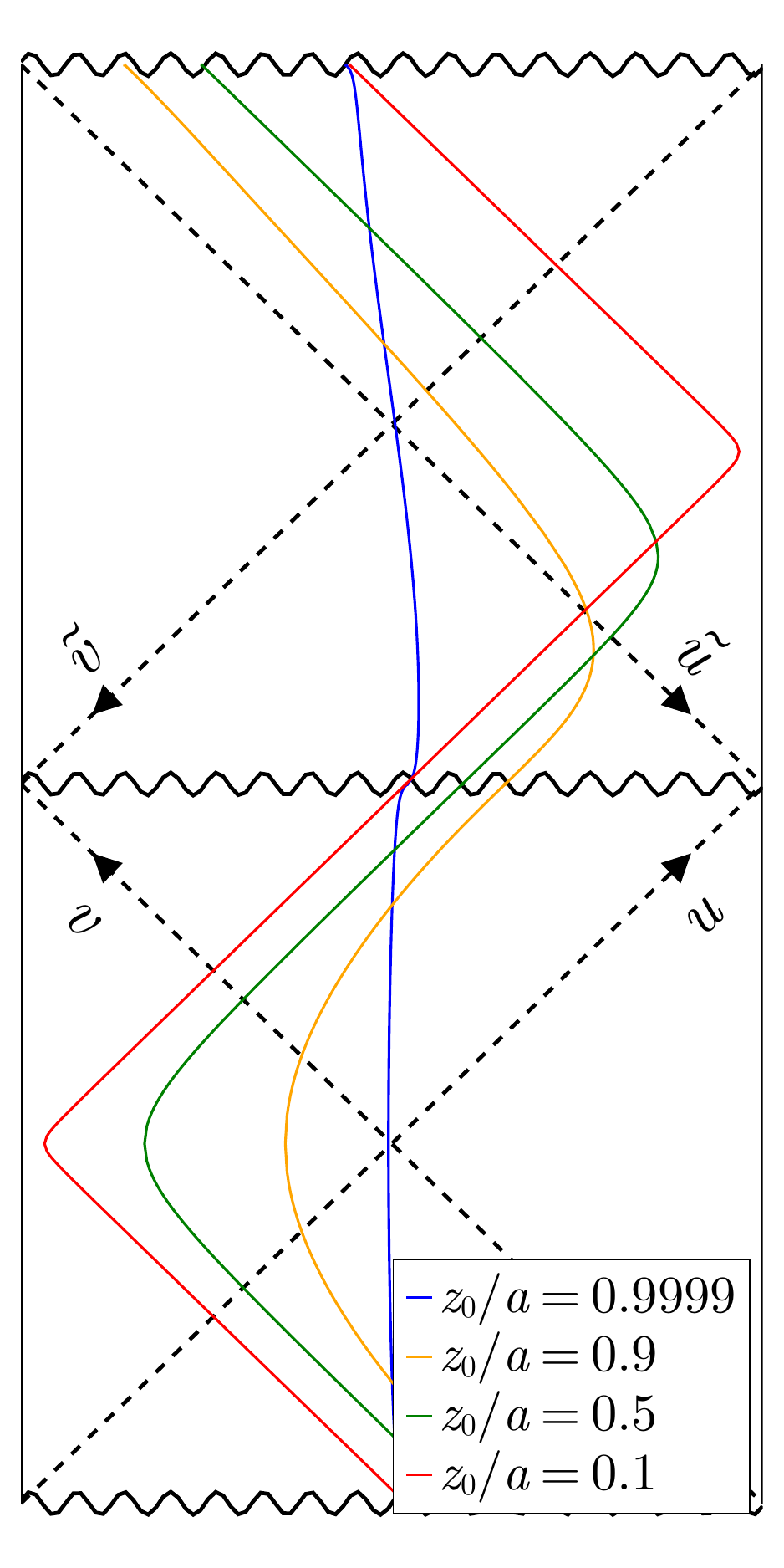}
    \caption{The EOW branes in the  $\D \p$ phase. It is conformally compactified using arctan. The lower square corresponds to the original BTZ black hole, while the upper one corresponds to the extended region.
    In this picture, the dual BCFT lives in the left asymptotic boundary of the lower BTZ. The EOW brane with $z_0/a= 1$, corresponding to $\D \p=0$, forms a vertical line, which is consistent with the configuration described in \cite{Hartman:2013qma}.}
    \label{fig:Lorentzian_EOWs_in_BTZ}
\end{figure}

Now, we introduce the same coordinates for the extended region adjacent to the original BTZ black hole by shifting the range of $\tau$ to $[\pi/2, 3\pi/2]$. For clarity, we use a tilde to distinguish the original coordinates and the extended ones. Although \eqref{eq:eow_of_btz_v_less} is the equation for the original region, it is important to emphasize that it still governs the EOW brane even outside of the original region, as we are using identical coordinates in the extended region. Instead of \eqref{eq:init_cond_of_btz_v_less}, however, we need to impose that the EOW brane is connected and smooth at the junction point $z=\infty$ (equivalently, $\tilde{z}=\infty$) from the perspective of the global AdS. Thus, we have
\begin{equation}
    \tilde{u}_{3} =\frac{1}{v_{\infty }^{2}}\frac{A(\tilde{z} /a)}{1+\tilde{z} /a} e^{B(\tilde{z} /a)} ,\quad \tilde{v}_{3} =-v_{\infty }^{2}\frac{1-\tilde{z} /a}{A(\tilde{z} /a)} e^{-B(\tilde{z} /a)},
\end{equation}
where
\begin{equation}
    v_{\infty } :=\lim _{z\rightarrow \infty } e^{B( z/a)} =\exp\left[\frac{\zeta_{0}}{\sqrt{\left( 1-\zeta_{0}^{2}\right)\left( 2-\zeta_{0}^{2}\right)}}\left( K\left(\frac{1}{2-\zeta_{0}^{2}}\right) -\Pi \left(\frac{1-\zeta_{0}^{2}}{2-\zeta_{0}^{2}} \middle|\frac{1}{2-\zeta_{0}^{2}}\right)\right)\right].
\end{equation}
Figure \ref{fig:Lorentzian_EOWs_in_BTZ} shows the result of the solutions.

Next, we will calculate the pseudo entropy using the HRT prescription. We take a half-line $x>0$ as the sub-region $A$. In the HRT prescription, the pseudo entropy at time $t$ is given by
\begin{equation}\label{eq:HRT-in-BTZ}
    S(t)=\max_{P_2\in Q_0}\frac{\ell(P_1,P_2)}{4G}.
\end{equation}
Here, $Q_0$ represents the set of points that are located on the EOW brane and have an $x$ coordinate of zero. $P_1$ is the endpoint of $A$ at time $t$ and $\ell(P_1,P_2)$ represents the space-like geodesic length between $P_1$ and $P_2$. Maximization is carried out only for points that are spacelike separated. The explicit form of $\ell(P_1,P_2)$ in the global coordinates is given by AdS-invariant function:
\begin{equation}\label{eq:geometric length of BTZ}
    \cosh\ell(P_1,P_2) =\sigma\left(P_1\middle |P_2\right):=\frac{\cos( \tau _{1} -\tau _{2}) -\sin \rho _{1}\sin \rho _{2}\cos( \th_{1} -\th_{2})}{\cos \rho _{1}\cos \rho _{2}}.
\end{equation}
In our setup, $P_1$ is $(\ep, t, 0)$ in the Poincar\'e coordinates, where $\ep$ is the UV cut-off scale of the CFT. Now, using eq.\eqref{eq:coord_change_from_poi_to_kru} and eq.\eqref{eq:coord_change_from_glo_to_kru}, we can calculate the entropy. Numerical results tell us that the point $P_2$, which maximizes $\sigma\left(P_1\middle |P_2\right)$, is located in the extended region. Then, we have
\begin{equation}
    \sigma =\frac{a}{\epsilon }\frac{-1+\tilde{u}\tilde{v} +e^{t/a}\tilde{u} -e^{-t/a}\tilde{v}}{1+\tilde{u}\tilde{v}},
\end{equation}
which becomes at late times $t\gg a$
\begin{equation}
   \sigma \simeq\frac{e^{t/a}}{2v_{\infty }^{2} \epsilon }\frac{A(\tilde{z} /a)}{\tilde{z} /a} e^{B(\tilde{z} /a)}.
\end{equation}
Therefore, we need to solve
\begin{equation}
    \frac{d\sigma }{d\tilde{z}} =\frac{e^{t/a} a}{2v_{\infty }^{2} \epsilon }\frac{e^{B(\tilde{z} /a)}}{\tilde{z}^{2} /a}( A^\prime (\tilde{z} /a) \cdot \tilde{z} /a+A(\tilde{z} /a) B^\prime (\tilde{z} /a) \cdot \tilde{z} /a-A(\tilde{z} /a)) =0.
\end{equation}
In fact, this equation can be solved analytically and the real solution $\tilde{z}$ that is greater than $z_0$ is
\begin{equation}
    \frac{\tilde{z}}{a} =\tilde{\zeta }_{0} :=\frac{\zeta _{0}}{\left( 1-\zeta _{0}^{2}\right)^{1/4}}.
\end{equation}
Finally, we obtain
\begin{equation}\label{eq:entropy_of_BTZ_v_less}
    S( t) =\frac{c}{6}\cosh^{-1} \sigma \simeq \frac{c}{6}\frac{1}{a} t+\frac{c}{6}\ln\frac{a}{\epsilon } +\log g,
\end{equation}
where
\begin{equation}
    \log g=\frac{c}{6}\ln\left(\frac{A(\tilde{\zeta }_{0}) e^{B(\tilde{\zeta }_{0})}}{v_{\infty }^{2}\tilde{\zeta }_{0}}\right).
\end{equation}
Since eq.\eqref{eq:entropy_of_BTZ_v_less}, excluding the $\log g$ term, is identical to the entropy of $\D \p=0$ if $a$ is fixed, we can identify $\log g$ --- which is determined only by $\D \p$ via $\zeta_0=z_0/a$ --- as the boundary entropy.

\begin{figure}[h]
 \begin{minipage}{0.3\hsize}
  \centering
  \includegraphics[width=1\linewidth]{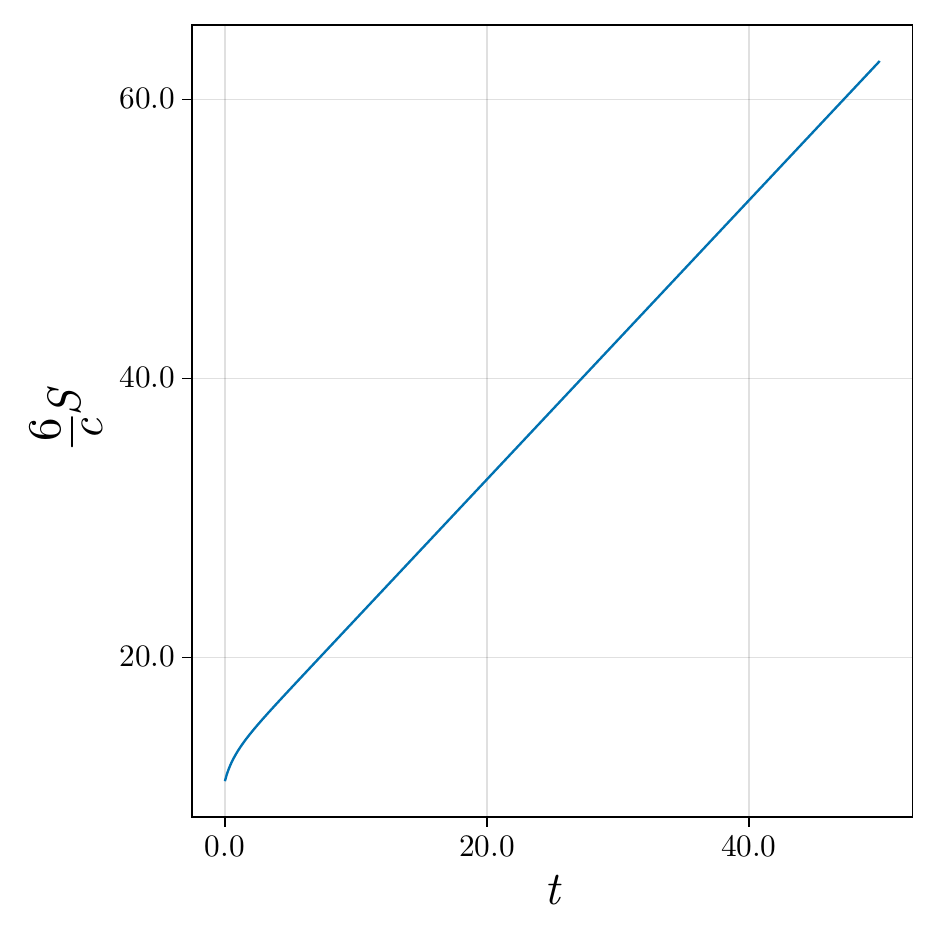}
 \end{minipage}
 \begin{minipage}{0.3\hsize}
  \centering
  \includegraphics[width=1\linewidth]{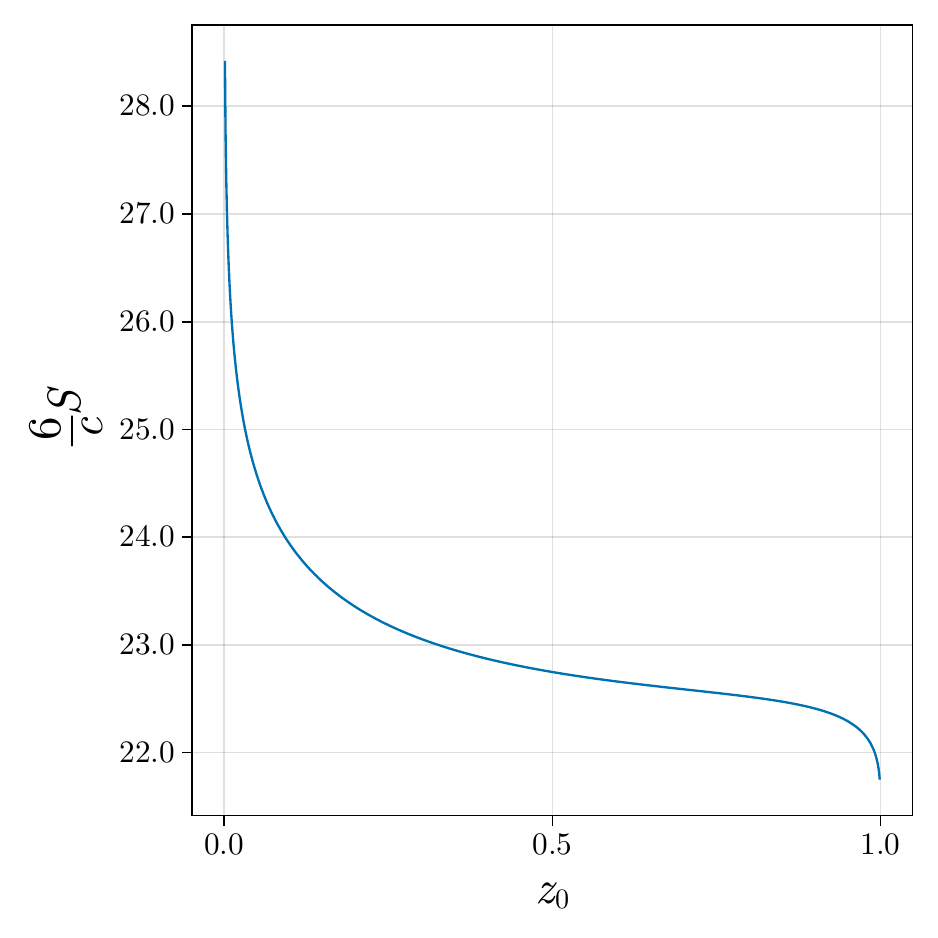}
 \end{minipage}
 \begin{minipage}{0.3\hsize}
  \centering
  \includegraphics[width=1\linewidth]{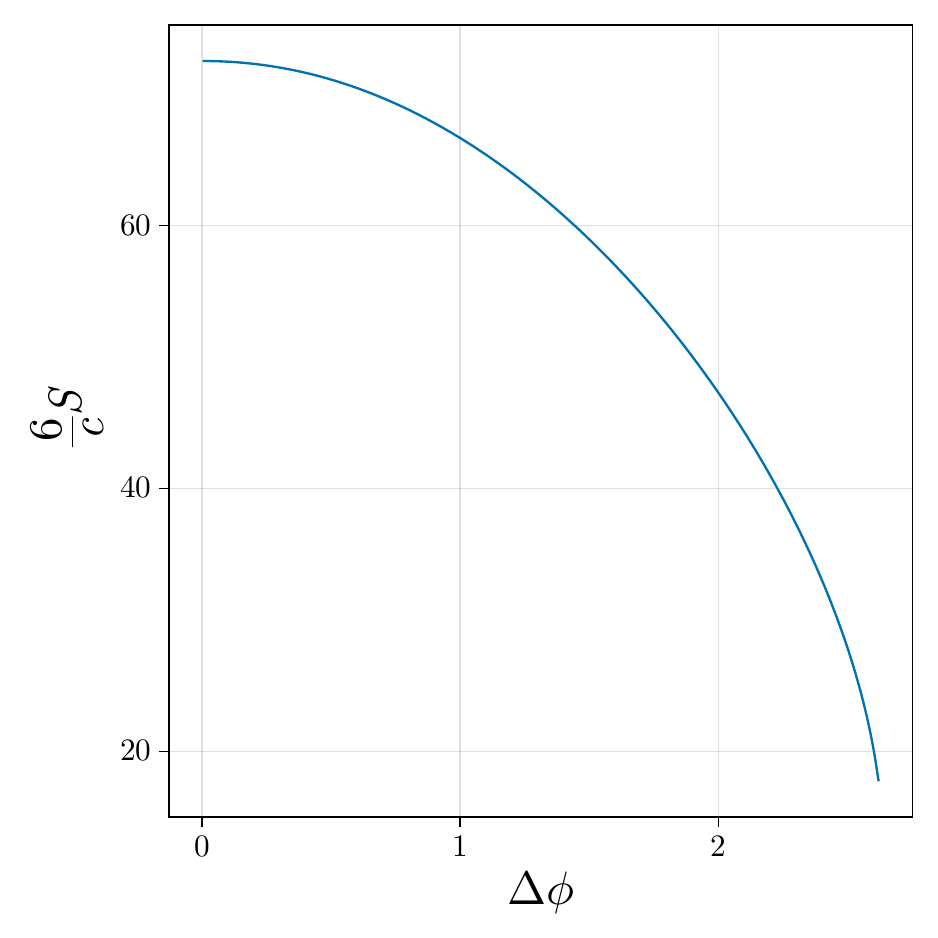}
 \end{minipage}
 \caption{Graphs of entropy behavior. Left: $t$-dependency, i.e., we fix $a=1$ and $z_0=0.5.$ Middle: $z_0$-dependency, i.e., fix $a=1$ and $t=10.$ Right: $\Delta\phi$-dependency, i.e., fix $\beta=1$ and $t=10.$ For all graphs, we set $\ep=10^{-5}.$}
 \label{fig:Extended_EOW_brane_in_Poincare}
\end{figure}

\subsection{Other profiles of EOW brane for   \texorpdfstring{$V(\phi) \neq 0$}{non-zero potential}} \label{sec:other_profile}

Before we move on, we would like to pause here to explore more solutions of EOW branes for $V(\phi) \neq 0$. 

\begin{figure}[h]
    \centering
    \includegraphics[width=0.5\linewidth]{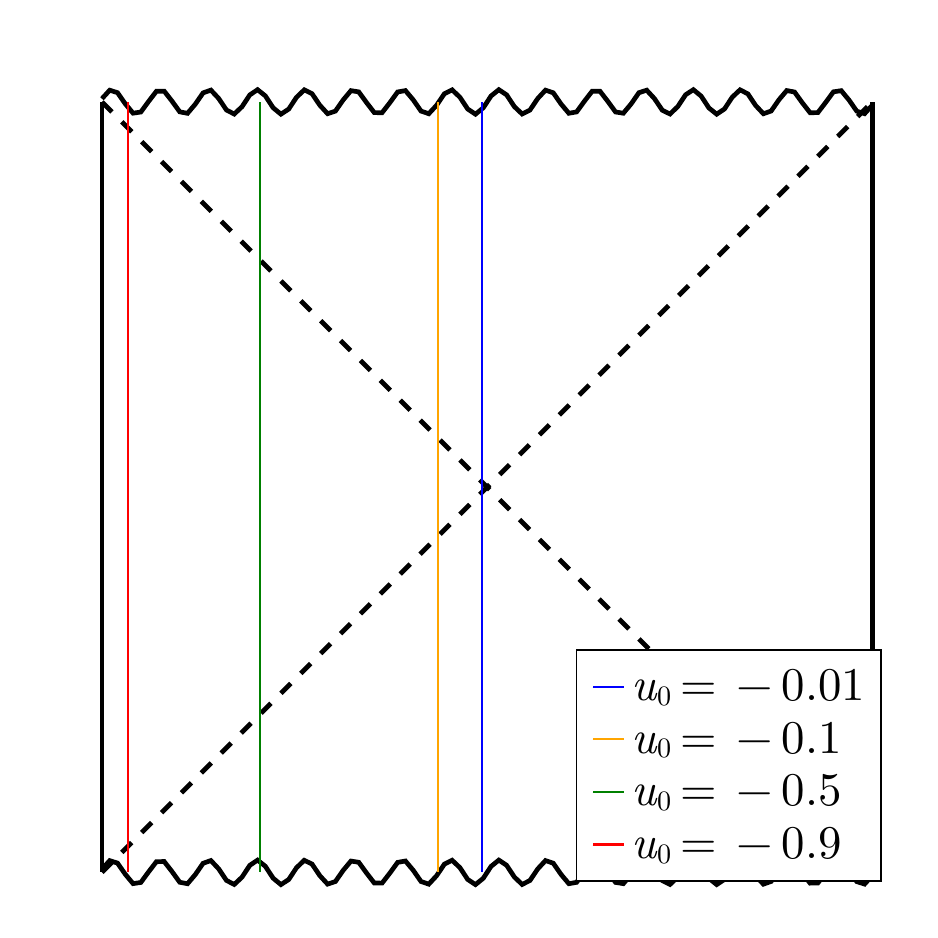}
    \caption{Brane configurations with $\phi^\prime=0.$ These correspond to constant tension branes without the brane scalar.}
    \label{fig:constant_brane_in_BTZ}
\end{figure}

First, the solution of $\phi^{\prime2}=0$, i.e. $\phi=$const. is
\begin{equation}\label{eq:constant tension sol in BTZ}
    \tan^{-1}v=\tan^{-1}u-2\tan^{-1}u_0,\quad V(\p)=\frac{2u_0}{1+u_0^2},\quad-1<u_0<1.
\end{equation}
where $u_0$ is a negative constant. Since $\phi^\prime=0$, this solution corresponds to the one without the brane scalar and the constant potential is the tension of the brane. In fact, the value of the potential is constrained within the range of $-1<V(\p)<1,$ and the shape of the brane becomes a straight line in the Kruskal coordinates like figure \ref{fig:constant_brane_in_BTZ}. These are the standard solutions in the pure gravity model of AdS/BCFT studied in \cite{Takayanagi:2011zk,Fujita:2011fp}.

\begin{figure}[h]
    \centering
    \includegraphics[width=0.5\linewidth]{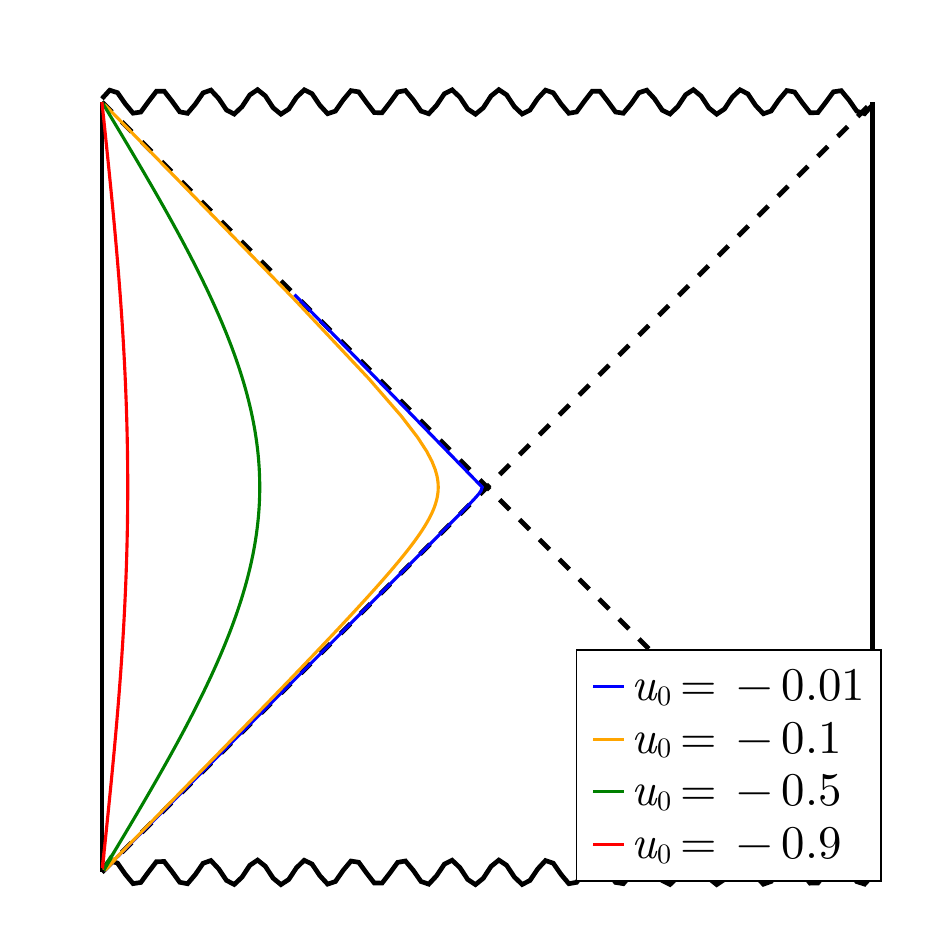}
    \caption{Brane profile of eq.\eqref{eq:constant_entropy_in_BTZ}. $uv=\rm{const.}$ means $z=\rm{const.}$}
    \label{fig:constant_brane_in_BTZ_2}
\end{figure}

There is another solution with the constant potential:
\begin{equation}\label{eq:constant_entropy_in_BTZ}
    vu=-u_0^2,\quad V(\p)=\frac{1+6u_0^2+u_0^4}{4u_0( 1+u_0^2 )},\quad-1<u_0<0.
\end{equation}
Unlike the solution in eq.\eqref{eq:constant tension sol in BTZ}, $\p^\prime$ is not zero, in fact,
\begin{equation}
    \phi^{\prime2}=-\frac{u_0}{u^2 \left(1+u_0^2\right)}>0,
\end{equation}
and $V(\p)\geq1.$ As we can see in figure \ref{fig:constant_brane_in_BTZ_2}, this brane does not extend across with the horizon.

We can analytically calculate the entanglement entropy. Since $u_2v_2=-u_0^2$, using eq.\eqref{eq:geometric length of BTZ}, we have
\begin{equation}
    \ell=\cosh^{-1}\left(1+2\frac{-\D u \D v}{\a}\right),\quad \a=\frac{2\ep(1-u_0^2)}{a+\ep}.
\end{equation}
To maximize $\ell$ with respect to $u_2$, we need to find the maximum value of $-\D u \D v$. We can easily find it as follows:
\begin{equation}
    -\D u \D v=-\sqrt{\frac{a-\epsilon }{a+\epsilon }}\left( e^{-t/a} v_{2} +e^{t/a}\frac{u_{0}^{2}}{v_{2}}\right) +\frac{a-\epsilon }{a+\epsilon } +u_{0}^{2} \leq \left( u_{0} +\sqrt{\frac{a-\epsilon }{a+\epsilon }}\right)^{2}.
\end{equation}
Namely, since $D$ is time-independent, the entanglement entropy is also a constant. The value of the entanglement entropy is
\begin{equation}
    S(t)=\frac{c}{6}\ln\left(\frac{( 1+u_{0})^{2}}{1-u_{0}^{2}}\frac{2a}{\epsilon }\right).
\end{equation}

This solution has unusual features compared to previous ones.
Since the EOW brane is located at $z=\rm{const.}$ in Lorentzian coordinates, through analytical continuation, we find that it remains constant, which means the EOW brane does not end on the asymptotic boundary even in Euclidean metric and the description of the path integral from some boundary state fails\footnote{We note that a similar situation, where the brane does not intersect the asymptotic boundary in the Euclidean signature, was also discussed in \cite{Akal:2020wfl}.}. The gravitational force toward the horizon direction is cancelled by the effect of time-dependent localized scalar field. As the left side and right side of the BTZ black hole are connected through the Euclidean region, the EOW brane exists also in the right side.

Finally let us mention possible CFT interpretations of this EOW brane. One possibility is that it is a certain thermofield double-like state without the IR degree of freedom. This may look similar to the prescription of the $T\bar{T}$ deformation\cite{McGough:2016lol}, where the finite radius boundary conceals the UV degree of freedom, the EOW brane is hovering. Unlike the $T\bar{T}$ deformation, however, the EOW brane is subject to the Neumann boundary condition. Another possibility is the final state projection \cite{Horowitz:2003he,Akal:2021dqt}. 

In order to think of the projection proposal, we consider a geometry which may be corresponding to the final state projection at a finite time:
\begin{equation}
    ( u+\alpha )( v-\alpha ) =-u_{0}^{2}.
\end{equation}

By the analytical continuation, we have the corresponding Euclidean brane:
\begin{equation}
    \sqrt{\frac{a-z}{a+z}} =\alpha \cos( \tau /a) +\sqrt{u_{0}^{2} -\alpha ^{2}\sin^{2}( \tau /a)}.
\end{equation}
This implies that $\beta = 2\pi a.$

For another topic, there exists a configuration that does not have the well-defined Euclidean brane. For instance, $z=z_0\exp(-t^2/2\sigma)$ poses no problem at least in Lorentzian space if we choose the parameters not to make the EOW brane space-like. On the other hand, in the Euclidean space, this EOW brane follows $z=z_0\exp(\tau^2/\sigma),$ which obviously encounters difficulties. Namely, this brane penetrates $z=a$ in the Euclidean space and has no periodicity.

\subsection{Complete analysis of holographic pseudo entropy at critical point\texorpdfstring{: $\Delta \phi=\Delta \phi_*$}{}}

\begin{figure}[h]
    \centering
    \includegraphics[width=0.5\linewidth]{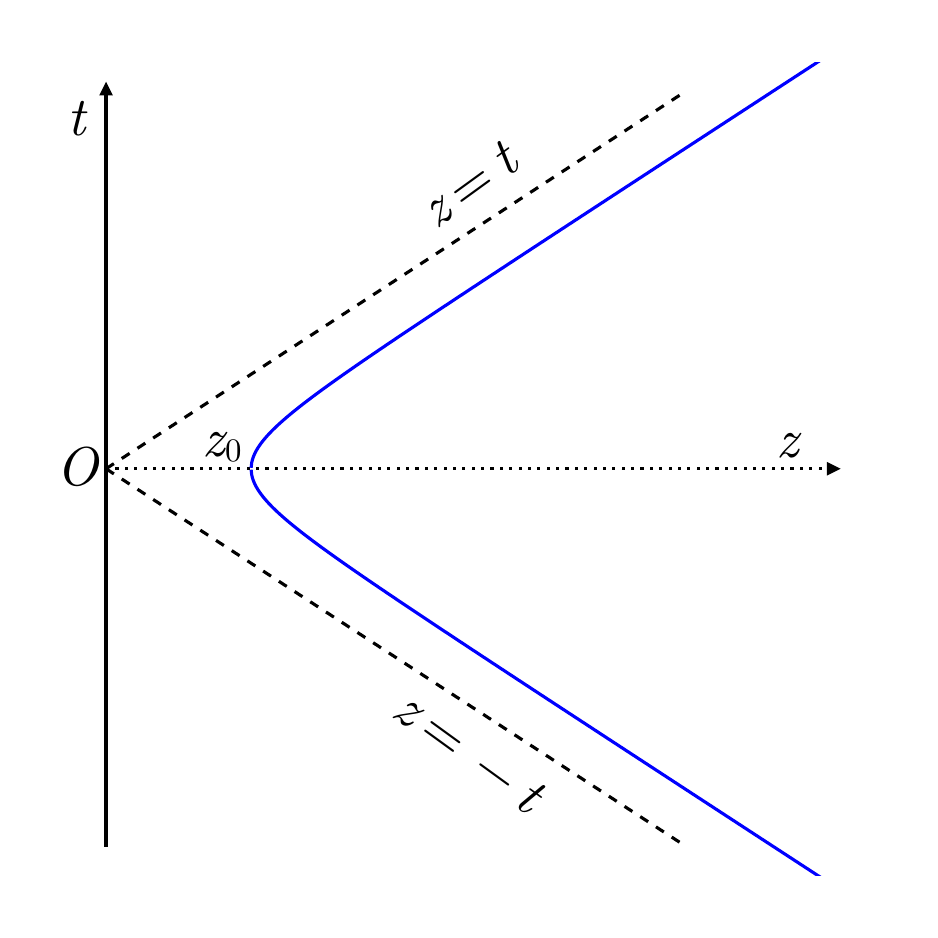}
    \caption{The blue line represents an EOW brane with $V(\p)=0$ in the Poincar\'e patch. At late times, this EOW brane moves at a speed approaching that of light.}
    \label{fig:EOW_in_poincare_patch}
\end{figure}

For the critical phase $\D \p=\D \p_*$, the corresponding geometry is the Poincar\'e AdS
\begin{equation}\label{eq:Poincare_metric}
    ds^2=\frac{dz^2-dt^2+dx^2}{z^2},
\end{equation}
and the EOW brane is determined by the following differential equation
\begin{equation}
    1-\dot{z}^{2} =\frac{z_{0}^{4}}{z^{4}},
\end{equation}
with the initial condition $z(t=0)=z_0$. The solution of this equation can be described by using the hypergeometric function as follows:
\begin{equation}
    | t| =z_{0} f( z/z_{0}),\quad f(x):=x\cdot {}_{2} F_{1}\left( -\frac{1}{4} ,\frac{1}{2} ;\frac{3}{4} ;\frac{1}{x^4}\right)-\a,\quad \a=\frac{\sqrt{\pi } \Gamma ( 3/4)}{\Gamma ( 1/4)}.
\end{equation}
As we can see in figure \ref{fig:EOW_in_poincare_patch}, this EOW brane will fall at a light speed at late time:
\begin{equation}
    z\simeq t+\alpha z_{0}.
\end{equation}
This implies that the EOW brane can be extended to the global AdS where the Poicar\'e patch is embedded.

The Poincar\'e AdS is embedded into the global AdS via $\mathbb{R}^{2,2}$ as follows:
\begin{gather}\label{eq:poincare_embed}
    X_0=\frac{z}{2z_{0}}\left( 1+\frac{1}{z^{2}}\left( x^{2} -t^{2} +z_{0}^{2}\right)\right)=\frac{\cos\tau}{\cos\rho},\\
    X_1=\frac{x}{z}=\tan\r\sin\p,\\
    X_2=\frac{z}{2z_{0}}\left( 1+\frac{1}{z^{2}}\left( x^{2} -t^{2} -z_{0}^{2}\right)\right)=\tan\r\cos\p,\\
    X_3=\frac{t}{z}=\frac{\sin\tau}{\cos\rho}.
\end{gather}
Thus, using the global coordinates, the Poincar\'e coordinates are represented as
\begin{equation}\label{eq:original_poincare}
    \frac{z}{z_0}=\frac{\cos \rho }{\cos \tau -\sin \rho \cos \theta } ,\ \frac{t}{z_0}=\frac{\sin \tau }{\cos \tau -\sin \rho \cos \theta } ,\ \frac{x}{z_0}=\frac{\sin \rho \sin \theta }{\cos \tau -\sin \rho \cos \theta }.
\end{equation}
The Poincar\'e AdS corresponds to the region where $\cos\t-\sin\r\cos\th>0.$  Now we define new coordinates for the region out of the Poincar\'e patch, namely $\cos\t-\sin\r\cos\th<0, $ as the following:
\begin{equation}\label{eq:new_poincare}
    \frac{\tilde{z}}{z_0}=\frac{\cos \rho }{\cos \tau -\sin \rho \cos \theta } ,\ \frac{\tilde{t}}{z_0}=\frac{\sin \tau }{\cos \tau -\sin \rho \cos \theta } ,\ \frac{\tilde{x}}{z_0}=\frac{\sin \rho \sin \theta }{\cos \tau -\sin \rho \cos \theta }.
\end{equation}
Note that $\tilde{z}$ is negative. Obviously, since eq.\eqref{eq:original_poincare} and eq.\eqref{eq:new_poincare} have the same form, new coordinates have the same metric with the original Poincar\'e metric (eq.\eqref{eq:Poincare_metric})and are embedded into the global AdS in the same way (eq.\eqref{eq:poincare_embed}). Therefore, the differential equation of the EOW brane outside of the Poincar\'e patch is also
\begin{equation}
    1-\dot{\tilde{z}}^{2} =\frac{z_{0}^{4}}{\tilde{z}^{4}}.
\end{equation}
Thus, we can find the solution
\begin{equation}
    | \tilde{t} -\tilde{t}_{0}| =z_{0} f( -\tilde{z} /z_{0}),\quad \tilde{t}_0=-2\a z_0,  \label{soldima}
\end{equation}
by imposing $\tilde{z}\simeq \tilde{t}+\alpha z_0$ for $\tilde{t}\to-\infty$.

\begin{figure}[h]
 \begin{minipage}{0.5\hsize}
  \centering
  \includegraphics[width=0.5\linewidth]{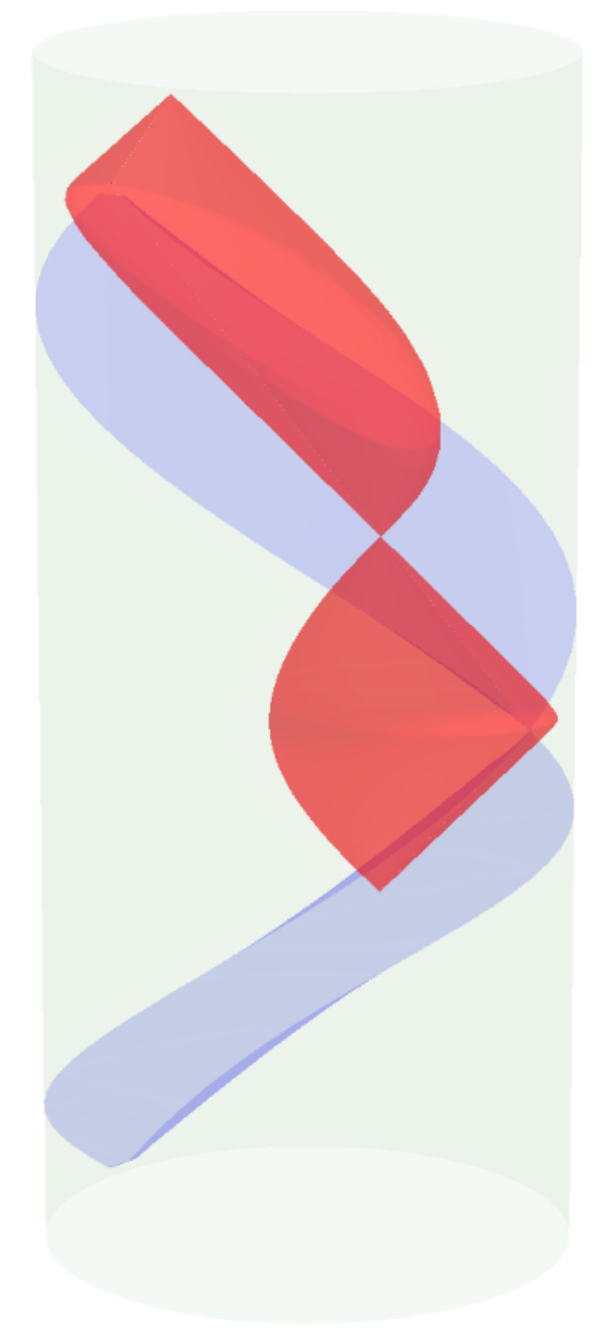}
 \end{minipage}
 \begin{minipage}{0.5\hsize}
  \centering
  \includegraphics[width=0.7\linewidth]{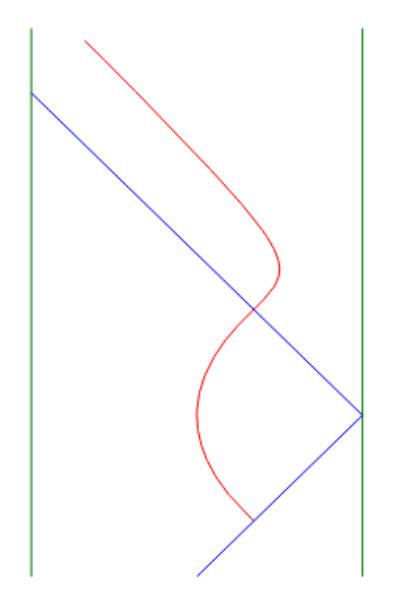}
 \end{minipage}
 \caption{Left: The EOW brane embedded in the global AdS is represented as a red surface. The original Poincar\'e AdS is the region surrounded by two blue Poincar\'e horizons. Right: The cross-sectional view of a cylinder cut in half.}
 \label{fig:Extended_EOW_brane_in_PoincareA}
\end{figure}

\begin{figure}[h]
  \centering
  \includegraphics[width=0.5\linewidth]{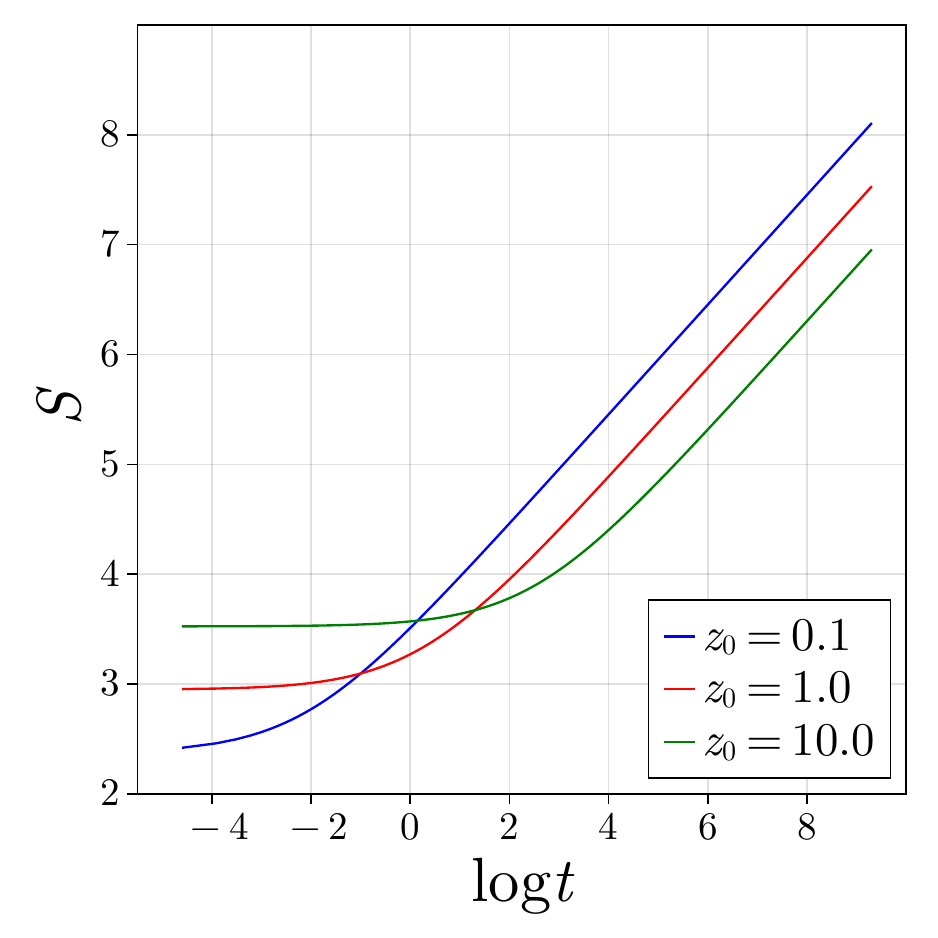}
 \caption{The graph of the entanglement entropy $S$ vs time $t$. We set the Newtonian constant $G=1$ and UV cut-off $\epsilon=10^{-5}.$}
 \label{fig:entropy_of_poincare}
\end{figure}

Figure \ref{fig:Extended_EOW_brane_in_PoincareA} shows the extended EOW brane. In a similar way in the small $\D \p$ phase, we can evaluate the entanglement entropy using HRT surface $\G$. We set subregion $A$ of a time slice of CFT, as a half line $x>0.$ Then, $\G$ ends $(z,t,x)=(\epsilon, t, 0)$ where $\epsilon$ is UV cut-off. Another endpoint of $\G$ is on the red line in the right side of figure \ref{fig:Extended_EOW_brane_in_PoincareA}. Therefore, to calculate the entanglement entropy, we need to find maximum value of the length of $\G$ for a fixed $t$.

Figure \ref{fig:entropy_of_poincare} tells us that for any $z_0$, the entanglement entropy grows logarithmically at late time.
\begin{equation}
    S\simeq\frac{1}{2G}\log\frac{t}{\ep}-\frac{1}{4G}\log\frac{z_0}{\ep}=\frac{c}{3}\log\frac{t}{\ep}-\frac{c}{6}\log\frac{z_0}{\ep}.
    \label{logteva}
\end{equation}
It is worth noting that taking the part of the spacetime beyond the Poincar\'e horizon into account is crucial to get the $\frac{c}{3}\log t$ term \cite{Shimaji:2018czt,Caputa:2019avh}. Typically, both the inside-Poincar\'e piece and the outside-of-Poincar\'e piece of the HRT surface contribute to the $\log t$ term and together give $\frac{c}{3}\log t$. The same behavior is also observed in the study of holographic joining quench \cite{Shimaji:2018czt,Caputa:2019avh}. 

Combining the results obtained so far, we have worked out the behavior of the entanglement phase transition shown in (\ref{EPTb}).

\subsection{Higher dimension}
Now we move on to the analysis of holographic entanglement phase transition in higher dimensions. Since the black hole and thermal AdS phase are very similar to the previous three dimensional setup, here we concentrate on the critical point at $\Delta\phi=\phi_*$. 

\begin{figure}[h]
 \begin{minipage}{0.3\hsize}
  \centering
  \includegraphics[width=0.8\linewidth]{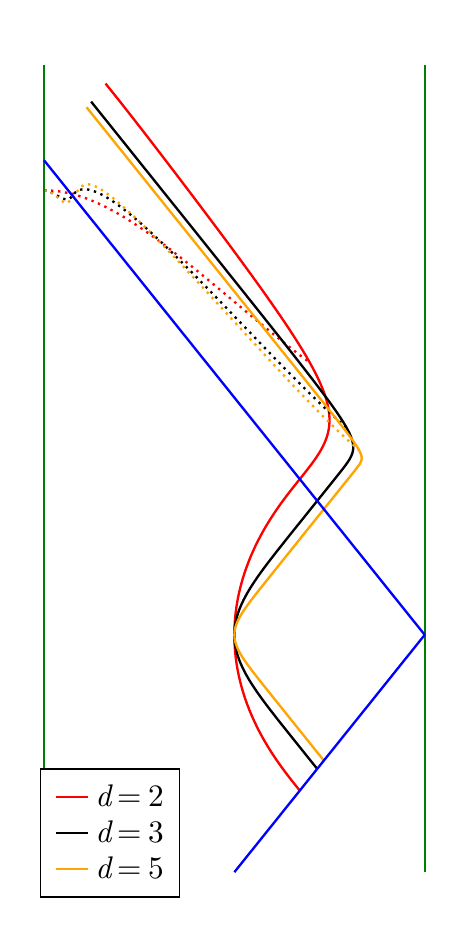}
 \end{minipage}
 \begin{minipage}{0.7\hsize}
  \centering
  \includegraphics[width=\linewidth]{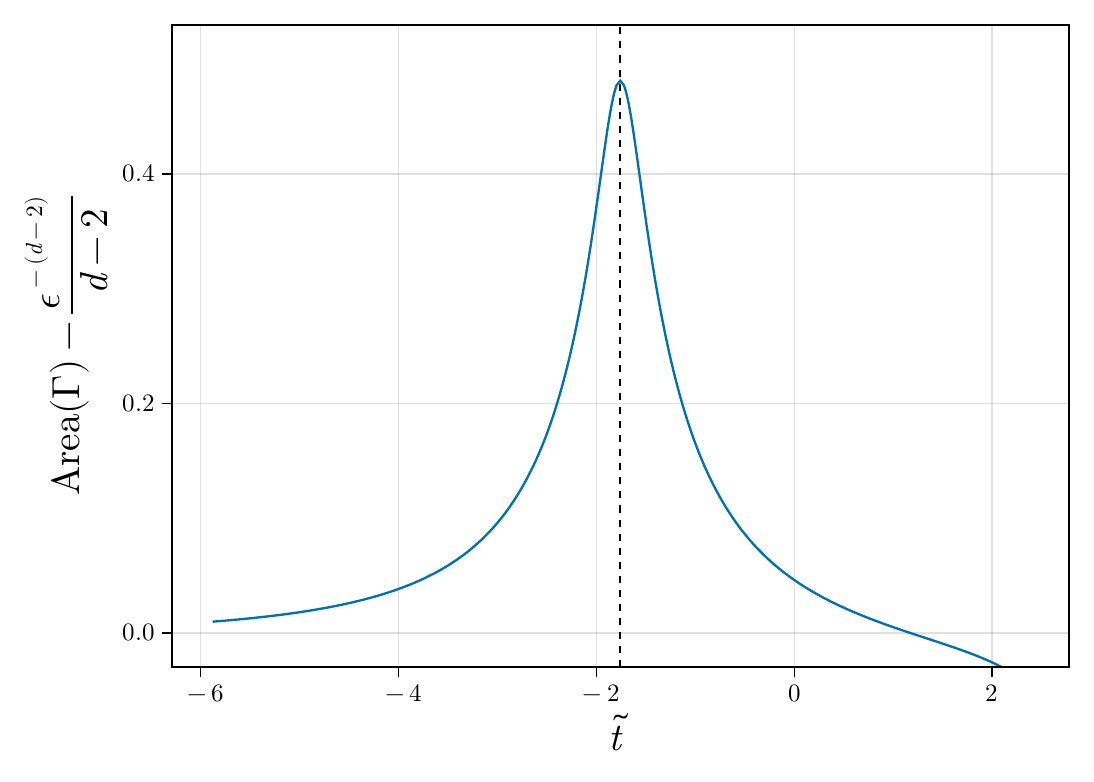}
 \end{minipage}
 \caption{Left: Solid lines and dotted lines represent EOW branes and HRT surface respectively at time $t$. Right: The horizontal axis is time coordinate of the region out of Poincar\'e patch. A dotted line corresponds to $\tilde{t}=\tilde{t}_0$. The vertical axis shows the area of the HRT surface in $d=4$ connecting $(\e, t)$ and $(\tilde{z}(\tilde{t}),\tilde{t})$. We set $t=10$, $z_0=1$ and $\e=10^{-5}$ in both graphs.}
 \label{fig:higher_dim_eow_branes}
\end{figure}

In a similar way to $d=2$, we can analyze the Poincar\'e case in higher dimension, in which the metric is
\begin{equation}
    ds^2=\frac{dz^2-dt^2+dx_1^2+dx_2^2+dx_{d-1}^2}{z^2},\quad (d\geq3).
\end{equation}
The differential equation of the EOW brane in a $d$-dimensional Poincar\'e space is
\begin{equation}
    1-\dot{z}^{2} =\left(\frac{z_{0}}{z}\right)^{4( d-1)},\quad z(t=0)=z_0,
    \label{highersceoma}
\end{equation}
and its solution is
\begin{equation}
    |t|=z_{0} f_{d}( z/z_{0}), \label{soldimb}
\end{equation}
where
\begin{equation}
    f_d(x):=x\cdot {}_{2} F_{1}\left(\frac{1}{2} ,\frac{-1}{4( d-1)} ;\frac{4d-5}{4( d-1)} ;\frac{1}{x^{4( d-1)}}\right)-\a_d,\quad \a_d:=\frac{\sqrt{\pi } \Gamma \left(\frac{4d-5}{4( d-1)}\right)}{\Gamma \left(\frac{2d-3}{4( d-1)}\right)}.
\end{equation}
For higher dimensions, we can also define new coordinates for the region outside of the Poincar\'e patch in a similar way. Then, outside of the horizon, the EOW brane follows
\begin{equation}
    | \tilde{t} -\tilde{t}_{0}| =z_{0} f_{d}( -\tilde{z} /z_{0}),\quad \tilde{t}_{0}=-2\alpha_d z_0.
\end{equation}
Figure \ref{fig:higher_dim_eow_branes} shows behaviors of EOW branes.

Next, we consider a subregion $A$ of a CFT time slice at a certain time $t$, defined as $A=\{(\e, t, x_1,\dots,x_{d-1})|x_1>0\}.$ Hence, the HRT surface $\G = \{(z,t(z),0,x_2,\dots,x_{d-1})\}$ ends on $\partial A$ and on the EOW brane. The HRT surface maximizes the area of $\G$,
\begin{equation}\label{eq:Area_function}
    \operatorname{Area} \Gamma =\int_\G \frac{\sqrt{1-( dt/dz)^{2}}}{z^{d-1}} dzdx_{2} \cdots dx_{d-1} =V_{d-2}\int_{i}^f \frac{\sqrt{dz^{2} -dt^{2}}}{z^{d-1}},
\end{equation}
where label $i$ and $f$ represent endpoints of $\G$. The HRT surface holds the Euler Lagrange equation with respect to the area function
\begin{equation}\label{eq:HRT_diff_eq}
\dot{z}^{2} =1+\left(\frac{t_{*}}{z}\right)^{2( d-1)},
\end{equation}
where $t_*$ is some positive constant and a solution of this equation is
\begin{equation}
    \frac{t_{f}}{t_*} -\frac{t_{i}}{t_*} =B_{d}\left( \frac{z_{f}}{t_{*}}\right) -B_{d}\left(\frac{z_{i}} {t_{*}}\right),\quad
    B_{d}( x) :=x\cdot {}_{2} F_{1}\left(\frac{1}{2} ,\frac{-1}{2( d-1)} ;\frac{2d-3}{2( d-1)} ;\frac{-1}{x^{2( d-1)}}\right).
\end{equation}
Furthermore, by substituting eq.\eqref{eq:HRT_diff_eq} into eq.\eqref{eq:Area_function}, we have
\begin{equation}
    \operatorname{Area}(\G)=V_{d-2}\int_{z_i}^{z_f}\frac{dz}{z^{d-1}\sqrt{1+( z/t_{*})^{2( d-1)}}}=\frac{V_{d-2}}{t_*^{d-2}}\left(A_d\left(\frac{z_f}{t_*}\right)-A_d\left(\frac{z_f}{t_*}\right)\right),
\end{equation}
where
\begin{equation}
    A_{d}( x) :=\frac{\sqrt{1+x^{2( d-1)}}}{( d-2) x^{d-2}}\left[ {}_{2} F_{1}\left( 1,\frac{d-2}{2( d-1)} ;\frac{2d-3}{2( d-1)} ;\frac{-1}{x^{2( d-1)}}\right) -1\right].
\end{equation}
Since the EOW brane locates outside of the Poincar\'e horizon, we need to modify the area of the EOW brane as follows:
\begin{align}
    \operatorname{Area}( \Gamma ) &=\frac{V_{d-2}}{t_{*}^{d-2}}\left( A_{d}( \infty ) -A_{d}\left(\frac{\epsilon }{t_{*}}\right)\right) +\frac{V_{d-2}}{t_{*}^{d-2}}\left( A_{d}( \infty ) -A_{d}\left( -\frac{\tilde{z}_{i}}{t_{*}}\right)\right)\\
    &=-\frac{V_{d-2}}{t_{*}^{d-2}}\left( A_d\left(\frac{\epsilon }{t_{*}}\right) +A_d\left( -\frac{\tilde{z}_{i}}{t_{*}}\right)\right)\label{eq:area_of_HRT_in_higher_dim}.
\end{align}
Here we use $A_d(\inf)=0$ and $\tilde{z}_i$ to represent the coordinates of the endpoint of $\G$ on the EOW brane. To compute the area, we need to know $t_*$. This is determined via
\begin{equation}
    B_{d}\left(\frac{\epsilon }{t_{*}}\right) +B_{d}\left( -\frac{\tilde{z}_{i}}{t_{*}}\right) =\frac{t}{t_{*}} -\frac{\tilde{t}_{i}}{t_{*}}.
\end{equation}

We can determine the position of the endpoint of $\G$ on the EOW brane numerically. The right side of figure \ref{fig:higher_dim_eow_branes} tells us that the HRT surface ends at $(\tilde{z}_i,\tilde{t}_i)=(-z_0,\tilde{t}_0).$

\begin{figure}[h]
  \centering
  \includegraphics[width=0.7\linewidth]{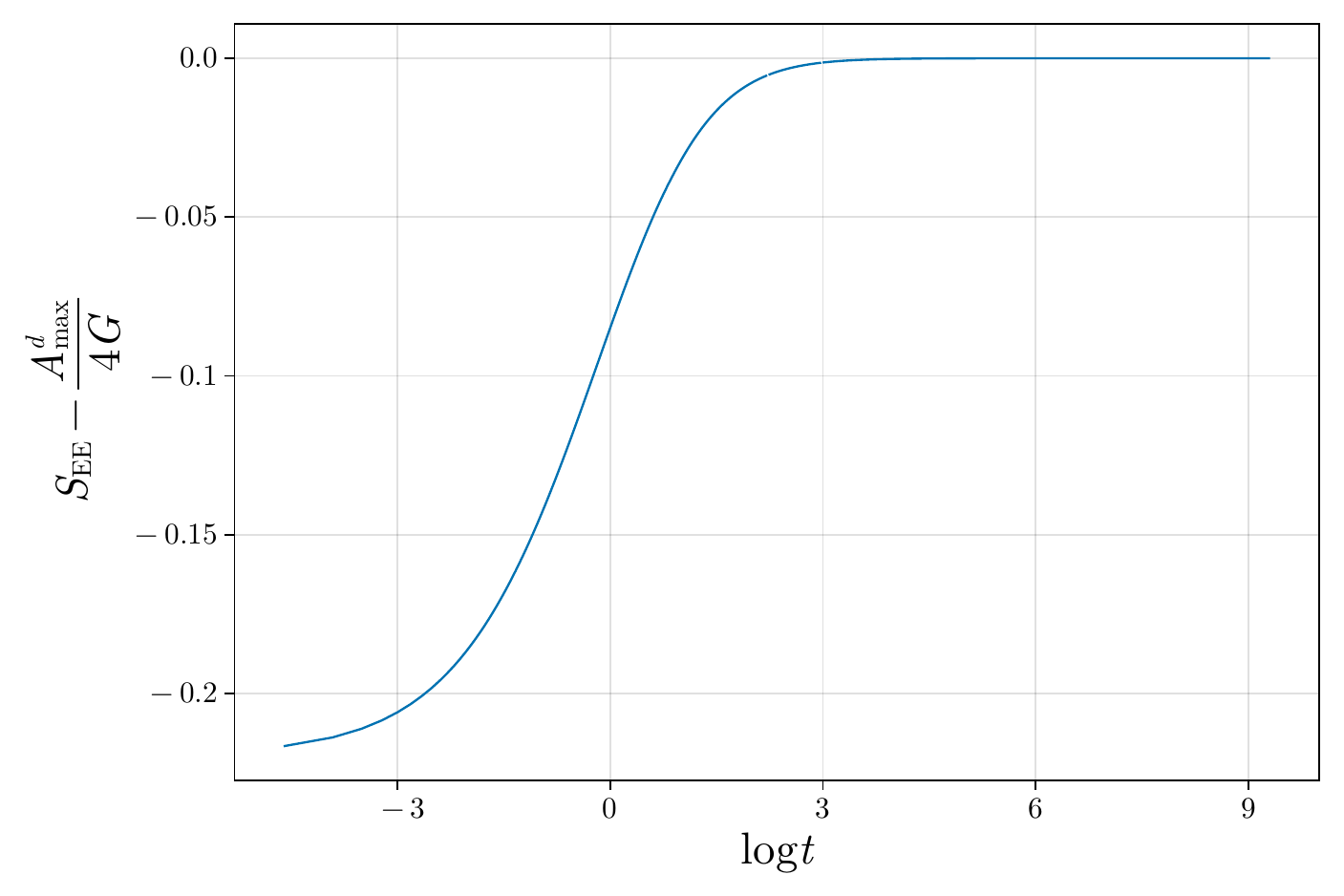}
 \caption{The graph of the entanglement entropy $S_{\rm{EE}}$ vs time $t$ in $d=4$ dimension. We set $G=1$, $z_0=1$ and $\epsilon=10^{-5}$.}
 \label{fig:entropy_of_poincare_in_d_4}
\end{figure}

Furthermore, at late time, $t_*$ behaves as
\begin{equation}
    t_{*} \simeq \frac{\sqrt{\pi }}{2\Gamma \left(\frac{d}{2d-2}\right) \Gamma \left(\frac{2d-3}{2d-2}\right)} t.
\end{equation}
Hence, both $\frac{\e}{t_*}$ and $-\frac{\tilde{z}_i}{t_*}$ decrease in the time evolution. Now, for small $x\ll 1$, $A_d(x)$ behaves as
\begin{equation}
    A_d(x)\simeq-\frac{1}{d-2} x^{-( d-2)}.
\end{equation}
Therefore, using eq.\eqref{eq:area_of_HRT_in_higher_dim}, we have
\begin{equation}
    \operatorname{Area}( \Gamma ) \simeq A_{\rm{max}}^d:=\frac{V_{d-2}}{d-2}\left(\frac{1}{z_{0}^{d-2}} +\frac{1}{\epsilon ^{d-2}}\right),
\end{equation}
namely, at late time, the leading term of the entanglement entropy is constant. Numerical calculation supports this result. See figure \ref{fig:entropy_of_poincare_in_d_4}. In this way, in the higher dimensional case, there is no logarithmic evolution of $S_A$ at the critical point, as opposed to the AdS$_3$ case.

\subsection{Null energy condition and the brane profile}

It is useful to reconsider our EOW profile in terms of null energy condition. Consider the Poincar\'e AdS$_{d+1}$: $ds^2=z^{-2}(dz^2-dt^2+\sum_{i=1}^{d-1}dx_i^2)$ and place a EOW brane with the profile $z=z(t)$. The physical region is taken to the boundary side of the EOW brane $z\leq z(t)$.
In this case, the extrinsic curvature reads
\ba
K_{tt}-h_{tt}K=-\frac{(d-1)\s{1-\dot{z}^2}}{z^2},\ \ \ \ 
K_{x_ix_i}-h_{xx}K=\frac{(d-1)(1-\dot{z}^2)-z\ddot{z}}{z^2(1-\dot{z}^2)^{3/2}}.
\ea

The null vector on this EOW brane is taken to be 
$(N^t,N^{x_1},N^{x_2},\ddd)=(1,\pm\s{1-\dot{z}^2},0,\ddd)$. Then the null energy condition reads
\ba
N^a N^b (K_{ab}-h_{ab}K)=-\frac{\ddot{z}}{z\s{1-\dot{z}^2}}\geq 0. \label{NECa}
\ea
This constrains the profile of the EOW brane. 

In our model of EOW brane with a localized scalar, the boundary energy stress tensor $T^{(Q)}_{ab}$ on the EOW brane, which is related to the extrinsic curvature via the Neumann boundary condition
$K_{ab}-h_{ab}K=T^{(Q)}_{ab}$, reads 
\ba
&& T^{(Q)}_{tt}=\dot{\phi}^2+\frac{V}{z^2}(1-\dot{z}^2),\no
&& T^{(Q)}_{x_ix_i}=\frac{\dot{\phi}^2}{1-\dot{z}^2}-\frac{V}{z^2}.
\ea
The null energy condition reads 
\ba
N^a N^b T^{(Q)}_{ab}=2\dot{\phi}^2\geq 0. \label{NECb}
\ea

The profile of our previous EOW brane solutions with the scalar (\ref{soldima}) and (\ref{soldimb}) violates this condition (\ref{NECa}). One quick way to see this is to note that the scalar field takes imaginary values in the Lorentzian solution and thus $\dot{\phi}^2<0$.
Since it has an imaginary value of scalar field, (\ref{NECb}) is also violated as expected. Thus the EOW brane solution which shows the entanglement phase transition violates the null energy condition. This is not surprising because, in order to realize this entanglement phase transition, we need a dissipative effect, which may lead to a non-unitary evolution, to suppress the entanglement growth. 

It might also be intriguing to consider the energy conservation in the gravity dual. By taking a derivative of $K_{ab}-h_{ab}K=T^{(Q)}_{ab}$, we obtain
\ba
\nabla^a (K_{ab}-h_{ab}K)=\nabla^a T^{(Q)}_{ab}.
\ea
It is straightforward from the differential geometry to see $\nabla^a (K_{ab}-h_{ab}K)=R_{nb}$, where $n$ is in the direction normal to the EOW brane. If the bulk dynamics is described by a pure Einstein equation (in any dimensions), it is clear that $R_{nb}\propto g_{nb}=0$. Thus the boundary energy stress tensor is conserved $\nabla^a T^{(Q)}_{ab}=0$, which means that the energy on EOW brane is preserved and there is no energy flux from the bulk to the brane. This is true in all setups we consider in this paper.

If there are matter fields in the bulk, they may lead to the bulk energy flux $T^{\rm (bulk)}_{nb}$, which is proportional to $R_{nb}$ via the Einstein equation. In this case we have $\nabla^a T^{(Q)}_{ab}=R_{nb}\propto T^{\rm (bulk)}_{nb}$ and thus there is energy flow between the EOW brane and the bulk.

\subsection{Critical behavior from a simple dissipating model}

Before we move on, we would like to point out that the critical logarithmic time evolution of entanglement entropy can be found in a simple dissipating model in a two dimensional CFT.

A simple modeling of quantum quenches is to use the BCFT \cite{Calabrese:2005in}. The time evolution under quantum quenches in a two dimensional CFT can be described by 
\ba
|\Psi(t)\lb\propto e^{-itH}e^{-\frac{\beta}{4}H}|B\lb,
\ea
where $|B\lb$ is the boundary state.  At late time, this state looks like a thermal state at the temperature $1/\beta$ as long as we examine coarse-grained observables. The evolution of entanglement entropy $S_A$, where the subsystem $A$ is a half space, can be found from a standard replica method calculation in the CFT, as follows \cite{Hartman:2013qma}.
\ba
S_A=\frac{c}{6}\log \left[\frac{\beta}{2\pi \ep}\cosh\left(\frac{2\pi t}{\beta}\right)\right],
\label{QEE}
\ea
up to a constant term. At late time, this grows linearly 
as $S_A\simeq \frac{\pi c}{3\beta}t$.

We can take into account the dissipating effect by introducing a damping factor $e^{-\gamma t H}$ under the time evolution, where 
$\gamma$ is the parameter which controls the amount of the dissipation. This leads to the quantum state 
\ba
|\Psi_{d}(t)\lb\propto e^{-itH}e^{-\gamma t H}e^{-\frac{\beta}{4}H}|B\lb.
\ea
To calculate the evolution of entanglement entropy, we can replace $\beta$ in (\ref{QEE}) with $\beta+4\gamma t$, which leads to 
\ba
S_A=\frac{c}{6}\log\left[\frac{\beta+4\gamma t}{2\pi\ep}\cosh\left(\frac{2\pi t}{4\gamma t+\beta}\right)\right].
\ea
At late time, this behaves like
\ba
S_A\simeq \frac{c}{6}\log\left[\frac{2\gamma t}{\pi \ep}\cosh\left(\frac{\pi}{2\gamma}\right)\right].
\ea
This shows the critical logarithmic behavior. Note that the coefficient of the logarithmic evolution gets halved when compared with the result in our holographic model (\ref{logteva}). Also note that, when such a damping factor is involved, the logarithmic time evolution is generic in two dimensional CFTs even if the starting point is not the boundary state \cite{Kusuki:2023ckn}.

\subsection{Comments on the analysis from the CFT side}

We have so far realized an entanglement phase transition in AdS/BCFT by introducing a brane localized scalar field on the gravity side. It is a natural question to consider the corresponding setup on the CFT side, i.e. a BCFT defined on a strip where the boundary conditions on the two boundaries are related by a marginal transformation. 

In appendix \ref{app:CFT_analysis}, we consider a slightly different but parallel setup and compute the pseudo R\'enyi entropy at the leading order of the marginal perturbation purely in the CFT language. Denoting the perturbation parameter as $\lambda$, our holographic computation implies that the leading order correction of the pseudo entanglement entropy is at order $\lambda^2$ and gives a negative linear $t$ contribution. In the CFT computation summarized in appendix \ref{app:CFT_analysis}, we find that, for even $n$, the $\lambda^2$ order correction of the $n$-th pseudo R\'enyi entropy is zero, while for odd $n$, it turns out to give a positive linear $t$ contribution. 

Although the quantities computed on the two sides are different and we can not compare them with each other, it would be an intriguing future question to understand how the analyses on the two sides are related with each other.


\section{Gauge fields on the brane}\label{sec:gauge_field}
As a natural generalization of matter fields localized on the EOW brane, we can also consider the gauge field instead of scalar field. Again, we aim to construct an AdS/BCFT solution where the BCFT is defined on a strip. The action we consider is the Einstein-Hilbert action with  gauge fields on the brane (we will work in Euclidean signature and set $V=0$ for simplicity)
\begin{align}
    S&=S_{\rm EH}+S_{\rm brane},\nn\\
    S_{\rm EH}&=-\frac{1}{16\pi G}\int d^{d+1}x \sqrt{g}(R+d(d-1)),\nn\\
    S_{\rm brane}&=-\frac{1}{8\pi G}\int \sqrt{h}\left(K-\frac{1}{4}h^{cd}h^{ef}F_{ce}F_{df}\right).
\end{align}
Taking a variation with respect to the induced metric, we obtain the equations of motion (or Neumann boundary condition) for the EOW brane
\be
K_{ab}-Kh_{ab}=-\frac{1}{4}h_{ab}h^{ce}h^{df}F_{cd}F_{ef}+F_{ac}F_{bd}h^{cd}.\label{bEOM}
\ee

\subsection{EOW brane with gauge field in \texorpdfstring{Poincar\'{e}}{Poincare} AdS}
 We start with the analysis of EOW brane placed in the $d+1$ dimensional Poincar\'{e} vacuum AdS, given by 
 \be
ds^2=\frac{dz^2+d\tau^2+\sum_{i=1}^{d-1}dx_i^2}{z^2}.
 \ee
 We specify the profile of the EOW brane by $z=z(\tau)$, assuming the translational invariance 
 along $x^i$ direction. 
 The induced metric reads
\be
ds^2=\frac{(1+\dot{z}^2)d\tau^2+\sum_{i=1}^{d-1}dx_i^2}{z^2}.
\ee
 and the extrinsic curvature on this brane 
 is computed as follows: 
\begin{align}
    K_{\tau\tau}-Kh_{\tau\tau}&=\frac{(d-1)\sqrt{1+\dot{z}^2}}{z^2},\nn\\
     K_{xx}-Kh_{xx}&=\frac{(d-1)(1+\dot{z}^2)+z\ddot{z}}{z^2(1+\dot{z}^2)^{\frac{3}{2}}}.
\end{align}
Here, we choose the physical region of AdS/BCFT setup to be $0<z<z(\tau)$ and the normal vector is pointed in the outer direction. Below, we will study the solution to (\ref{bEOM}) in each dimension.

\subsubsection{\texorpdfstring{AdS$_3$}{AdS3}}
As the simplest example, consider the AdS$_3$ case. 
The equations of motion for the gauge field 
\be
\de_a(\s{h}F^{ab})=0,  \label{maxwell}
\ee
can be solved as 
\be
F_{\tau x}=E_x\cdot \frac{\s{1+\dot{z}^2}}{z^2},
\ee
where $E_x$ is a constant. The boundary condition (\ref{bEOM}) leads to
\begin{align}
    2\sqrt{1+\dot{z}^2}&=F_{\tau x}^2 z^4,\nn\\
2(1+\dot{z}^2+z\ddot{z})&=F_{\tau x}^2 z^4\sqrt{1+\dot{z}^2},
\end{align}
which ends up with
\be
\ddot{z}=0.
\ee
Therefore, the brane profile is solved as
\be
z=\s{\frac{4}{E_x^4}-1}\cdot \tau+\mbox{const}.
\ee
Indeed, we can see that the above equation of motion is identical to the known vacuum solution \cite{Takayanagi:2011zk} with the nonzero potential (or constant tension) given by $V=-\frac{E_x^2}{2}$. Since the brane profile is now a straight line, the solution does not include the gravity dual of BCFT on a strip, as opposed to the localized scalar field case.

\subsubsection{\texorpdfstring{AdS$_4$}{AdS4}}
In the AdS$_4$ case, there are three components of gauge flux and by solving (\ref{maxwell}) we can set them in the following form:
\be
F_{\tau x_i}=E_i\cdot \frac{\s{1+\dot{z}^2}}{z}, \ \ \  F_{x_1 x_2}=B_{12},
\ee
where $E_1, E_2$ and $B_{12}$ are all constants. On the other hand, (\ref{bEOM}) lead to
\begin{align}
    \frac{2\sqrt{1+\dot{z}^2}}{z^2} &= \frac{z^2}{2} \left( F_{\tau x_1}^2 + F_{\tau x_2}^2 \right) - \frac{z^2(1 + \dot{z}^2)}{2} \, F_{x_1 x_2}^2 ,\nn\\
    \frac{2(1+\dot{z}^2)+z\ddot{z}}{z^2(1+\dot{z}^2)^{\frac{3}{2}}} &=- \frac{z^2}{2(1 + \dot{z}^2)}\left( F_{\tau x_2}^2 - F_{\tau x_1}^2 \right) + \frac{z^2}{2} \, F_{x_1 x_2}^2,\\
     \frac{2(1+\dot{z}^2)+z\ddot{z}}{z^2(1+\dot{z}^2)^{\frac{3}{2}}} &=- \frac{z^2}{2(1 + \dot{z}^2)}\left( F_{\tau x_1}^2 - F_{\tau x_2}^2 \right) + \frac{z^2}{2} \, F_{x_1 x_2}^2.
\end{align}

\begin{figure}[t]
    \centering
    \includegraphics[width=10cm]{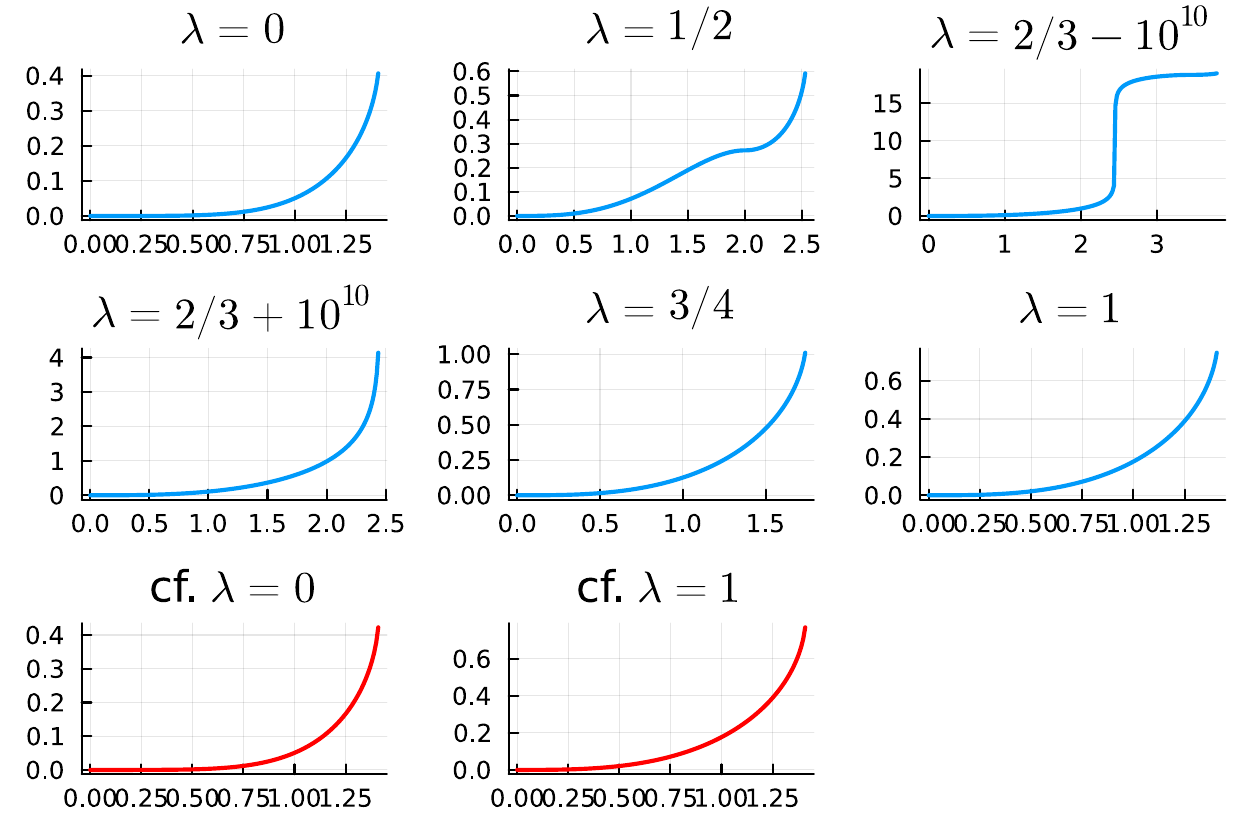}
    \caption{The profiles of the EOW brane with the gauge fields $E=\sqrt{\lambda}, B=1-\lambda$.
    For comparison, the results using expressions involving hypergeometric functions for $\lambda=0$ and $\lambda=1$ are shown in the bottom.}
    \label{fig:AdS4gauge_BraneProfile}
\end{figure}

In the presence of both electric and magnetic fields, these equations reduce to
\begin{align}
    \begin{cases}
        \displaystyle\frac{2}{z^2\sqrt{1 + \dot{z}^2}} &= \displaystyle E^2 - \frac{B^2z^2}{2}\\
        4(1 + \dot{z}^2) + z\ddot{z} &= E^2 z^2 (1 + \dot{z}^2)^{3/2}
    \end{cases}, 
\end{align}
where $E= E_1=E_2$ and $B=B_{12}$.
These can be solved as
\begin{align}
    \pm \t  &= \int_0^z \cfrac{ 
\z^2 |2E^2 - B^2\z^2| }{ \sqrt{16 - \z^4(2E^2 - B^2\z^2)^2} } \, d\z + \text{const.}.
\end{align}
The value of the integral on the right-hand side behaves as depicted in figure \ref{fig:AdS4gauge_BraneProfile}, depending on the values of \(E\) and \(B\).
For \(E^2 / |B| > 2\), i.e. \(\lambda > 2/3\) in figure \ref{fig:AdS4gauge_BraneProfile}, the smallest positive root of the denominator is \(\z = \sqrt{E^2 - \sqrt{E^4 - 4B^2}} / |B|\), and the shape of the EOW brane is convex.
On the other hand, for \(E^2 / |B| < 2\), i.e. \(\lambda < 2/3\), the smallest positive root of the denominator is \(\z = \sqrt{E^2 + \sqrt{E^4 + 4B^2}} / |B|\), and the shape has a plateau at \(\z = \sqrt{2}|E| / |B|\).

\begin{figure}[t]
    \centering
    \includegraphics[width=8cm]{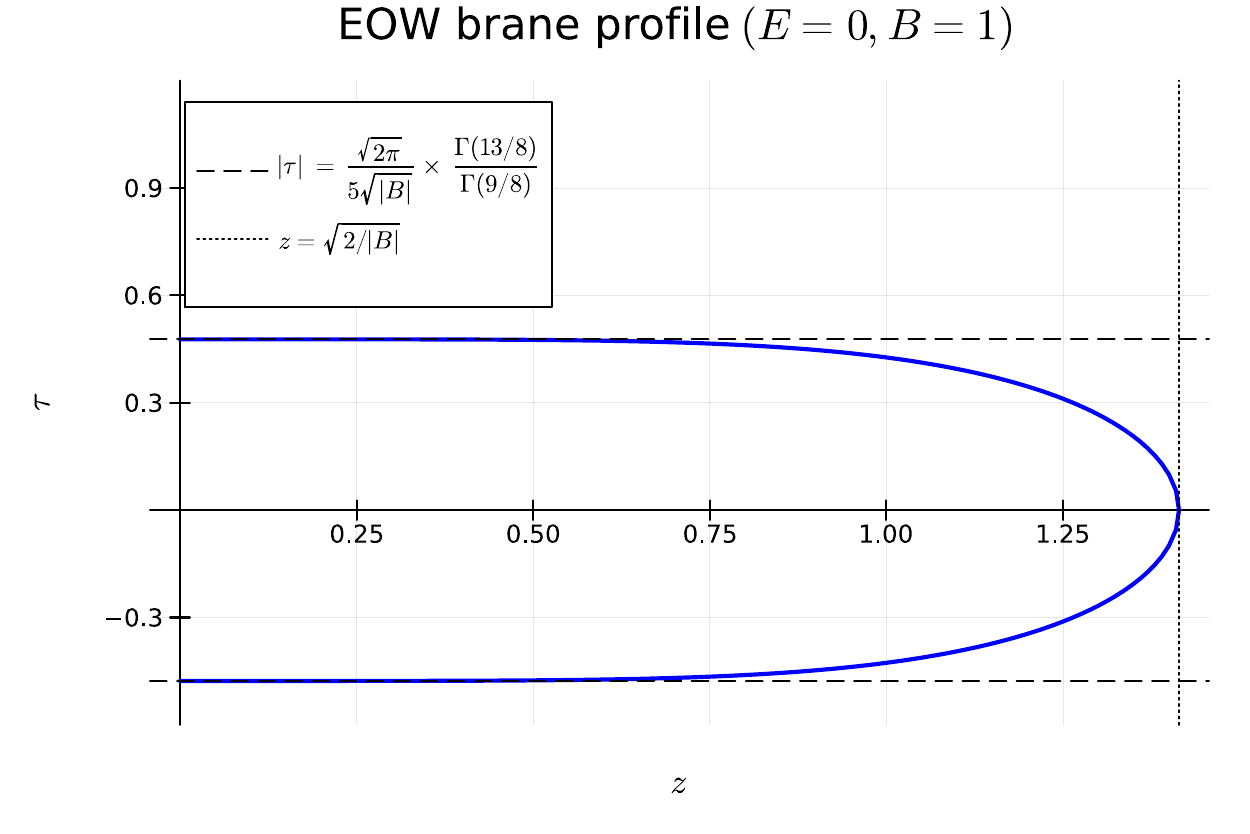}
    \includegraphics[width=8cm]{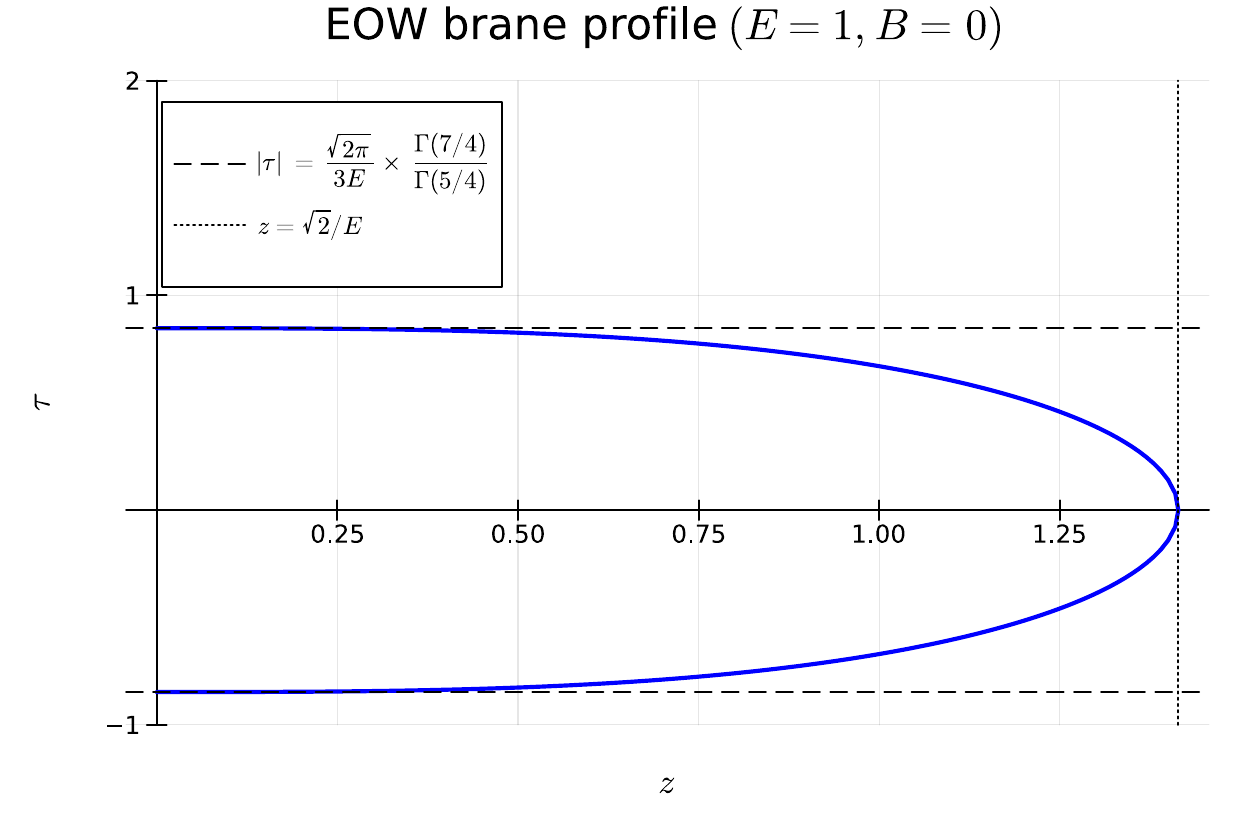}
    \caption{Left: the profile of the EOW brane with $E=0, B=1$.
    Right: the profile of the EOW brane with $E=1, B=0$.}
    \label{fig:AdS4gauge_BraneProfile_EB}
\end{figure}

Especially, in the absence of magnetic/electric fields, we obtain connected solutions:
\begin{align}
    \pm \t &= \frac{B^2z^5}{20} \, {}_2F_1 \left( \frac{1}{2}, \frac{5}{8}; \frac{13}{8}; \frac{B^4z^8}{16} \right) - \frac{\sqrt{2\pi}}{5\sqrt{|B|}} \cdot \frac{\Gamma(13/8)}{\Gamma(9/8)} \qquad (E=0),\\
    \pm \t &= \frac{E^2z^3}{6} \, {}_2F_1 \left( \frac{1}{2}, \frac{3}{4}; \frac{7}{4}; \frac{E^4z^4}{4} \right) - \frac{\sqrt{2\pi}}{3E} \cdot \cfrac{\Gamma(7/4)}{\Gamma(5/4)} \qquad (B=0).
\end{align}
The profiles of the EOW branes in these cases are shown in figure \ref{fig:AdS4gauge_BraneProfile_EB}.

\subsubsection{\texorpdfstring{AdS$_{d+1}$}{AdS(d+1)}}

Finally, we analyze the higher dimensional solution in Poincar\'{e} AdS$_{d+1}$. For simplicity, we turn on only electric fields and they take the form so that they satisfy (\ref{maxwell}):
\be
F_{\tau x_i}=E_i\cdot \frac{\s{1+\dot{z}^2}}{z^{4-d}}.
\ee
On the other hand, (\ref{bEOM}) leads to
\begin{align}
    \frac{2(d-1)\sqrt{1+\dot{z}^2}}{z^4}&=\sum_i F_{\tau x_i}^2,\nn\\
    \frac{(d-1)(1+\dot{z}^2)+z\ddot{z}}{z^4\sqrt{1+\dot{z}^2}}&=-\sum_j \frac{F_{\tau x_j}^2}{2}+F_{\tau x_i}^2.
\end{align}

These can be solved by setting 
\ba
&& E_i=E,\ \ \ \ \ (i=1,2,\ddd,d-1),\no
&& \dot{z}^2=\frac{4}{E^4}z^{8-4d}-1.  \label{scemo}
\ea

Note that the differential equation for the brane profile (\ref{scemo}) is the same as the previous scalar case with replacing $d-2$ with $d-1$ as can be seen from the Lorentzian version (\ref{highersceoma}). This suggests the following observation: the gauge fixing kills degrees of freedom for one dimension. We can state that the scalar field on the EOW brane in the vacuum AdS$_{d}$ describes the same configuration of the gauge field on the EOW brane in the vacuum AdS$_{d+1}$.

\subsection{Localized gauge field on EOW brane in BTZ phase }
Here, we consider the EOW brane solution with the localized gauge field in the BTZ phase. Again, we work with the BTZ metric in the Euclidean signature, obtained by setting $\tau=it$ in (\ref{BTZm})
\be
ds^2=\frac{h(z)d\tau^2}{z^2}+\frac{dz^2}{h(z)z^2}+\frac{dx^2}{z^2},
\ee
where $h(z)=1-\frac{z^2}{a^2}$. We choose the brane profile in the form 
$z=z(\tau)$ as before. The induced metric on the brane reads
\be
ds^2=\frac{h(z)^2+\dot{z}^2}{h(z)z^2}d\tau^2+\frac{dx^2}{z^2},
\ee
where $\dot{f}=\frac{df}{d\tau}$ and $f'=\frac{df}{dz}$. The extrinsic curvature reads
\begin{align}
    K_{xx}-Kh_{xx}&=\frac{2h^3-zh^2h'-3zh'\dot{z}^2+2h\dot{z}^2+2hz\ddot{z}}{2z^2z(h+\frac{\dot{z}^2}{h})^{\frac{3}{2}}},\nn\\
     K_{\tau\tau}-Kh_{\tau\tau}&=\frac{h\sqrt{h+\frac{\dot{z}^2}{h}}}{z^2},
\end{align}
where $f'=\frac{df}{dz}$.

\subsubsection{Construction of Solution}

The Neumann boundary condition (\ref{bEOM}) gives
\begin{align}
    K_{xx}-Kh_{xx}&=\frac{hz^2 F_{\tau x}^2}{2(h^2+\dot{z}^2)},\nn\\
    K_{\tau\tau}-Kh_{\tau\tau}&=\frac{F_{\tau x}^2z^2}{2}.
\end{align}
The equations of motion of gauge fields (\ref{maxwell}) are solved as 
\ba
F_{\tau x}=\frac{f}{z^2}\s{\frac{h(z)^2+\dot{z}^2}{h(z)}},
\ea
where $f$ is a constant described as the strength of gauge flux.

\begin{figure}[h]
    \centering
    \includegraphics[width=8cm]{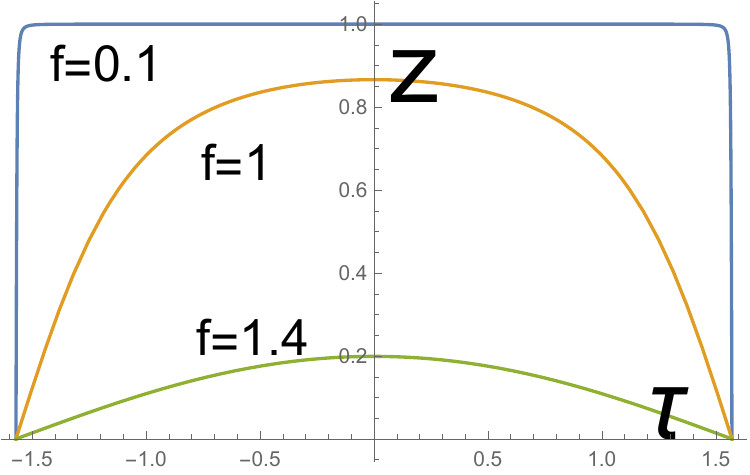}
    \caption{The profiles $z=z(\tau)$ of the EOW branes with the gauge field in BTZ geometry. The horizontal and vertical axis describes $\tau$ and $z$, respectively and we set $a=1$. The blue, orange and green curves correspond to the values $f=0.1, 1$ and $1.4$.}
    \label{fig:BTZprofile}
\end{figure}

Combining the above form, we obtain the differential equation for the brane profile
\be
\dot{z}^2=\frac{4h(z)^3}{f^4}-h(z)^2.
\ee
This is solved as follows
\be
\frac{z}{a}=\frac{\s{1-\frac{f^4}{4}}\cos\frac{\tau}{a}}{\s{1-\left(1-\frac{f^4}{4}\right)\sin\frac{\tau}{a}^2}}, 
\ \ \ \ \ \ F_{\tau x}=\frac{2h(z)}{fz^2}.
\label{tadsbrane}
\ee
for the range $-\frac{\pi}{2}a\leq \tau\leq \frac{\pi}{2}a$.
Refer to figure \ref{fig:BTZprofile} for the plot.

Note that compared with the EOW brane solution with the localized scalar \cite{Kanda:2023zse}, our solution with the gauge field takes a much simpler form. In particular, the width of the strip is given by $\pi a$, which coincides with a half of the black hole inverse temperature.

\subsubsection{Free energy in BTZ phase}

Now let us evaluate the free energy i.e. the on-shell action for this BTZ solution with the EOW brane $Q$. This is dual to a BCFT on $\Sigma$, which is a cylinder. The action we consider is
\begin{align}
  I_{\rm BTZ}&=I_{\rm EH}+I_{\rm Q}+I_{\rm \Sigma}-I_{\rm c.t.},  \nn\\
      I_{\rm EH}&=-\frac{1}{16\pi G}\int_\mathcal{M}\sqrt{g}(R+2),
  \nn\\
  I_{\rm brane}&=-\frac{1}{8\pi G}\int_Q\sqrt{h}(K-\frac{1}{4}F_{ab}F^{ab}),\nn\\
  I_{\rm \Sigma}&=-\frac{1}{8\pi G}\int_\Sigma\sqrt{h}K,\nn\\
  I_{\rm c.t.}&=-\frac{1}{16\pi G}\int_\mathcal{M}\sqrt{g^{\rm (vac)}}(R+2)-\frac{1}{8\pi G}\int_Q\sqrt{h^{\rm (vac)}}K-\frac{1}{8\pi G}\int_\Sigma\sqrt{h^{\rm (vac)}}K,
  \label{actiononep}
  \end{align}
  where $I_{\rm c.t}$ is the same action evaluated on the vacuum AdS (i.e. Poicate AdS$_3$) with the same asymptotic AdS boundary $\Sigma$, which provides the standard holographic counter term.
  
  Now, we would like to evaluate $I_{\rm BTZ}$ for our AdS/BCFT solution surrounded by the EOW brane (\ref{tadsbrane}) with the gauge flux. We choose the UV cut off to be $z\geq\ep$ and the length in $x$ direction to be $L$ via the compactification $x\sim x+L$, so that $\Sigma$ looks like a cylinder.
    Each term in the on-shell action is evaluated as follows:
\begin{align}
I_{\rm EH}&=\frac{1}{4\pi G_N}\int{g}=
\frac{L}{2\pi G_N}\int^{\frac{\pi}{2}a-\delta}_{0}d\tau
\left[\frac{1}{2\ep^2}-\frac{1}{2z^2}\right],
  \nn\\
  I_{\rm Q}&=\frac{3f^2}{16\pi G_N}\int_Q\s{h}=
\frac{3L}{4\pi G_N}\int^{\frac{\pi}{2}a-\delta}_0d\tau\frac{h(z)}{z^2},\nn\\
  I_{\rm \Sigma}-I_{\rm c.t}&=-\frac{(\pi\ti{a}-2\delta)L}{8\pi G_N\ep^2},
  \end{align}
where $\ti{a}$ is the rescaled value of $a$ to adjust the BTZ metric and the Poincar\'{e} AdS$_3$ at $z=\ep$ defined by $\ti{a}=a\s{1-\ep^2/a^2}$. The infinitesimal parameter $\delta$ is the cut off $\tau$ which corresponds to the asymptotic boundary $z=\ep$ on the EOW brane, which leads to the relation $\delta=\ep/\s{4/f^4-1}$.
The evaluation of  $I_{\rm \Sigma}-I_{\rm c.t}$ goes very similarly as in our earlier work \cite{Kanda:2023zse} of EOW brane with the localized scalar field.

By combining these estimations, we eventually find
\ba
I_{\rm BTZ}&=&\frac{L}{2\pi G_N}\int^{\frac{\pi}{2}a-\delta}_{0}\frac{d\tau}{z(\tau)^2}-\frac{5L}{16G_N a}\no
&\simeq & \frac{L}{2\pi G_N\ep\s{\frac{4}{f^4}-1}}-\frac{L}{16 G_N a}.
\ea
In the zero flux limit $f\to 0$, we find $I_{\rm BTZ}\to -\frac{L}{16 G_N a}$ and this reproduces the known result of the AdS/BCFT without the gauge flux in \cite{Takayanagi:2011zk,Fujita:2011fp}.

\subsection{Localized gauge field on EOW brane in TAdS phase}

Now we look at the EOW brane solution with the localized gauge field in thermal AdS$_3$ (TAdS), which is the other phase for BCFT on a cylinder.  The metric is given by
\be
ds^2=\frac{dz^2}{k(z)z^2}+\frac{d\tau^2}{z^2}+\frac{k(z)dx^2}{z^2},
\ee
where $k(z)=1-\frac{z^2}{b^2}$. We compactify $x$ as $x\sim x+L$ as before. The requirement of smooth geometry fixes $L$ to be 
\ba
L=2\pi b.  \label{valL}
\ea

\subsubsection{Construction of solution}

For the EOW brane defined by $z=z(\tau)$, the induced metric reads
\be
ds^2=\frac{k(z)+\dot{z}^2}{k(z)z^2}d\tau^2+\frac{k(z)dx^2}{z^2}.
\ee
The extrinsic curvature is found to be
\begin{align}
     K_{\tau\tau}-Kh_{\tau\tau}&=\frac{z(2k^2-zk'\dot{z}^2+2k\dot{z}^2+2kz\ddot{z})}{2z^2(k+\dot{z}^2)^{\frac{3}{2}}},\nn\\
     K_{xx}-Kh_{xx}&=\frac{(2k-zk')\sqrt{k+\dot{z}^2}}{2z^2k}.
\end{align}
The equations of motion for brane (\ref{bEOM}) reads
\begin{align}
     K_{xx}-Kh_{xx}&=\frac{kz^2F_{\tau x}^2}{2(k+\dot{z}^2)},\nn\\
     K_{\tau\tau}-Kh_{\tau\tau}&=\frac{F_{\tau x}^2z^2}{2k}.
\end{align}
The gauge field equation of motion (\ref{maxwell}) is solved to be
\ba
F_{\tau x}=f \frac{\s{k(z)+\dot{z}^2}}{z^2}.
\ea
After short algebra, we obtain the differential equation for the brane profile
\be
\dot{z}^2=\frac{4}{f^4}-k(z),
\ee
which is solved as 
\be
z=C_1e^{\frac{\tau}{b}}+C_2e^{-\frac{\tau}{b}},\label{gsolzc}
\ee
where $C_1$ and $C_2$ are arbitrary constants with the 
constraint $4C_1C_2=-b^2(\frac{4}{f^4}-1)$. However, this is not an expected connected solution dual to a BCFT on a strip. Therefore in the thermal AdS$_3$ case, the only available dual solution is the two disconnected EOW branes localized 
at $z=\pm\frac{\pi}{2}a$.

\begin{figure}[h]
    \centering
    \includegraphics[width=8cm]{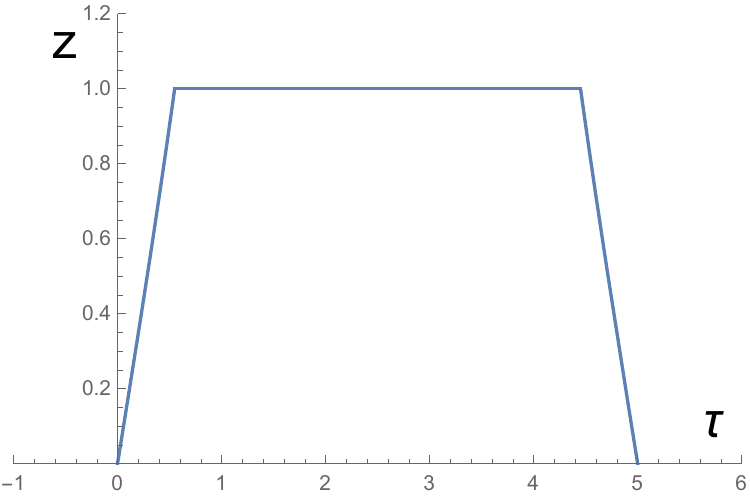}
    \caption{The profiles $z=z(\tau)$ of the EOW branes with the gauge field in TAdS geometry. The horizontal and vertical axis describes $\tau$ and $z$, respectively. We choose $b=1$, $\pi a=5$ and $f=1$. }
    \label{fig:BTZprofileA}
\end{figure}

As before, we are considering a BCFT on a cylinder whose width is $\pi a$ and the circumference is $L=2\pi b$. Since the EOW brane can end on the tip $z=b$  in the TAdS geometry, we choose the two disconnected EOW branes from the general solution (\ref{gsolzc}):
\ba
z=b\s{\frac{4}{f^4}-1}\sinh\frac{\tau}{b},
\ea
and
\ba
z=b\s{\frac{4}{f^4}-1}\sinh\frac{\pi a-\tau}{b},
\ea
which is plotted in figure \ref{fig:BTZprofileA}. We introduce the cut off $\delta$ again such that $z=\ep$ at $\tau=\delta$, which leads to $\delta=\frac{\ep}{\s{4/f^4-1}}$.

We introduce a positive constant $\tau_0$ such that $\s{\frac{4}{f^4}-1}\sinh\frac{\tau_0}{b}=1$. Then the existence of such a disconnected solution is possible only when 
\ba
\tau_0<\frac{\pi}{2a}, \label{regionQQ}
\ea
to avoid an intersection of EOW branes.

\subsubsection{Free energy in TAdS phase}

The evaluation of free energy in the TAdS case can be conducted similarly by computing each term in (\ref{actiononep}) as follows:
\begin{align}
I_{\rm EH}&=\frac{1}{4\pi G_N}\int{g}=
\frac{L}{2\pi G_N}\int^{\frac{\pi}{2}a}_{\delta}d\tau
\left[\frac{1}{2\ep^2}-\frac{1}{2z^2}\right],
  \nn\\
  I_{\rm Q}&=\frac{3f^2}{16\pi G_N}\int_Q\s{h}=
\frac{3L}{4\pi G_N}\int^{\tau_0}_\delta d\tau\frac{1}{z^2},\nn\\
  I_{\rm \Sigma}-I_{\rm c.t}&=-\frac{(\pi a-2\delta)\ti{L}}{8\pi G_N\ep^2},
  \end{align}
where we defined $\ti{L}=L\s{1-\ep^2/\b^2}$ and $I_{\rm c.t}$ is the action for the vacuum Poincar\'{e} configuration with two disconnected straight branes. Notice also that $L$ is determined by $b$ as in (\ref{valL}).

By combining all terms, we eventually find
\ba
I_{\rm TAdS}&=&\frac{L}{2\pi G_N}\int^{\tau_0}_{\delta} \frac{d\tau}{z(\tau)^2}-\frac{L}{4\pi G_N b^2}\left(\frac{\pi}{2}a-\tau_0\right)
+\frac{La}{16\pi G_Nb^2}\no
&\simeq & \frac{L}{2\pi G_N\ep\s{\frac{4}{f^4}-1}}-\frac{L}{\pi G_N b}\cdot \frac{f^2}{4-f^4}+\frac{L}{4\pi G_N b^2}\left(-\frac{\pi}{4}a+\tau_0\right).
\ea
In the zero flux limit $f\to 0$, we obtain $I_{\rm TAdS}\to -\frac{La}{16 G_N b^2}$ which reproduces the known result of AdS/BCFT without the gauge field \cite{Takayanagi:2011zk,Fujita:2011fp}.

\subsection{Phase transitions between BTZ and TAdS phase}

In summary, the finite part of free energies of EOW brane background with the localized gauge field in the BTZ and TAdS phase read
\ba
&& I^{\rm Finite}_{\rm BTZ}=-\frac{\pi b}{8 G_N a},\no
&& I^{\rm Finite}_{\rm TAdS}=-\frac{2f^{2}}{(4-f^4)G_N}-\frac{\pi a}{8G_N b}+\frac{\tau_0}{2 G_N b},
\ea
where we set $L=2\pi b$ using  (\ref{valL}).

These two solutions are possible ones for the gravity dual of a BCFT on a cylinder with the width $\pi a$ and the circumference
$2\pi b$. In general we expect that when $a\ll b$ the BTZ phase is favored, while the TAdS phase does when $b\ll a$.

\begin{figure}[h]
    \centering
    \includegraphics[width=5cm]{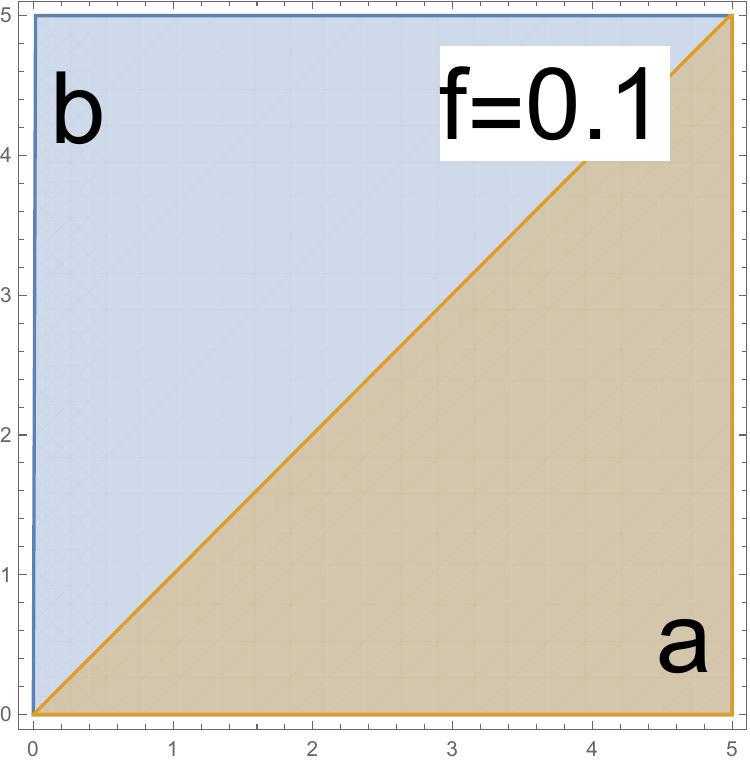}
\includegraphics[width=5cm]{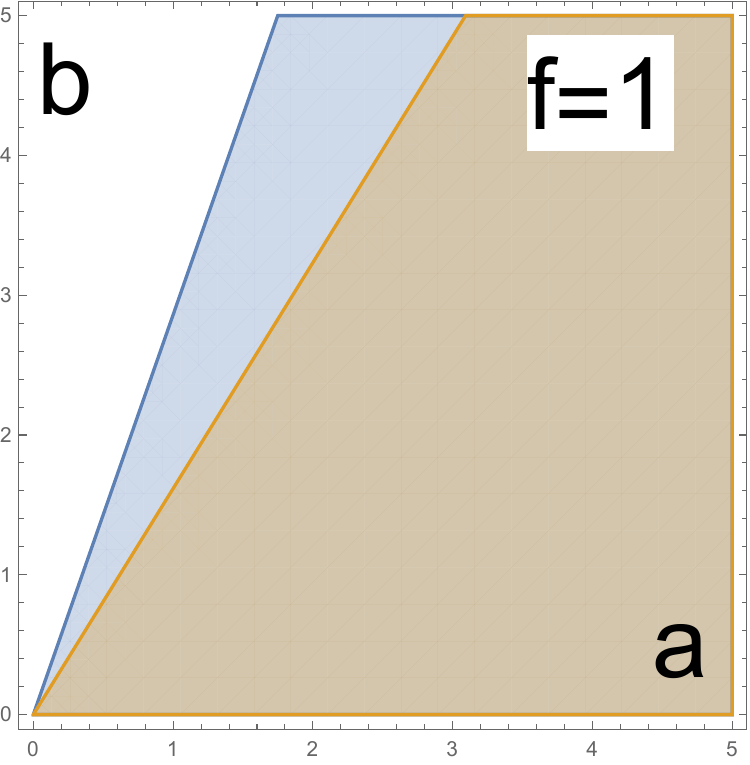}
    \includegraphics[width=5cm]{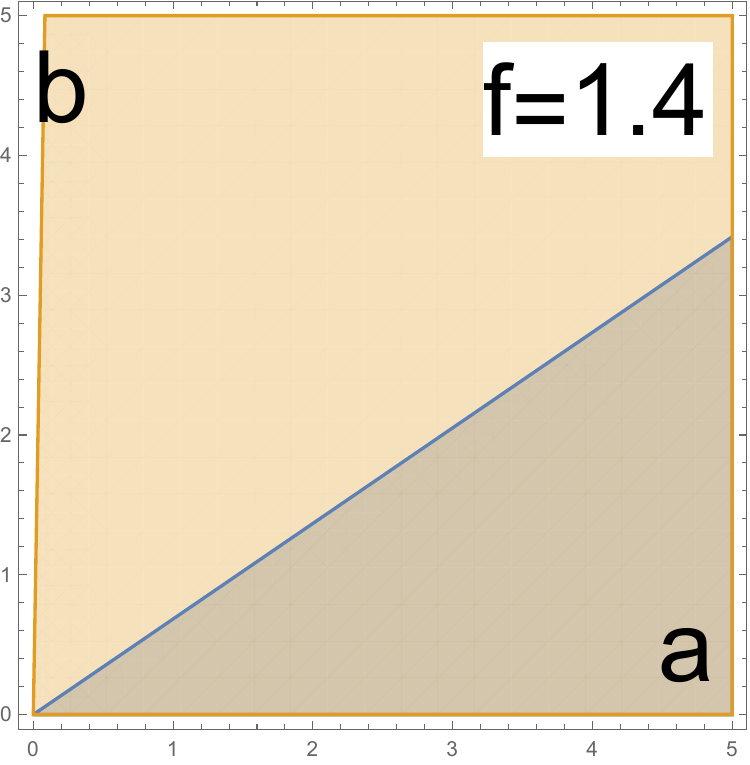}
    \caption{Comparison of free energy between BTZ and TAdS phases of the EOW brane with gauge field. The blue regions satisfy the condition (\ref{regionQQ}) where the EOW brane solution in TAdS exists. The orange region describes the region where the TAdS solution is favored over the BTZ one i.e. $I_{\mathrm{TAdS}}<I_{\mathrm{BTZ}}$. The left, middle and right panels correspond to $f=0.1, f=1$ and $f=1.4$, respectively. Note that at $f=0$ we recover the known result $b<a$.}
    \label{fig:GaugePhase}
\end{figure}

The phase structure is plotted in figure \ref{fig:GaugePhase} for various values of $f$. The TAdS phase becomes dominant when the blue and orange colored regions overlap. In the zero flux limit $f\to 0$, the TAdS is realized when $b<a$, which reproduces the known result in \cite{Takayanagi:2011zk,Fujita:2011fp}. As the flux gets larger, the TAdS phase gets squeezed as can be seen from figure \ref{fig:GaugePhase}. It is also intriguing that there is not Poincar\'e AdS$_3$ phase for an EOW brane with the localized gauge flux, as opposed to the one with the localized scalar field.

\section{Soft wall model with bulk scalar}\label{sec:Janus}
\par So far we have discussed the pseudo entropy defined by the transition matrix constructed from two different boundary states. In the bulk gravity side, we consider the brane-localized field which interpolates these boundary states with thin branes. It is also natural to consider the bulk field analogue of this interpolating brane-localized field. In other words, we have so far discussed the hard wall model and there should also be the corresponding soft wall model. To this end, we consider a bulk dilaton field which interpolates two different field values at the conformal boundaries. In the conformal field theory side, this is nothing but interface CFTs \cite{Bak:2003jk,Clark:2004sb,Bak:2007jm}. See figure \ref{fig: Boundary localized field model} and figure \ref{fig:bulk field model} for the boundary path integral dual to the previous AdS/BCFT model and that for our current bulk field model, respectively. 

In our present model, the path integral prepares a transition matrix defined by two vacuum states for two CFTs with different marginal deformation. More concretely, we consider the two CFT with Hamiltonian $H_{\pm} = H_0\pm \phi_{\pm} \int O$ and each of the Hamiltonian has their own vacuum, say $\ket{0_{\pm}}$. Here $\phi_\pm$ are boundary values of the dilaton field which is dual to the brane deformation. Then we consider a constant Euclidean time slice $\tau=\tau_0$, where the transition matrix is defined. In this setup, the transition matrix $\mathcal{T}$ for $\tau_0>0$ can be expressed as 
\begin{equation}
\begin{split}
    \mathcal{T}(\tau_0>0) &= \mathcal{N}\cdot \lim_{T\to\infty}e^{-\tau_0 H_+}e^{-T H_-}\ket{0_-}\bra{0_+}e^{-H_+(T-t_0)}\\ &= \mathcal{N}\cdot e^{-\tau_0 H_+}\ket{0_-}\bra{0_+},
\end{split}
\end{equation}
where $\mathcal{N}$ is a normalization factor. Similarly for $\tau_0<0$, we have
\begin{equation}
     \mathcal{T}(\tau_0<0) = \mathcal{N}\cdot \ket{0_-}\bra{0_+} e^{\tau_0 H_-}.
\end{equation}
One difference from the usual holographic model of the interface CFTs is that we put the defect on the time slice and compute the pseudo entropy of the sub-region which is parallel to this defect. 

\begin{figure}[h]
    \centering
    \begin{minipage}[h]{0.45\linewidth}
        \centering
       
       \begin{tikzpicture}
 \fill[fill=lightgray](0,0)rectangle(4,6);
 \draw[ultra thick,red](0,0)--(4,0);
 \draw(2,0)node[below]{$\ket{B(\phi_1)}$};
  \draw[ultra thick,red](0,6)--(4,6);
 \draw(2,6)node[above]{$\bra{B(\phi_2)}$};
 \draw[<-](-1,7)--(-1,-1);
 \draw(-1,7)node[left]{$\tau$};
 \draw[<->](4.5,0)--(4.5,6);
 \draw[dashed](0,3)--(4,3);
  \draw[orange](1,4)--(3,4);
 \draw(2,4)node[above]{\textcolor{orange}{$A$}};
 \draw[orange,dashed](-1.2,4)--(1,4);
 \draw(-1.2,4)node[left]{\textcolor{orange}{$\tau_0$}};
 \draw(-1.2,3)--(-0.8,3);
 \draw(-1.2,3)node[left]{$\tau=0$};
 \draw(4.6,3)node[right]{$\frac{\beta}{2}$};
 \end{tikzpicture}
        \caption{{The model with different boundary states discussed in the above sections.}}
 \label{fig: Boundary localized field model}
    \end{minipage}
\hspace{0.04\columnwidth} 
    \begin{minipage}[h]{0.45\linewidth}
        \centering
\begin{tikzpicture}
 \fill[fill=lightgray](0,-1)rectangle(4,7);
 
 \draw[<-](-1,7)--(-1,-1);
 \draw(-1,7)node[left]{$\tau$};
 \draw[blue,ultra thick](0,3)--(4,3);
 \draw[orange](1,4)--(3,4);
 \draw(2,4)node[above]{\textcolor{orange}{$A$}};
 \draw[orange,dashed](-1.2,4)--(1,4);
 \draw(-1.2,4)node[left]{\textcolor{orange}{$\tau_0$}};
 \draw(-1.2,3)--(-0.8,3);
 \draw(-1.2,3)node[left]{$\tau=0$};
  \draw(2,4.5)node[above]{\Huge{CFT{$_{\Phi_2}$}}};
    \draw(2,0.5)node[above]{\Huge{CFT{$_{\Phi_1}$}}};
    \draw(4,3)node[right]{\LARGE defect};
 \end{tikzpicture}
        \caption{The model with bulk fields. The bulk fields interpolate two CFTs, CFT$_{\Phi_1}$ and CFT$_{\Phi_2}$. In the CFT side, there is a defect lies on $\tau=0$.}
        \label{fig:bulk field model}
    \end{minipage}
\end{figure}

  The solution of the Einstein equation with back-reacting dilaton field is known as the Janus solution \cite{Bak:2003jk,Clark:2004sb,Bak:2007jm}. Here we focus on the AdS$_3$ case \cite{Bak:2007jm}. Explicitly, the Janus solution with the defect on Euclidean time slice $\tau=0$ can be written as follows,
  \begin{equation}\label{eq:Euclid Janus solution}
    \begin{split}
         ds^2 &= dr^2 + f(r)\frac{dx^2+d\xi^2}{\xi^2},\quad 
         f(r) = \frac{1}{2}(1+\sqrt{1-2\gamma^2}\cosh{2r}),\\
        \Phi(r) &=  \frac{1}{\sqrt{2}}\log\qty(\frac{1+\sqrt{1-2\gamma^2}+\sqrt{2}\gamma\tanh{r}}{1+\sqrt{1-2\gamma^2}-\sqrt{2}\gamma\tanh{r}}),
    \end{split}
\end{equation}
where we adopt the hyperbolic slice coordinate $(r,\xi,x)$. Here $\gamma$ is a constant related to the strength of deformation by spacelike defect as usual defect. Especially, when $\gamma=0$, there is no defect on the CFT side.  In this case, the gravity dual is the vacuum AdS$_3$. We consider the range of gamma $\abs{\gamma}<\gamma_c=\frac{1}{\sqrt{2}}$.  Also the $\gamma$ is related to the asymptotic value of the dilaton field $\Phi$ as 
\begin{equation}
    \Phi(r \to \pm \infty) =   \frac{1}{\sqrt{2}}\log\qty(\frac{1+\sqrt{1-2\gamma^2}\pm\sqrt{2}\gamma}{1+\sqrt{1-2\gamma^2}\mp\sqrt{2}\gamma}).
\end{equation}
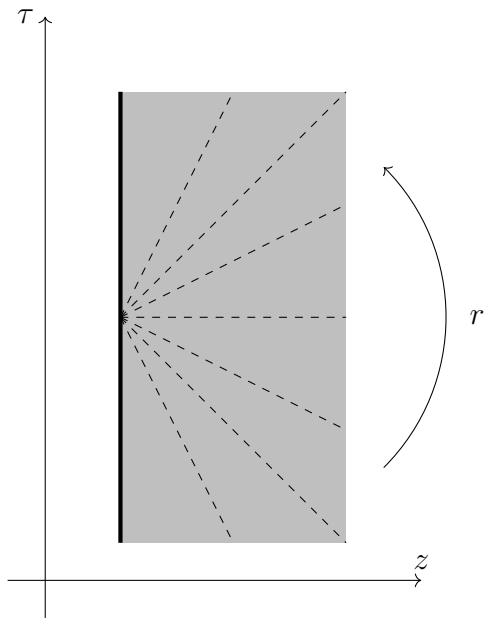
\begin{figure}[h]
    \centering
    \begin{tikzpicture}
        \fill[fill=lightgray](0,0)rectangle(3,6);
        \draw[ultra thick](0,0)--(0,6);
        \draw[dashed](0,3)--(1.5,6);
        \draw[dashed](0,3)--(3,6);
        \draw[dashed](0,3)--(3,4.5);
        \draw[dashed](0,3)--(3,3);
        \draw[dashed](0,3)--(3,1.5);
        \draw[dashed](0,3)--(1.5,0);
        \draw[dashed](0,3)--(3,0);
         \draw[<-](-1,7)--(-1,-1);
 \draw(-1,7)node[left]{$\tau$};
  \draw[->](-1.5,-0.5)--(4,-0.5);
 \draw(4,-0.5)node[above]{$z$};
 \draw[->](3.5,1)to[out=45,in=-45](3.5,5);
  \draw(4.5,3)node[right]{$r$};
    \end{tikzpicture}
    \caption{Figure for the Janus solution (\ref{eq:Euclid Janus solution}). The dashed line denotes the constant $r$ surfaces, $r=\text{const.}$}
    \label{fig:enter-label}
\end{figure}
\subsection{Pseudo entropy from Euclidean Janus solution}
 Let us move on to the computation of the pseudo entropy in our setup. As in figure \ref{fig:bulk field model}, we take the subsystem $A$ to be parallel to the defect. Note that this is different from situations explored in \cite{Azeyanagi:2007qj,Karch:2021qhd} where the subsystem is taken to be perpendicular to the defect. We compute both the Euclidean and the Lorentzian time evolution of the pseudo entropy with the same trick as above. That is, we take the subsystem $A$ as the interval $[-l,l]$ on the constant Euclidean time slice $\tau= \text{const.}$ and perform the Wick rotation after all the computations. Thus our computation of the pseudo entropy is basically entropy of the transition matrix obtained from the two distinct vacuum states and its Lorentzian time evolution. Note that in our computation, the bulk geometry is not AdS spacetime itself, the choice of the time coordinate is not the same as the usual one. Here we relate the bulk time coordinate and the boundary time coordinate by using the analogue of the coordinate transformation in the vacuum AdS$_3$. We see this point in detail later. 
 \par Finally we move to the computation of the holographic pseudo entropy. What we need to do is to compute the geodesic length connecting the entangling surface $\partial A$ in the Janus geometry (\ref{eq:Euclid Janus solution}).
 
\paragraph{The geodesic equation}~\par
To compute the geodesic $x=x(r),\xi=\xi(r)$ and its length, we consider the Lagrangian
\begin{equation}\label{eq:Euclid Lagrangian}
    \mL = \sqrt{1+f(r)\frac{\Dot{x}^2+\Dot{\xi}^2}{\xi^2}},
\end{equation}
where the dots denote the $r$ derivative. To compute the geodesic, it is useful to utilize the symmetries in this setup. The Lagrangian (\ref{eq:Euclid Lagrangian}) has two symmetries,
\begin{enumerate}
    \item AdS$_2$ isometry
    \begin{equation}
        x\to x+\lambda x,\quad \xi\to \xi+\lambda\xi.
    \end{equation}
    The corresponding Noether charge is 
    \begin{equation}\label{eq
:QAdS}
        Q_{\mathrm{AdS}_2}= \frac{f(r)}{\mL}\frac{\xi\Dot{\xi}+x\Dot{x}}{\xi^2}.
    \end{equation}
    \item $x$-translation
    \begin{equation}
        x\to x+c.
    \end{equation}
        The corresponding Noether charge is 
        \begin{equation}\label{eq:Euclid QT}
            Q_T= \frac{f(r)}{\mL}\frac{\Dot{x}}{\xi^2}.
        \end{equation}
\end{enumerate}
By dividing (\ref{eq
:QAdS}) by (\ref{eq:Euclid QT}), we have
\begin{equation}\begin{split}
    \frac{Q_{\mathrm{AdS}_2}}{Q_T} &= \frac{1}{2}\frac{d}{dx}\qty(\xi^2(x)+x^2),\\
    \xi^2(x)+x^2 &= 2\frac{Q_{\mathrm{AdS}_2}}{Q_T}x+A.
\end{split}
\end{equation}
where $A$ is an integration constant. Thanks to the $\mathbb{Z}_2$ symmetry $x\to-x$, we find $ Q_{\mathrm{AdS}_2}=0$. Thus, we have    \begin{equation}\label{eq:Euxlid xi(x)}
        \xi^2+x^2=A,
    \end{equation}
    with some constant $A>0$. Note that to derive (\ref{eq:Euxlid xi(x)}), we do not use the concrete form of $f(r)$. 
By combining (\ref{eq:Euclid Lagrangian}), (\ref{eq:Euxlid xi(x)}) and (\ref{eq:Euclid QT}), we obtain geodesic equations, 
\begin{equation}\label{eq:geodesic eq1}
    \begin{split}
        \left(\frac{\sqrt{A}\Dot{x}}{A-x^2}\right)^2 &= \frac{\alpha}{f(r)(f(r)-\alpha)},\\
        \mL^2 &= \frac{f(r)}{f(r)-\alpha}, \quad 
        \alpha = AQ_T^2.
    \end{split}
\end{equation}
From the first equation of (\ref{eq:geodesic eq1}) we see
\begin{equation}\label{eq:sol geodesic}
    \begin{split}
       \frac{x(r)}{\sqrt{A}} &=\tanh{\qty(-\sqrt{\frac{-k}{\chi}}\EllipticF{G(r)}{k}+B)},\\
        G(r) &:=\Arcsin{\qty(\frac{1}{2} \sqrt{\frac{(2 \alpha -\chi -1) \left(\chi \cosh{2r}+1 \right)\text{csch}^2{r}}{\alpha  \chi }})},\\
        k&=-\frac{4 \alpha  \chi }{(1-\chi)(2\alpha-\chi-1)},
    \end{split}
\end{equation}
where $\chi = \sqrt{1-2 \gamma^2}$, $B$ is an integration constant and $-\frac{\pi}{2}< \Arcsin{\qty(\cdot)} < \frac{\pi}{2}$. $\EllipticF{z}{m}$ denotes the incomplete elliptic integral of the first kind. See for details of the elliptic integrals in appendix \ref{appendix: elliptic integrals}. 

\paragraph{Cutoff surface}~\par 
In AdS/CFT correspondence, we need to regularize the physical quantities of the boundary theory by introducing a cutoff surface in the bulk gravity. For asymptotically AdS spacetime, the regularization procedure is based on the Fefferman-Graham expansion of the asymptotically AdS spacetime\cite{Bianchi:2001kw,Skenderis:2002wp}. However, it is known that for the Janus geometry dual to the interface CFT we cannot construct the FG coordinates which cover all spacetime\cite{Bak:2016rpn,Estes:2014hka}. In other words, there may be several ways for defining the coordinate transformation from the H$_2$ slice coordinate $(r,\xi,x)$ to the Poincar\'e-like coordinate $(z,t,x)$. There are several considerations about ways of putting the cutoff surface for these geometries\cite{Gutperle:2016gfe}. Here, we define the AdS (regularized) boundary by 
\ba\label{eq:cutoff Euclid}
\xi=\ep\s{f(r)}.
\ea
This definition is motivated by two reasons. One is that, in the pure AdS case, $\gamma=0$ and $\chi=1$, this surface completely mathcs $z=\ep$ in the Poincar\'e coordinates. We can easily see this from (\ref{eq:tf poincare amd ads2}). The second reason is that, by properly defining time $\tau$, we can set the induced metric of the cutoff surface which will be the CFT metric as the flat metric.
Indeed, from (\ref{eq:cutoff Euclid}), the induced metric reads 
\ba
ds^2|_{\mathrm{ind}}&=&\frac{dx^2}{\ep^2}+\frac{1}{\ep^2}\left(1+\frac{4f}{f'^2}\right)d\xi^2.
\ea
We now introduce the boundary time $\tau$ by setting
\ba
d\tau^2=\left(1+\frac{\ep^2}{\xi'^2}\right)d\xi^2= \qty(1+\frac{\qty(\frac{\xi}{\ep})^2}{-\chi^2+\qty(2\qty(\frac{\xi}{\ep})^2-1)^2})d\xi^2\label{wwq},
\ea
so that the boundary metric looks like
\ba
ds^2=\frac{dx^2+d\tau^2}{\ep^2}.
\ea
We can find $\tau=\tau(\xi)$ by solving (\ref{wwq}). Note that for $\tau\gg O(\ep)$, we find $\tau\sim \xi$, but for $\tau\sim O(\ep)$ there is a bit difference from $\xi$.
\begin{figure}
       \includegraphics[scale=0.34]{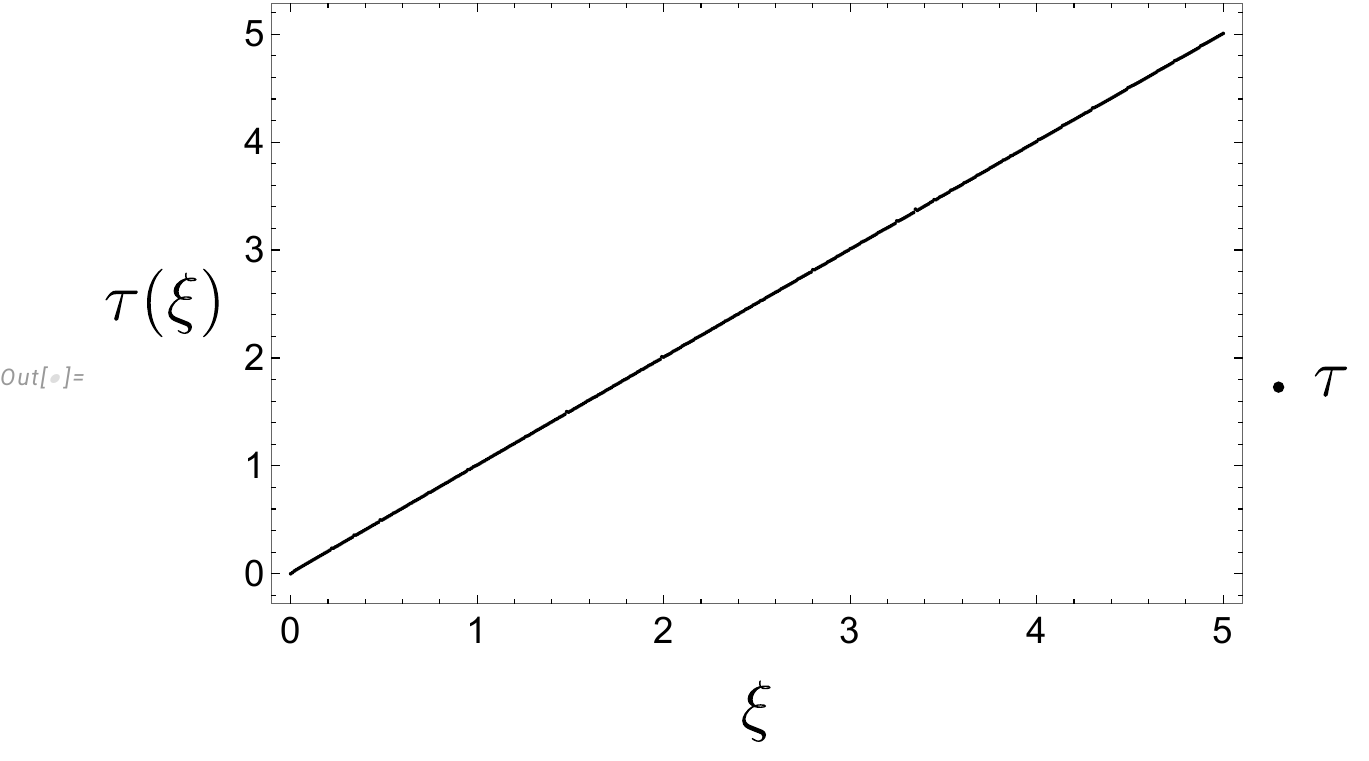}
        \includegraphics[scale=0.34]{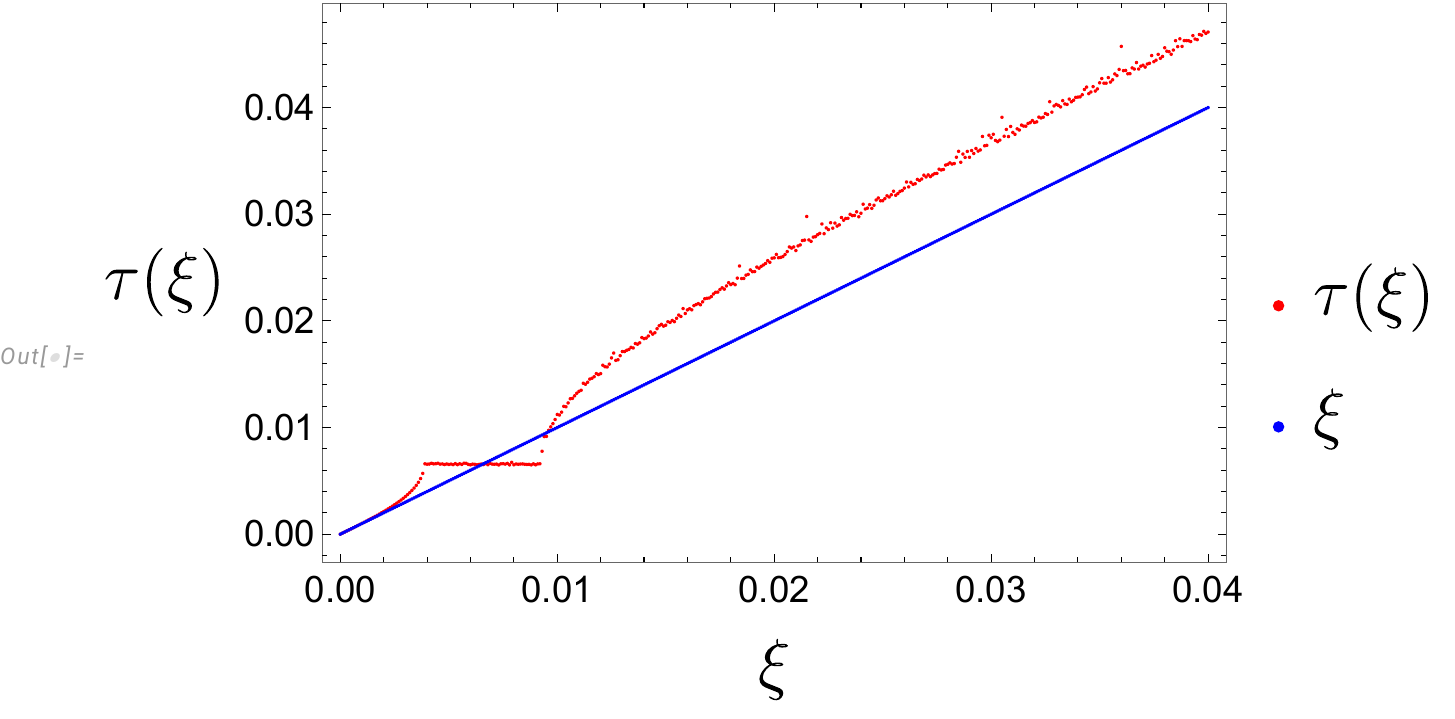}
    \caption{Plots for the $\tau(\xi)$ with $\chi=0.7,\ep=0.01$. For $\tau_0\gg\ep$, we confirm that $\tau(\xi)\sim \xi$. The right figure shows the plots for the time $\tau_0=O(\ep)$.}
\end{figure}

\paragraph{Determination of the constant $A, B,\alpha$}~\par Firstly let us determine $A$. As mentioned above, (\ref{eq:Euxlid xi(x)}) is derived just by the AdS$_2$ structure of the metric and it does not depend on the specific form of $f(r)$. Thus we suppose the value of $A$ is universal and is the same as the vacuum case, \textit{i.e.}  
\begin{equation}
    A= l^2 + \tau_0^2.
\end{equation} \par
This will be valid at least when $\tau_0\gg \ep$. The constant $B$ can be determined from the boundary condition at the turning point $x=0,r=r_t$. At the turning point, $\Dot{x}(r)$ should be divergent. From the first equation of (\ref{eq:geodesic eq1}), we find that  $f(r_t)=\alpha$, \textit{i.e.}
\begin{equation}\label{eq:turning point}
    1+\chi \cosh{2r_t}=2\alpha,
\end{equation}
holds. By using this identity, we also find 
\begin{equation}
    G(r_t)= \frac{\pi}{2}.
\end{equation}
Finally, by evaluating (\ref{eq:sol geodesic}) at the turning point $r=r_t, x=0$, we obtain 
\begin{equation}
    B= \sqrt{\frac{-k}{\chi}}K(k),
\end{equation}
where $\EllipticK{k}=\EllipticF{\frac{\pi}{2}}{k}$ denotes the complete elliptic integral of the first kind. \par 
To determine the constant $\alpha$ we introduce the value of $r_{\infty}$ where geodesic pass through the cutoff surface (\ref{eq:cutoff Euclid}). By 
\begin{equation}\label{eq:cut off surface}
   \frac{\xi(r_\infty)}{\sqrt{f(r_\infty)}} =\ep
\end{equation}
The constant $\alpha$ is determined with (\ref{eq:sol geodesic}) at the cutoff surface $r\to r_\infty$ and $x\to l$ 
\begin{equation}\label{eq:Euclid alpha}
    \frac{l}{\sqrt{A}}= -\tanh{\qty(\sqrt{\frac{-k}{\chi}}\qty(\EllipticF{G(r_\infty)}{k}-\EllipticK{k}))}.
    \end{equation}
\paragraph{The geodesic length}~  \par
Next we evaluate the geodesic length. 
This is done with an analogy of the coordinate change between the hyperbolic coordinate and the usual Poincar\'e coordinate in the vacuum AdS$_3$\cite{Bak:2016rpn}. See the appendix \ref{appendix:Hyperbolic Coordinate in AdS}. 

\begin{figure}[h]
    \centering
    \includegraphics[scale=0.5]{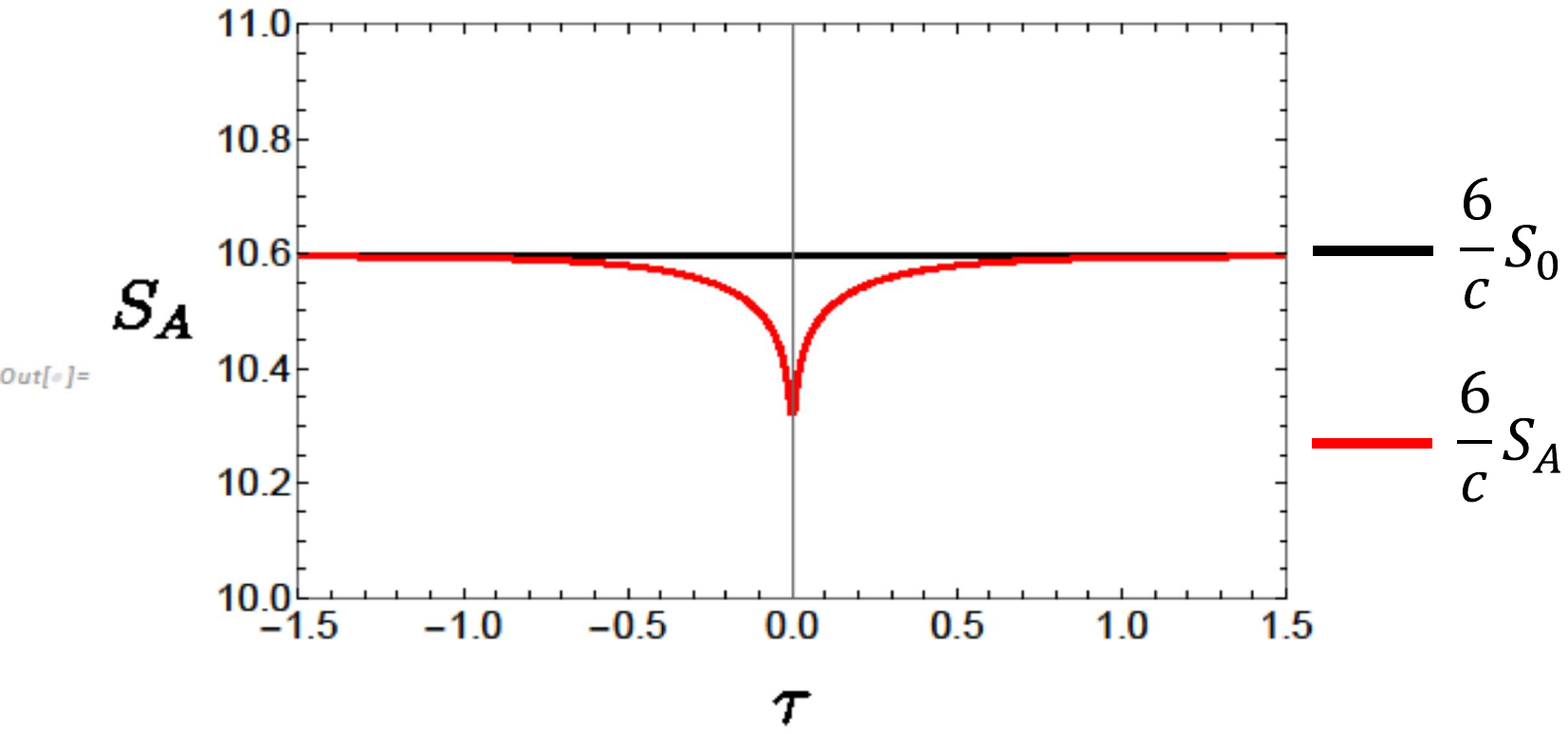}
    \caption{The plots of the pseudo entropy in Euclidean time evolution $S_A(\tau_0)$. We set $\gamma= 0.35, \ep= 0.01,l =1$. The black line expresses the entanglement entropy for undeformed CFT $S_0 = \frac{c}{3}\log\frac{2l}{\ep}$.}
    \label{fig:Euclidean Model PE Euclid time}
\end{figure}

\begin{figure}[h]
    \centering
    \begin{minipage}[h]{0.49\linewidth}
        \centering
        \label{tab:leftfigure}
        \includegraphics[width=\linewidth]{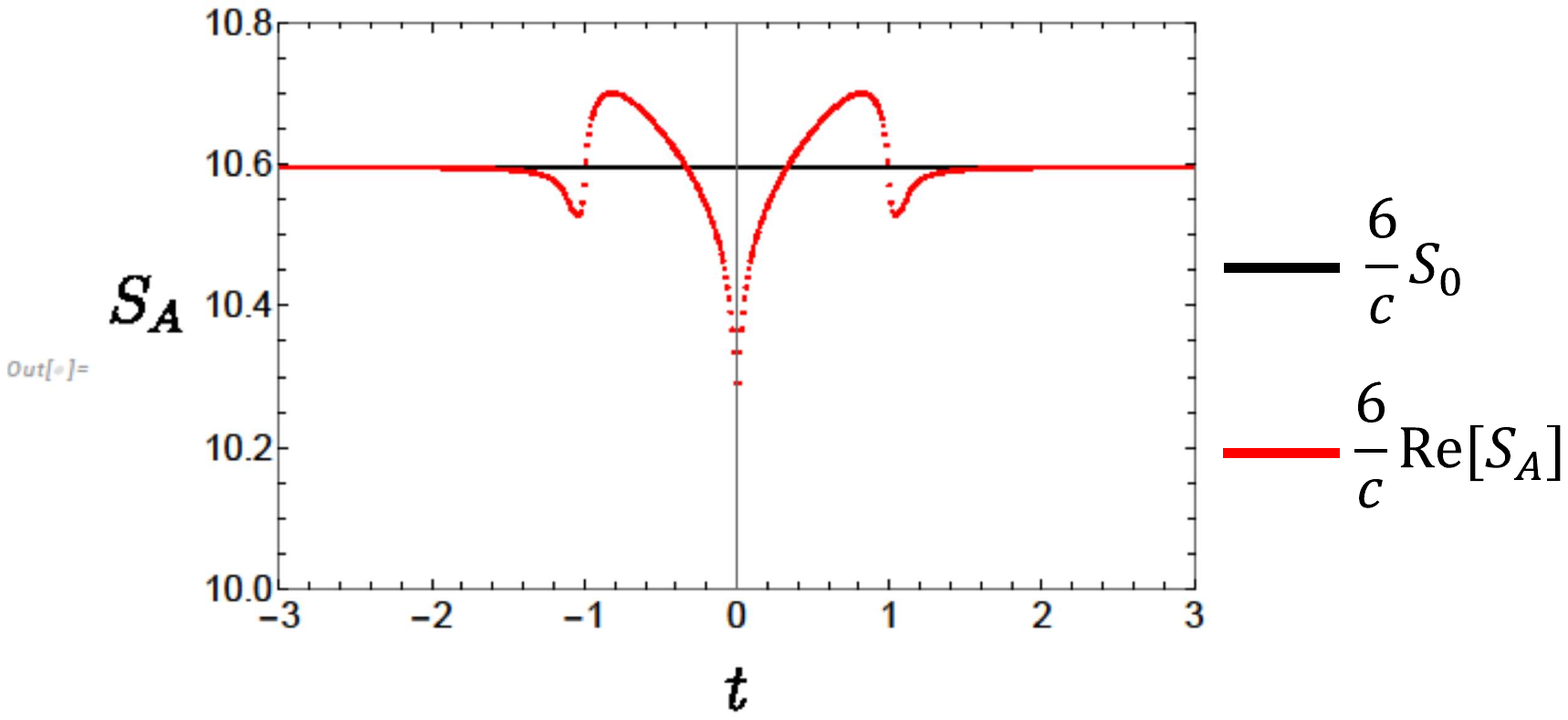}
    \end{minipage}
    \begin{minipage}[h]{0.49\linewidth}
        \centering
        \label{tab:rightfigure}
         \includegraphics[width=\linewidth]{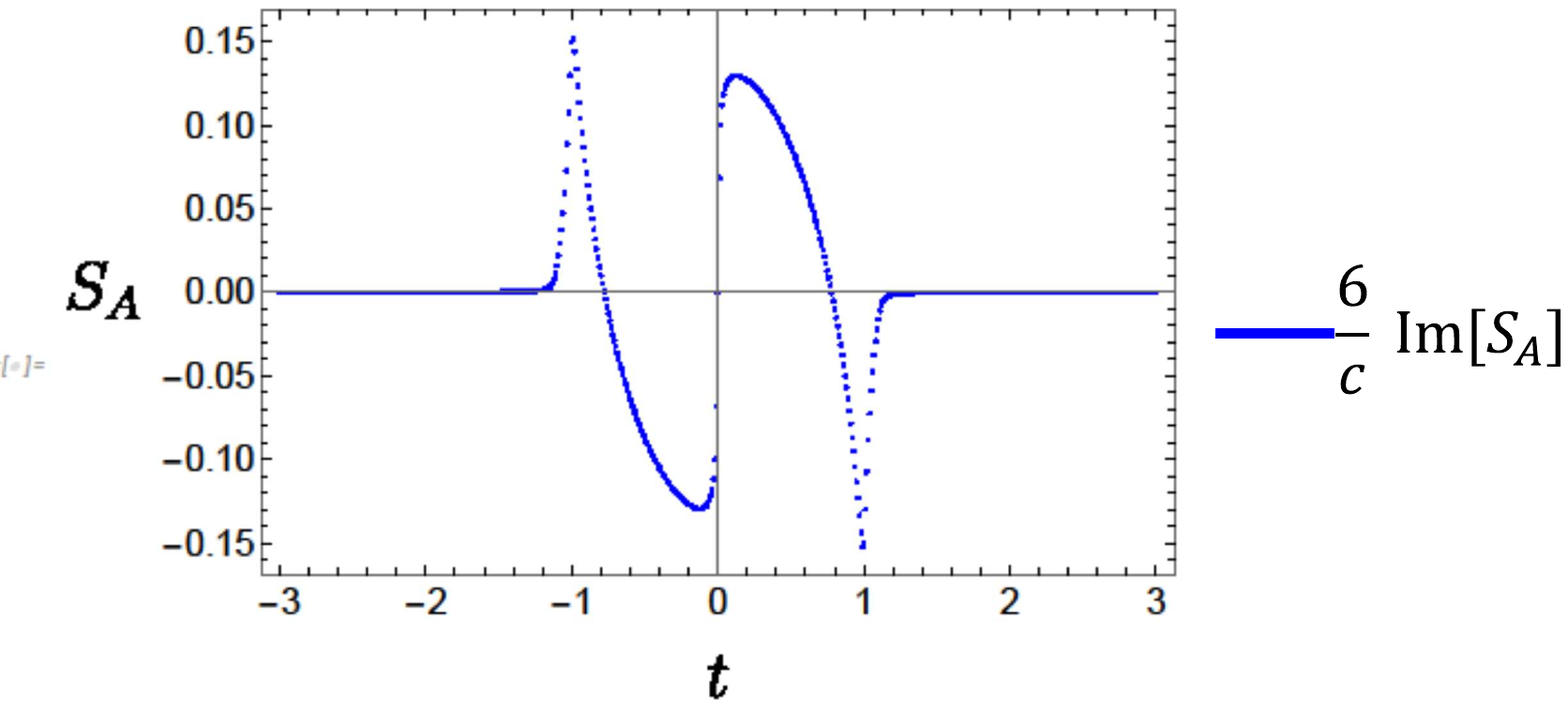}
    \end{minipage}
    
    \caption{The plots of the pseudo entropy in Lorentzian time evolution after the Wick rotation. the first picture is plot with $S_A(t_0)$. $\gamma= 0.4, \ep= \ep'=0.01,l =1$ in $t_0\in[0,5]$. The first figure shows the real part of the PE and the second one is its imaginary part.}
    \label{fig:Euclidean Model PE Lorentz time}
\end{figure}

As we can solve the geodesic equation in terms of the elliptic integrals, we can also evaluate the geodesic length in an explicit manner. Indeed, the length of the HRT surface $\Gamma_A$ is given by 
\begin{equation}\label{eq:geodesic length 1}
\begin{split}
    \mathrm{Area}[\Gamma_A] &= 2 \int_{r_t}^{r_\infty}dr\sqrt{\frac{f(r)}{f(r)-\alpha}}\\ 
    &= \qty[\sqrt{\frac{m}{\alpha \chi}}\qty((\chi+1) \EllipticF{H(r)}{m}+(2 \alpha -\chi -1)\EllipticPi{n}{H(r)}{m})]^{r_\infty}_{r_t},
\end{split}
\end{equation}
where $\EllipticPi{n}{z}{m}$ is the incomplete elliptic integral of the third kind and we introduce 
\begin{equation}\label{eq:H(r)}
    \begin{split}
         H(r)&:=\Arcsin{\left(\frac{1}{2} \sqrt{\frac{(\chi
   +1) (-2 \alpha +\chi  \cosh (2 r)+1) \text{csch}^2(r)}{\alpha  \chi }}\right)},\\m&:=\frac{4 \alpha  \chi }{(\chi+1) (2 \alpha
   +\chi -1)},\quad n:= \frac{2\alpha}{\chi+1}.
    \end{split}
\end{equation}\\
By using some identity involving the elliptic integrals and the formula for the cutoff (\ref{eq:cut off surface}) and the turning point (\ref{eq:turning point}), we can simplify the expression of the geodesic length. At the turning point (\ref{eq:turning point}), we can easily see $H(r_t)=0$ and 
\begin{equation}
    \EllipticF{H(r_t)}{m} = 0,\quad \EllipticPi{n}{H(r_t)}{m}=0.
\end{equation}
Next, we try to simplify the $r=r_\infty$ part of (\ref{eq:geodesic length 1}) in the specific range. To this end, let us recall the following formula \cite{elipticPi}.
\begin{equation}\label{eq:Elliptic Pi formula}
\begin{split}
    \EllipticPi{n}{z}{m} &= \EllipticF{z}{m}-\EllipticPi{\frac{m}{n}}{z}{m} \\ &+ \frac{1}{2\sqrt{\frac{(n-m)(n-1)}{n}}}\log\left(\frac{\sqrt{\frac{(n-m)(n-1)}{n}} \tan{z}+\sqrt{1-m\sin^2{z}}}{-\sqrt{\frac{(n-m)(n-1)}{n}} \tan{z}+\sqrt{1-m\sin^2{z}}}\right).
\end{split}
\end{equation}
Finally we have an expression for the geodesic length,
\begin{equation}\label{eq:geodesic length exact}
\begin{split}
    \mathrm{Area}[\Gamma_A] &= \sqrt{\frac{m}{\alpha \chi}}\qty(2\alpha \EllipticF{H(r_\infty)}{m}-(2\alpha - \chi -1) \EllipticPi{\frac{m}{n}}{H(r_\infty)}{m})\\
         &+\log\left(\frac{\sqrt{\frac{(n-m)(n-1)}{n}} \tan{H(r_\infty)}+\sqrt{1-m\sin^2{H(r_\infty)}}}{-\sqrt{\frac{(n-m)(n-1)}{n}} \tan{H(r_\infty)}+\sqrt{1-m\sin^2{H(r_\infty)}}}\right).
\end{split}
\end{equation}
 The pseudo entropy $S_A$ is given by the usual HRT formula (\ref{RTF}). As similar to the discussion in the previous subsection, we also consider the analytic continuation to Lorentzian path integral at $\tau_0=0$. Formally, this is done by replacing $\tau_0= i t_0+\ep$.  
 We show that the Euclidean and Lorentzian time evolution of the pseudo entropy $S_A$ in figure \ref{fig:Euclidean Model PE Euclid time} and figure \ref{fig:Euclidean Model PE Lorentz time} respectively\footnote{We should take care about the branch of $\sqrt{\frac{(n-m)(n-1)}{n}}$.}. We see that the pseudo entropy grows mildly as $S_A \sim \log{\frac{\tau_0}{l}}$ in the Euclidean time, and in Lorentzian time at time up to the half of the subsystem size $l$. At late time, the pseudo entropy reaches the value of vacuum entanglement entropy $S_0= \frac{c}{3}\log\frac{2l}{\ep}$. For the Euclidean case, this is because the large Euclidean time evolution makes the states to the vacuum of the Hamiltonian. For the Lorentzian time case, we may interpret that the reduced transition matrix forgets about the information, that is, one of the defects. However, of course, the global transition matrix contains the information. The pseudo etropies take small value around the defect. This is because the two vacuum states are different states for $\gamma\neq 0$.
 For the Lorentzian time evolution, the pseudo entropy moves up and down around $t_0 \sim l$. This behavior should be checked from the boundary side. We leave this as a future problem. \par 
Let us then introduce a parameter $\delta$ as 
\begin{equation}
    \delta= \frac{2(2\alpha-\chi-1)}{\chi}e^{-2r_\infty} = 2(2\alpha-\chi-1)\qty(\frac{\ep}{2\tau_0})^2.
\end{equation}
This quantity is motivated by the formula for the definition of $r_\infty$. At least for the time scale $\ep \ll \abs{\tau_0}$, we can safely suppose that the parameter $\delta$ is small. Since $H(r_\infty) \approx\Arcsin{\sqrt{\frac{1-\delta}{n}}}$ for small $\delta$, we obtain 
\begin{equation}
    \begin{split}
        \EllipticPi{n}{H(r_\infty)}{m} &= \EllipticF{H_\infty}{m}-\EllipticPi{\frac{m}{n}}{H_\infty}{m}
        + \frac{\log \left(
        \frac{4}{\delta}\left(\frac{n}{n-1}+\frac{m}{n-m}\right)^{-1}\right)}{2 \sqrt{\frac{(n-1) (n-m)}{n}}}+O\left(\delta ^1\right),
    \end{split}
\end{equation}
where we introduce $H_\infty = \Arcsin{\sqrt{\frac{1}{n}}}$. Also note that $\EllipticF{H_\infty}{m}$ and $\EllipticPi{\frac{m}{n}}{H_\infty}{m}$ is $O(\delta^0)$.
\par Finally we obtain the formula for the geodesic length as follows
\begin{equation}\label{eq:Janus EE}
    \begin{split}
        \mathrm{Area}[\Gamma_A] &= \sqrt{\frac{m}{\alpha \chi}}\qty(2\alpha \EllipticF{H_\infty}{m}-(2\alpha - \chi -1) \EllipticPi{\frac{m}{n}}{H_\infty}{m})\\
         &+ \log \left(
        \frac{4}{\delta}\left(\frac{n}{n-1}+\frac{m}{n-m}\right)^{-1}\right) +O(\delta^1).
    \end{split}
\end{equation}
Note that the pseudo entropy has the logarithmic cut-off dependence similar to the vacuum entanglement entropy in CFT$_2$. Note that this formula only works when $\abs{\tau_0}\gg \ep$,\textit{i.e.} $\delta \ll1$.

\subsubsection{Late time behavior of the pseudo entropy}
\paragraph{Late Euclidean time $\tau_0 \gg l$}~\par
 Since $\tau_0 \to \infty$ is a limit away from the defect, we naturally expect that the situation is close to the vacuum. When $\gamma=0$, $\alpha= 1+\frac{\tau_0^2}{l^2}$, we expect $\alpha \to \infty$ as $\tau \to \infty$. To confirm this fact, we take $\tau_0\gg l$ and $\alpha \gg 1$ limit of (\ref{eq:Euclid alpha}). After a short algebra we find that
 \begin{equation}
     \begin{split}
         (\text{l.h.s})&= \frac{l}{\tau_0} +O\qty(\qty(\frac{l}{\tau_0})^3),\\
         (\text{r.h.s})&= \sqrt{\frac{1}{\alpha}}+O\qty(\frac{1}{\alpha^{\frac{3}{2}}}).
     \end{split}
 \end{equation}
Thus we can conclude $\alpha = \qty(\frac{\tau_0}{l})^2+\cdots$ for general $\gamma$. Also we can estimate the geodesic length (\ref{eq:Janus EE}) when $\tau_0 \gg l \gg \ep$. Then we see the pseudo entropy has the late Euclidean time behavior,
\begin{equation}
    S_A = \frac{c}{3}\log\qty(\frac{2l}{\ep})+ O\qty(\qty(\frac{l}{\tau_0})^2).
\end{equation}
As a result, the pseudo entropy approaches the vacuum entanglement, which matches our expectation.
\paragraph{Late Lorentzian time $t_0 \gg l$}~\par
With Wick rotation $\tau_0=it+\ep',t>0$, we find the real time dynamics with the quantum states prepared from the Janus path integral. Following the same procedure as above, we can analyze the late time behavior of the pseudo entropy. However, we should take care of the branch of the argument a bit. Similar to the Euclidean analysis, we expect $\alpha \sim \frac{\tau_0^2}{l^2} \sim -\frac{t^2}{l^2}\qty(1-\frac{2 i \ep'}{t})$. Then, choosing $y= \frac{1+\chi}{2\alpha}$, we find
\begin{equation}
\begin{split}
    -\tanh{\qty(\sqrt{\frac{-k}{\chi}}\qty(\EllipticF{\Arcsin{\sqrt{1-y}}}{k}-\EllipticK{k}))} &= -\sqrt{\frac{2 y}{1+\chi}}+O(y^{\frac{5}{2}})\\
    \frac{1}{\sqrt{1+\frac{\tau_0^2}{l^2}}} &= \frac{1}{\sqrt{-\frac{t_0^2}{l^2}}}+O\qty(\frac{l^3}{t_0^3}),
\end{split}
\end{equation}
in $\alpha \to -\infty, t_0 \to \infty$. Thus we confirm $\alpha = -\frac{t_0^2}{l^2}+\cdots$. Next, we evaluate the pseudo entropy at late time. First, let us evaluate the inside of the square root in the (\ref{eq:H(r)}) at $r=r_\infty$,
\begin{equation}
    q:= \frac{(\chi+1)(-2\alpha+\chi \cosh{(r_\infty)}+1)\text{csch}^2(r_\infty)}{2\alpha \chi} = -\frac{1+\chi}{2t^2}(l^2-\ep^2)\qty(1+i \frac{4l^2}{l^2-\ep^2}\frac{\ep'}{t})+\cdots .
\end{equation}
Since $q$ is a complex number, we should take care of the branch of the square root\footnote{$\sqrt{q}= e^{\frac{1}{2}\log{\abs{q}}+\frac{i}{2}\Arg{q}+i n\pi}= \pm \sqrt{\abs{q}}e^{\frac{i}{2}\Arg{q}}$, where $\Arg{\cdot}$ runs from $-\pi$ to $\pi$.}. We take the positive sign here. The negative sign branch just gives the pseudo entropy with the opposite sign of the positive one. Note that 
\begin{equation}
    \begin{split}
        \sqrt{\abs{q}}= \frac{l}{t}\sqrt{\frac{1+\chi}{2}}\sqrt{\qty(1-\frac{\ep^2}{l^2})\qty(1+\qty(\frac{4}{1-\frac{\ep^2}{l^2}}))^2\qty(\frac{\ep'}{t})^2}+\cdots
    \end{split}
\end{equation}
by recalling that $\Arg{q}= -\pi + \frac{2\ep'}{t}$ for $t>0$. In this case,
\begin{equation}
    \begin{split}
        \sqrt{q}  = -i\frac{l}{t}\sqrt{\frac{1+\chi}{2}}\sqrt{\qty(1-\frac{\ep^2}{l^2})\qty(1+\qty(\frac{4}{1-\frac{\ep^2}{l^2}}))^2\qty(\frac{\ep'}{t})^2} e^{i \frac{2\ep'}{t}}+\cdots,
    \end{split}
\end{equation}
\begin{equation}
    \begin{split}
        H(r_\infty) = \Arcsin{\sqrt{q}}= -i \frac{\sqrt{l^2-\ep^2}}{\sqrt{2}t}\sqrt{1+\chi}\sqrt{1+\qty(\frac{4l^2}{l^2-\ep^2})^2\qty(\frac{\ep'}{t})^2}e^{i\frac{2\ep'}{t}}+\cdots.
    \end{split}
\end{equation}
Similarly, we take the same branch for $\sqrt{\alpha}$,
\begin{equation}
    \begin{split}
        \sqrt{\alpha}=  \sqrt{\abs{\alpha}}e^{\frac{i}{2}\Arg{\alpha}} = i\frac{t}{l}\sqrt{1+\frac{4\ep'^2}{t^2}}e^{-i\frac{2\ep'}{t}}.
    \end{split}
\end{equation}
For small $z$, it holds
\begin{equation}
    2\alpha\EllipticF{z}{m}-(2\alpha-\chi-1)\EllipticPi{\frac{m}{n}}{z}{m} = (1+2\alpha +2\alpha \chi)z + O(z^3).
\end{equation}
By using this expansion, the first term of (\ref{eq:geodesic length exact}) has the series expansion
\begin{equation}
   \sqrt{\frac{m}{\alpha \chi}}\qty( 2\alpha\EllipticF{H(r_\infty)}{m}-(2\alpha-\chi-1)\EllipticPi{\frac{m}{n}}{H(r_\infty)}{m})= 2(1+\chi) +\cdots. 
\end{equation}
Similarly we see
\begin{equation}
    \log{\qty(\frac{ \sqrt{\frac{(n-m)(n-1)}{n}}\tan{H(r_\infty)}+\sqrt{1-m \sin^2{H(r_\infty)}}}{-\sqrt{\frac{(n-m)(n-1)}{n}}\tan{H(r_\infty)}+\sqrt{1-m \sin^2{H(r_\infty)}}})} = \log{\qty(\frac{4l^2}{\ep^2})}+ O\qty(\frac{l^2}{t_0^2},\frac{\ep^2}{t^2}).
\end{equation}
From this, we can confirm that the pseudo entropy approaches the vacuum entanglement entropy in the Lorentzian late time in this branch.

\subsubsection{Early time behavior}
Here we consider the behavior of the pseudo entropy in the period of  $\ep \ll \abs{\tau_0} \ll l$. Again, we consider the subsystem $A$ with the length $2l$ (i.e. $A=[-l,l]$) at the specific time $\tau_0\simeq \xi(\infty)$. 
We find the relation between the value of $r$ at the turning point $r_t$ and $L$:
\ba
\frac{l}{\xi(\infty)}=\sinh \left[\int^{\infty}_{r_t}\frac{dr}{\s{f(r)\left(\frac{f(r)}{f(r_t)}-1\right)}}\right].
\ea
Let us consider the limit $r_t\to 0$ or equally $\frac{L}{\xi(\infty)}\gg 1$ . By using the fact that the dominant contribution for the integration is around $r\sim r_t$ and for rough calculation we can obtain $f(r)$ to $f(r_t)$ and since $r_t$ is small $f(r_t)=\alpha = \frac{1+\chi}{2}$, we obtain
\ba
\int^{\infty}_{r_t}\frac{dr}{\s{f(r)\left(\frac{f(r)}{f(r_t)}-1\right)}}
\simeq -\frac{1}{\chi}\log\left[\s{\frac{1-\chi}{1+\chi}}\frac{r_t}{4}\right]+O(r_t^2).
\ea
Thus, also by using $\tau_0\simeq \xi(\infty)$, we find
\ba
\frac{L}{\tau_0}\simeq \frac{1}{2}\left(\s{\frac{1-\chi}{1+\chi}}\frac{r_t}{4}\right)^{-\frac{1}{\s{\chi}}}.
\ea
The UV cutoff can be estimated for $\tau_0\gg\ep$ by $\ep\simeq \frac{\tau_0}{\s{f(r_\infty)}}$ i.e.
\be
r_\infty=\log\left(\frac{2}{\s{\chi}}\frac{\tau_0}{\ep}\right).
\ee
Now we can estimate the area of $\mathrm{\Gamma_A}$ in a similar way. We can evaluate its behavior in the early time limit $r_t\to 0$ as follows
\ba
\mathrm{Area}\qty[\Gamma_A]&=& 2 \int_{r_t}^{r_\infty}dr\sqrt{\frac{f(r)}{f(r)-\alpha}}\no &\simeq& 2r_\infty-\frac{\s{2(1+\chi)}}{\chi}\log \left(\s{\frac{1-\chi}{1+\chi}}\frac{r_t}{4}\right)+O(1),\no
&\simeq& -2\log\ep+\left(2-\frac{\s{2(1+\chi)}}{\chi}\right)\log\tau_0
+\frac{\s{2(1+\chi)}}{\chi}\log l.
\ea
The first term comes from the integral around $r_\infty$. Here $f(r_\infty)$ is large and integral for $\mathrm{Area}\qty[\Gamma_A]$ is evaluated in this large $r_\infty$ limit. From this result, we find that the pseudo entropy grows logarithmically in the early time region for both Euclidean and Lorentzian time evolution.

\subsubsection{Entanglement entropy on the defect; At \texorpdfstring{$\tau_0=t_0=0$}{tau0=t0=0}}
 From the time-reversal symmetry at $\tau_0=t_0=0$, we can easily obtain the geodesic formula for the subsystem on the defect. Especially we see that the geodesic is on the $r=0$ surface. Fortunately, the metric on the $r=0$ surface takes the same form as that of the constant time slice of the Poincar\'e AdS$_3$, \textit{i.e.}
 \begin{equation}
     ds^2|_{r=0}= f(0)\frac{d\xi^2+dx^2}{\xi^2}.
 \end{equation}
  The cutoff surface is given by $\xi= \ep f(0)$. Thus the entanglement entropy is given by the one of the vacuum entanglement entropy with the central charge $c_{\mathrm{eff}}= \frac{2l_\mathrm{AdS}}{3G_N} \sqrt{f(0)}$ and cutoff $\ep_{\mathrm{eff}}=\ep f(0)$. As a result, we have
  \begin{equation}
      S_A(\tau_0=t_0=0) = \frac{c_{\mathrm{eff}}}{3}\log{\frac{2l}{\ep_{\mathrm{eff}}}}= \sqrt{\frac{1+\chi}{2}}\frac{c}{3}\log\qty(\frac{2}{1+\chi}\frac{2l}{\ep}).
  \end{equation}
  Note that the coefficient $c_{\mathrm{eff}}$ does not depend on the choice of the cutoff. The effective central charge is a monotonically decreasing function for the deformation strength $\gamma$.  
  From this fact we may say that the defect plays a role of the measurement. 
  

\subsection{Pseudo entropy from time-like Janus solution}\label{sec:Lorentzian Model; Final State Projection}
 \subsubsection{The setup}
  Here we treat the Lorentzian path integral directly in the holographic CFT dual to a spacelike brane on $t=0$, instead of performing analytical continuation from the Euclidean setup. The path integral again gives the transition matrix of two distinct vacuum states of the two CFTs. More specifically, on a constant time slice $t=t_0$, we prepare the transition matrix defined through two states 
  \begin{equation}
  \begin{split}
      \mathcal{T}(t_0) &= \lim_{T\to\infty}  e^{-i(T-\abs{t_0})H_-}\ket{0_-}\bra{0_+}e^{+iT H_+}e^{i\abs{t_0}H_-}\\
       &=\ket{0_-}\bra{0_+} e^{+i\abs{t_0} H_-}.
  \end{split}\label{ljanup}
  \end{equation}
 The second expression is the same as the analytic continuation of the Euclidean model and as we see later we have the same plots of the pseudo entropy. Note that if we prepare quantum states for initial and final states which are not the eigenstates of the Hamiltonians, then we will have a different transition matrix from the one obtained from the analytic continuation. See figure \ref{fig:Euclidean vs Lorentzian} for a sketch. As we will discuss from now on, this path integral is holographically dual to the Lorentzian Janus solution which gives a thick spacelike brane in the gravity side. 
    
\begin{figure}[h]
        \centering
        \begin{tikzpicture}
 \fill[fill=lightgray](0,-1)rectangle(4,7);
 \draw[<-](-1,7)--(-1,-1);
 \draw(-1,7)node[left]{$t$};
 \draw[blue,ultra thick](0,3)--(4,3);
 \draw[red](1,1)--(3,1);
 \draw[red,dashed](1,1)--(-1.2,1);
 \draw(2,0)node[above]{\textcolor{red}{$A$}};
 \draw(-1.2,3)--(-0.8,3);
 \draw(-1.2,3)node[left]{$t=0$};
  \draw(-1.2,1)node[left]{$t=t_0$};
    \draw(4,3)node[right]{\LARGE spacelike defect};
    \draw(-1.4,7)node[left]{Lorentzian Time};
 \end{tikzpicture}
        \label{fig:Janus/Loretain model}
        \caption{Sketch for the path integral of the spacelike brane model for the measurement. We expect that the spacelike defect decreases entanglement like measurement. The left figure shows the vacuum process and the right figure shows the multiple measurement and quantum quench. In this process, there are entanglement creation from quench and decreasing from defect. So we expect an entanglement transition occurs.}
\end{figure}
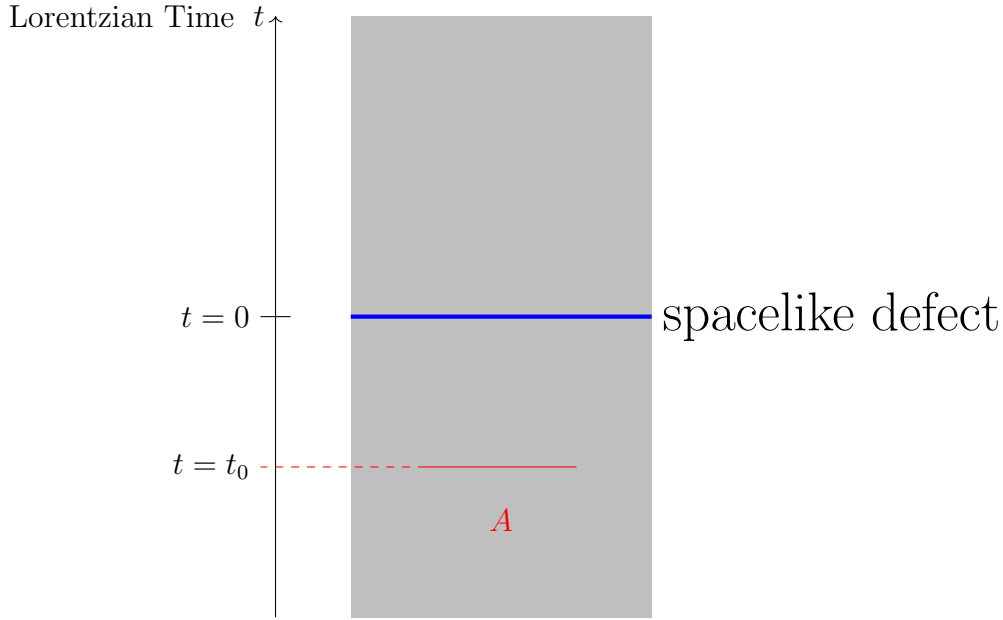
\begin{figure}[h]
    \centering
    \begin{minipage}[h]{0.45\linewidth}
        \centering
        \begin{tikzpicture}
\draw[blue](0,0)--(0,2);
\draw[blue](0,2)--(-2,2);
\draw[blue](-2,2.3)--(0,2.3);
\draw[blue](0,2.3)--(0,2.5);
\draw[blue,dashed](-2,2.3)--(-2.3,2.3);
\draw[blue,dashed](-2,2)--(-2.3,2);
\draw[red](0,2.5)--(0,4);
\draw(-1,2.7)node[above]{$t_0$};
\draw[black,<->](0,2.6)--(-2,2.6);
\draw(0,0)node[below]{$\ket{0_-}$};
\draw(0,4)node[above]{$\bra{0_+}$};
 \end{tikzpicture}
    \end{minipage}
    \begin{minipage}[h]{0.45\linewidth}
        \centering
        \begin{tikzpicture}
\draw[blue,dashed](0,0)--(0,0.3);
\draw[blue](0,0.3)--(2,0.3);
\draw[black](2,0)--(2,0.6);
\draw[blue](2,0.3)--(2.5,0.3);
\draw[red](2.5,0.3)--(4,0.3);
\draw[red,dashed](4,0.3)--(4,0.6);
\draw(1,0.5)node[above]{$T-\abs{t_0}$};
\draw[black,<->](0,0.5)--(2,0.5);
\draw(2.25,0.5)node[above]{$\abs{t_0}$};
\draw[black,<->](2,0.5)--(2.5,0.5);
\draw(3,0.5)node[above]{$T$};
\draw[black,<->](2.5,0.5)--(4,0.5);
\draw(0,0)node[below]{$\ket{0_-}$};
\draw(4,0.6)node[above]{$\bra{0_+}$};
 \end{tikzpicture}
    \end{minipage}
     \caption{The path integrals in the complex time plane for two models in this section. The vertical direction is Lorenztian and the parallel direction shows the Euclidean. The left shows the path integral for the analytic continuation of the Euclidean model in the last subsection. The right shows the path integral for the current Lorenzian model. If the initial states and final states are the vacuum states then these two are equivalent.}
     \label{fig:Euclidean vs Lorentzian}
\end{figure}
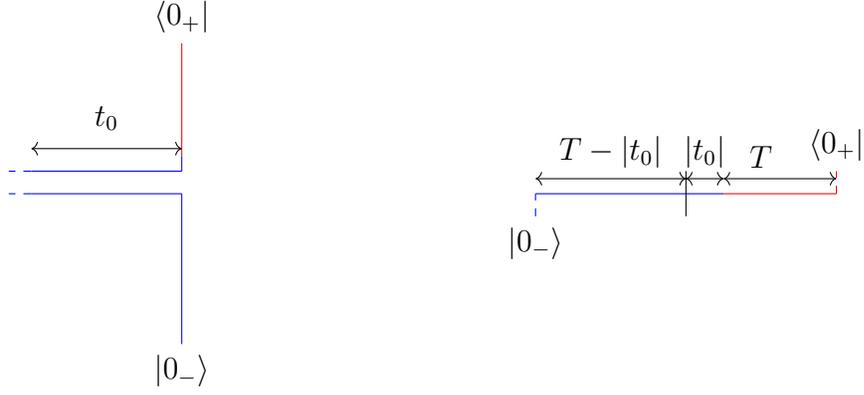
We know that the vacuum AdS$_3$ with the planar conformal boundary can be written in the Poincar\'e coordinate. 
 \begin{equation}
     ds^2 = \frac{-dt^2+dz^2+dx^2}{z^2}
 \end{equation}
We can introduce new coordinate $(r,\xi,x)$ by the following coordinate transformation from the Poincar\'e coordinate,
 \begin{equation}
 \begin{split}
      z &= \frac{\xi}{\cos{r}},\quad t= \xi \tan{r},\\
      \sin{r} &= \frac{t}{z},\quad z^2-t^2 =\xi^2.
 \end{split}
 \end{equation}
 In this coordinate, the metric takes the form of 
 \begin{equation}
     ds^2= -dr^2 +\cos^2{r}\frac{d\xi^2 +dx^2}{\xi^2}.
 \end{equation}
 If we suppose that $(r,\xi)$ is defined on the region such that
 \begin{equation}
     -\frac{\pi}{2}\leq r \leq \frac{\pi}{2},\quad \xi \geq 0,
 \end{equation}
 then this region is the restricted region in the Poincar\'e patch
 \begin{equation}
     -z \leq t \leq z.
 \end{equation}
  We call this region as region $\gretwo$.
 To cover the other regions, we can analytically continue $(r,\xi)$ to the complex value. We have two other regions. 
 The $t\geq z$ region, called region $\greone$, is defined by
 \begin{equation}
 \begin{split}
          r &= -i\rho +\frac{\pi}{2},\quad \rho\geq 0, \quad \xi = i\eta,\quad \eta \geq 0\\
          z  &= \frac{\eta}{\sinh{\rho}},\quad t = \eta \coth{\rho}.
 \end{split}
 \end{equation}
 Similarly, the $t\leq -z$ region, called region $\grethree$, is defined by
 \begin{equation}
 \begin{split}
       r &= i \rho -\frac{\pi}{2},\quad \rho\geq 0,\quad \xi = i \eta,\quad \eta \geq 0\\
     z  &= \frac{\eta}{\sinh{\rho}},\quad t = -\eta \coth{\rho}.
 \end{split}
 \end{equation}
 
 \begin{figure}[h]
     \centering
\includegraphics[width=0.7\linewidth]{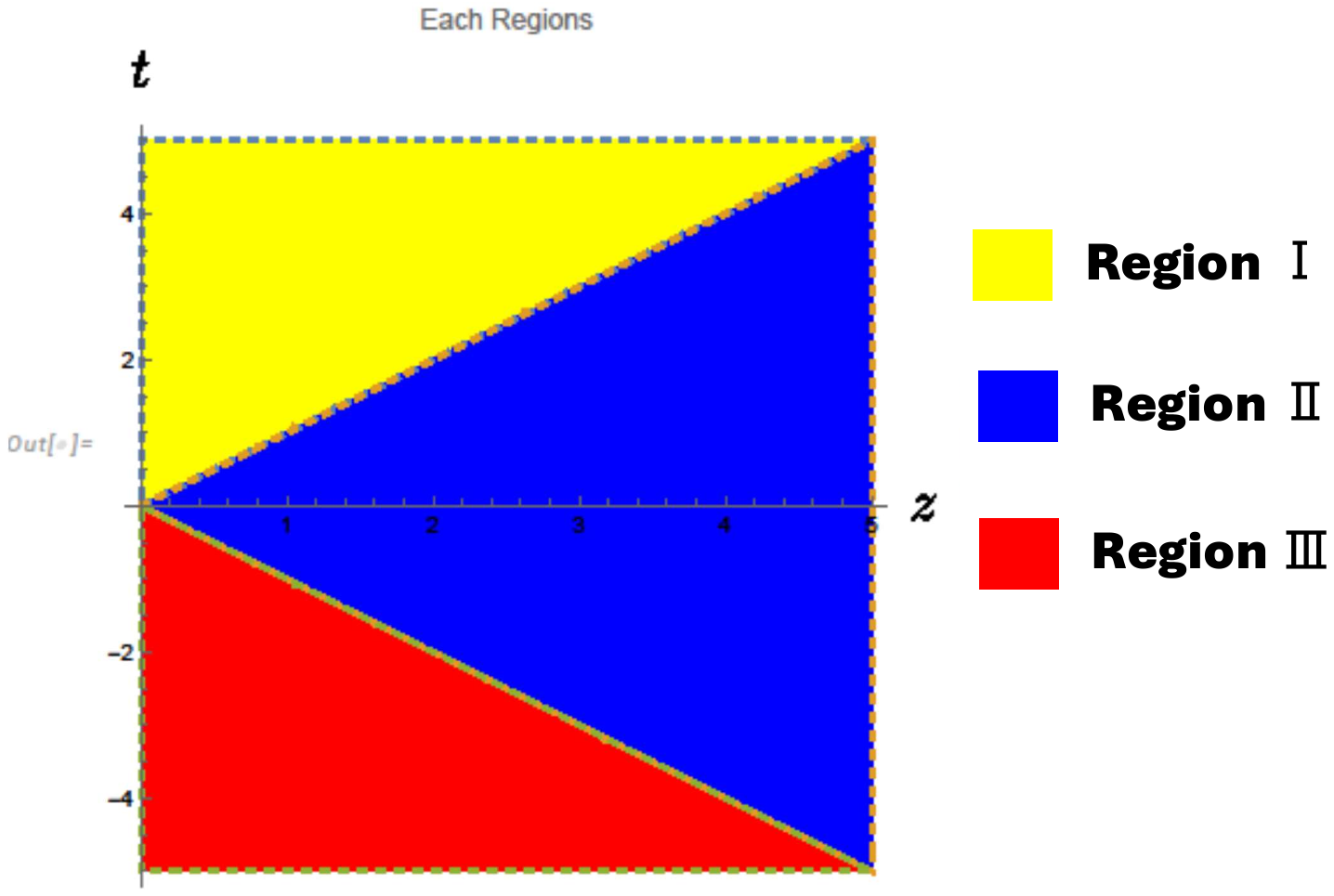}
\caption{Three regions in the vacuum AdS solution with dS$_2$/H$_2$ slice coordinate. The blue region represents the region $\gretwo$, the yellow represents the region $\greone$ and the red $\grethree$.}
\label{fig:Penrose Diagram Lorentzian Model}
 \end{figure}

\begin{figure}[h]
            \centering \includegraphics[width=0.5\linewidth]{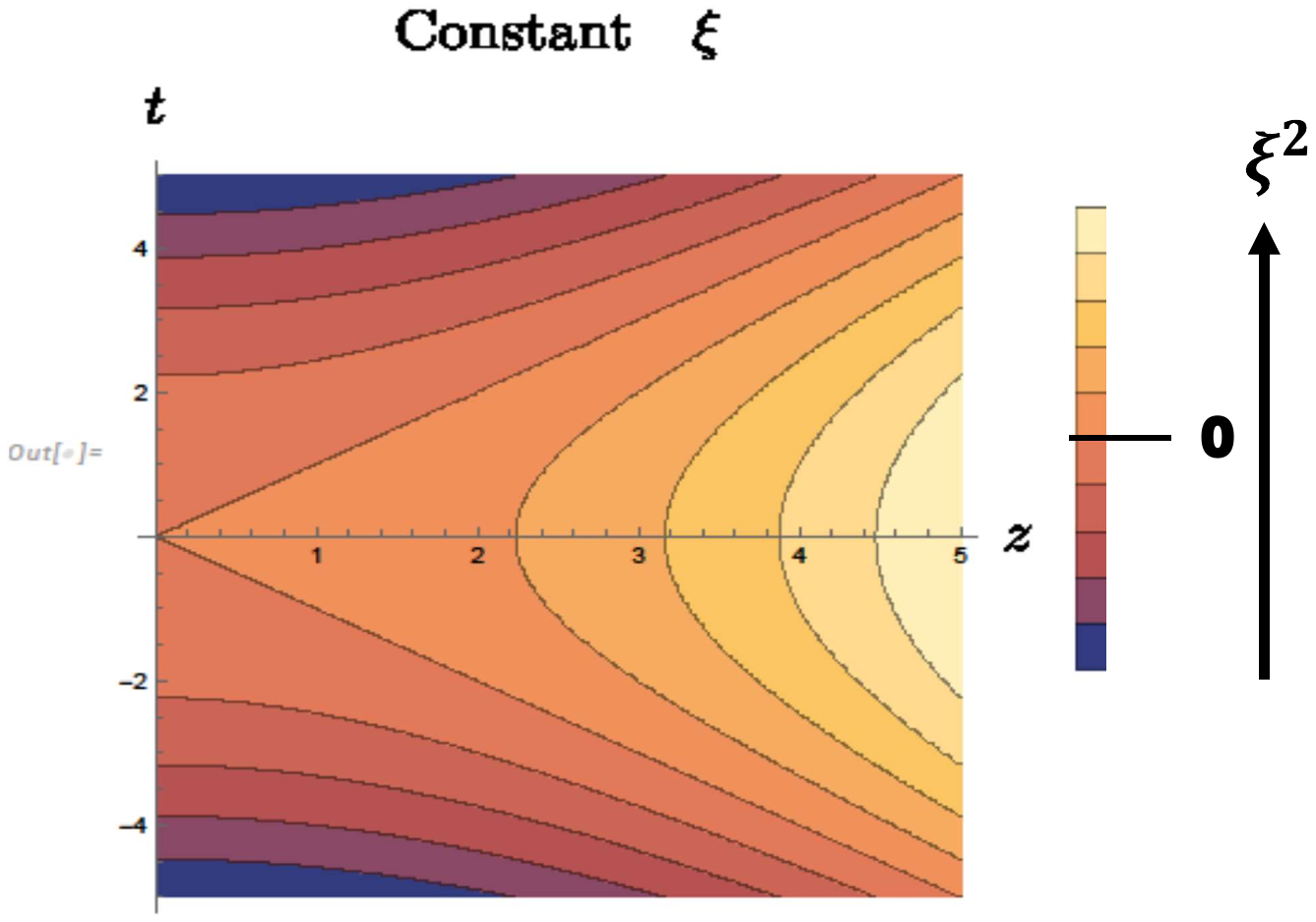}
            \includegraphics[width=0.4\linewidth]{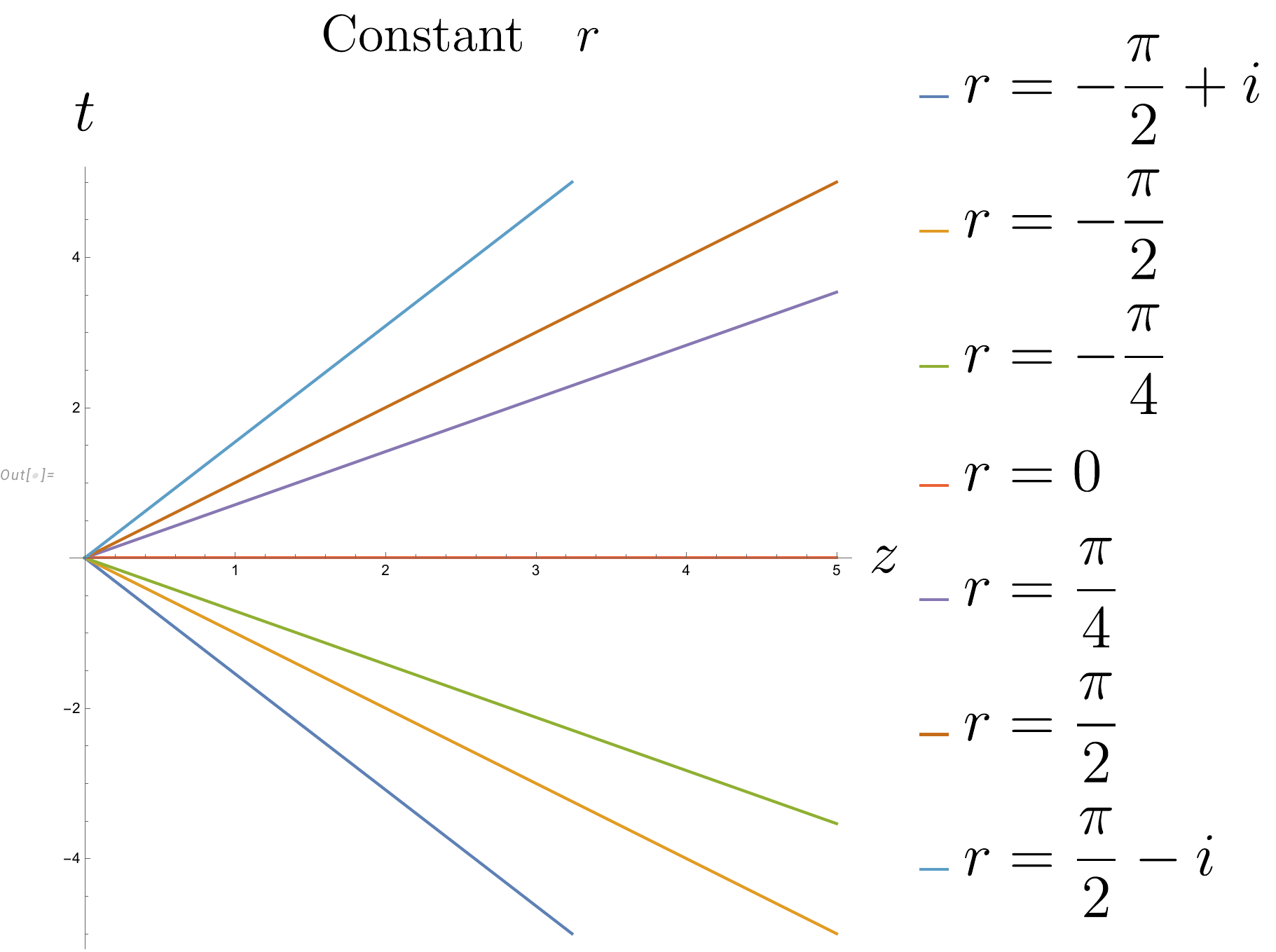}
            \caption{The contour maps of the constant $\xi^2$ surface (left) and the constant $r$  surface (right).}
            \label{fig:Constant xi}
\end{figure}

Now, the gravity dual of the Lorentzian time evolution of the transition matrix (\ref{ljanup}) is obtained by solving the Einstein equation with the dilaton field. The solution can also be obtained from the double Wick rotation of the original timelike Janus solution \cite{Bak:2003jk}. The final solution looks like
  \begin{equation}
        \text{Region $\gretwo$}: \quad 
      \begin{aligned}
          ds^2 &= -dr^2 + f(r)\frac{d\xi^2 +dx^2}{\xi^2},\\
          f(r) &= \frac{1}{2}\qty(1+\sqrt{1-2\gamma^2}\cos{2r}),\\
          \Phi(r) &= \frac{1}{\sqrt{2}}\log\qty(\frac{1+\sqrt{1-2\gamma^2}+i\sqrt{2}\gamma\tan{r}}{1+\sqrt{1-2\gamma^2}-i\sqrt{2}\gamma\tan{r}}).
      \end{aligned}
  \end{equation}
  If we focus on the region $\gretwo$, where $r$-coordinate is real, and 
  if $\gamma$ is real and $\abs{\gamma}<\frac{1}{\sqrt{2}}$, then the dilaton field is purely imaginary,
  \begin{equation}\label{eq:region one}
      \begin{split}
           \Phi(r) &= i \frac{2 \gamma \tan{r}}{1+\sqrt{1-2\gamma^2}}.
      \end{split}
  \end{equation}
  Note that this solution matches the wick rotated geometry of our Euclidean model (\ref{eq:Euclid Janus solution}) with peculiar Wick rotation $r \to i r$. By doing the appropriate analytic continuation, we can obtain the solution for the other two regions, 
\begin{equation}
\text{Region $\greone$}: \quad 
      \begin{aligned}
          ds^2 &= d\rho^2 -f(\rho)\frac{-d\eta^2 +dx^2}{\eta^2},\\
          f(\rho) &= \frac{1}{2}\qty(1-\sqrt{1-2\gamma^2}\cosh{2\rho}),\\
          \Phi(\rho) &= \frac{1}{\sqrt{2}}\log\qty(\frac{1+\sqrt{1-2\gamma^2}+\sqrt{2}\gamma\coth{\rho}}{1+\sqrt{1-2\gamma^2}-\sqrt{2}\gamma\coth{\rho}}),
      \end{aligned}
  \end{equation}
   \begin{equation}\label{eq:region three}
   \text{Region $\grethree$}: \quad 
      \begin{aligned}
          ds^2 &= d\rho^2 -f(\rho)\frac{-d\eta^2 +dx^2}{\eta^2},\\
          f(\rho) &= \frac{1}{2}\qty(1-\sqrt{1-2\gamma^2}\cosh{2\rho}),\\
          \Phi(\rho) &= \frac{1}{\sqrt{2}}\log\qty(\frac{1+\sqrt{1-2\gamma^2}-\sqrt{2}\gamma\coth{\rho}}{1+\sqrt{1-2\gamma^2}+\sqrt{2}\gamma\coth{\rho}}).
      \end{aligned}
  \end{equation}
 We have two comments on this metric.
 One is that there exist lines $\rho=\rho_c$ such that 
 \begin{equation}
      f(\rho_c)= 0,
 \end{equation}
 where the dS$_2$ slices shrink to a point. Note that on this critical line, 
  \begin{align}
      1+\sqrt{1-2\gamma^2}-\sqrt{2}\gamma\coth{\rho_c}=0,
  \end{align}
  holds and the dilaton field diverges. These show that the solution becomes singular at $\rho=\rho_c$. However, we proceed further by assuming that, since we consider a non-Hermitian evolution, it may have such an exotic feature in its gravity dual.  Moreover, for $0\leq\rho<\rho_c$, $f(\rho)>0$ and $\Phi(\rho)$ takes a complex value, which looks like a feature also common to our brane localized scalar model. The second comment is on the boundary where CFTs live. Because of the Wick rotation, in region $\gretwo$, the warped factor for the $H_2$ slices does not blow up for every $r$. On the other hand, for the region $\greone$ and $\grethree$, at the $\rho\to\infty$ limit, 
  the factor $-f(\rho)$ in (\ref{eq:region one}, \ref{eq:region three}) for the dS$_2$ blows up exponentially. Thus each region of the spacetime (\ref{eq:region one}) (\ref{eq:region three}) is asymptotically AdS. Thus we consider that the holographic theory lives on these asymptotic boundaries.
   \begin{figure}
     \centering
   \begin{tikzpicture}
   \draw(0,0)--(0,4);
   \draw(0,2)--(4,4);
    \draw[dashed](0,2)--(5,4.4);
    \draw(5.2,4.4)node[right]{$\rho=0,r=\frac{\pi}{2}$};
      \draw[dashed](0,2)--(5,-0.6);
    \draw(5,-0.6)node[right]{$\rho=0,r=-\frac{\pi}{2}$};
    \draw[dashed](0,2)--(3,5);
    \draw(3,5)node[above]{$\rho=\rho_c$};
    \draw[dashed](0,2)--(3,-1);
    \draw(3,-1)node[below]{$\rho=\rho_c$};
   \draw(0,2)--(4,0);
   \draw(0,2)--(2,0);
   \draw(0,2)--(2,4);
   \fill[red](0,2)--(0,0)--(2,0)--cycle;
   \fill[blue](0,2)--(2,4)--(0,4)--cycle;
   \fill[blue!30!white](0,2)--(2,4)--(4,4)--cycle;
   \fill[red!30!white](0,2)--(2,0)--(4,0)--cycle;
   \fill[yellow](0,2)--(4,4)--(4,0)--cycle;
   \draw[<->](5,1)to[out=45,in=-45](5,3);
   \draw(5.9,2)node{$\rho,r$};

   \draw(1,4)node[above]{{\textcolor{blue}{$f(\rho)>0$}}};
   \draw(3.3,4)node[above]{{\textcolor{blue}{$f(\rho)<0$}}};
   \draw(1,0)node[below]{{\textcolor{red}{$f(\rho)>0$}}};
   \draw(3.3,0)node[below]{{\textcolor{red}{$f(\rho)<0$}}};

   \draw[green,thick](0.2,0)--(0.2,4);
   \end{tikzpicture}
     \caption{Sketch for the Janus spacetime duals to the CFT with spacelike interface with nonzero $\gamma$. The red and pink region denotes the region $\grethree$, the yellow region denotes the region $\gretwo$ and the blue and the light blue region denotes $\greone$. The green line denotes the sketch of the cutoff surface we are introducing.}
     \label{fig:Janus Penrose}
 \end{figure}
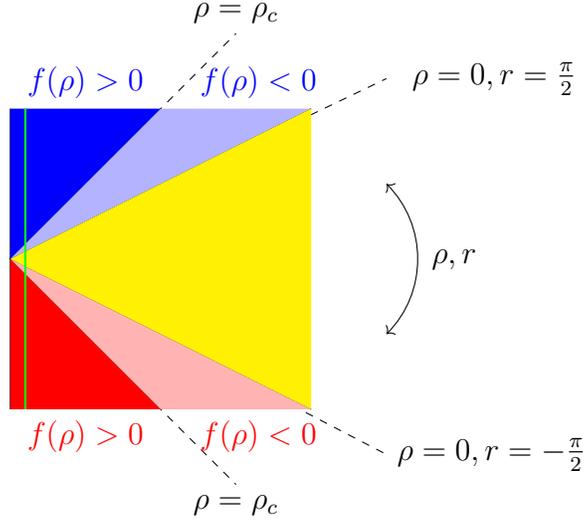

 \subsubsection{Time evolution of pseudo entropy}
\begin{figure}[h]
      \centering
\includegraphics[width=0.7\linewidth]{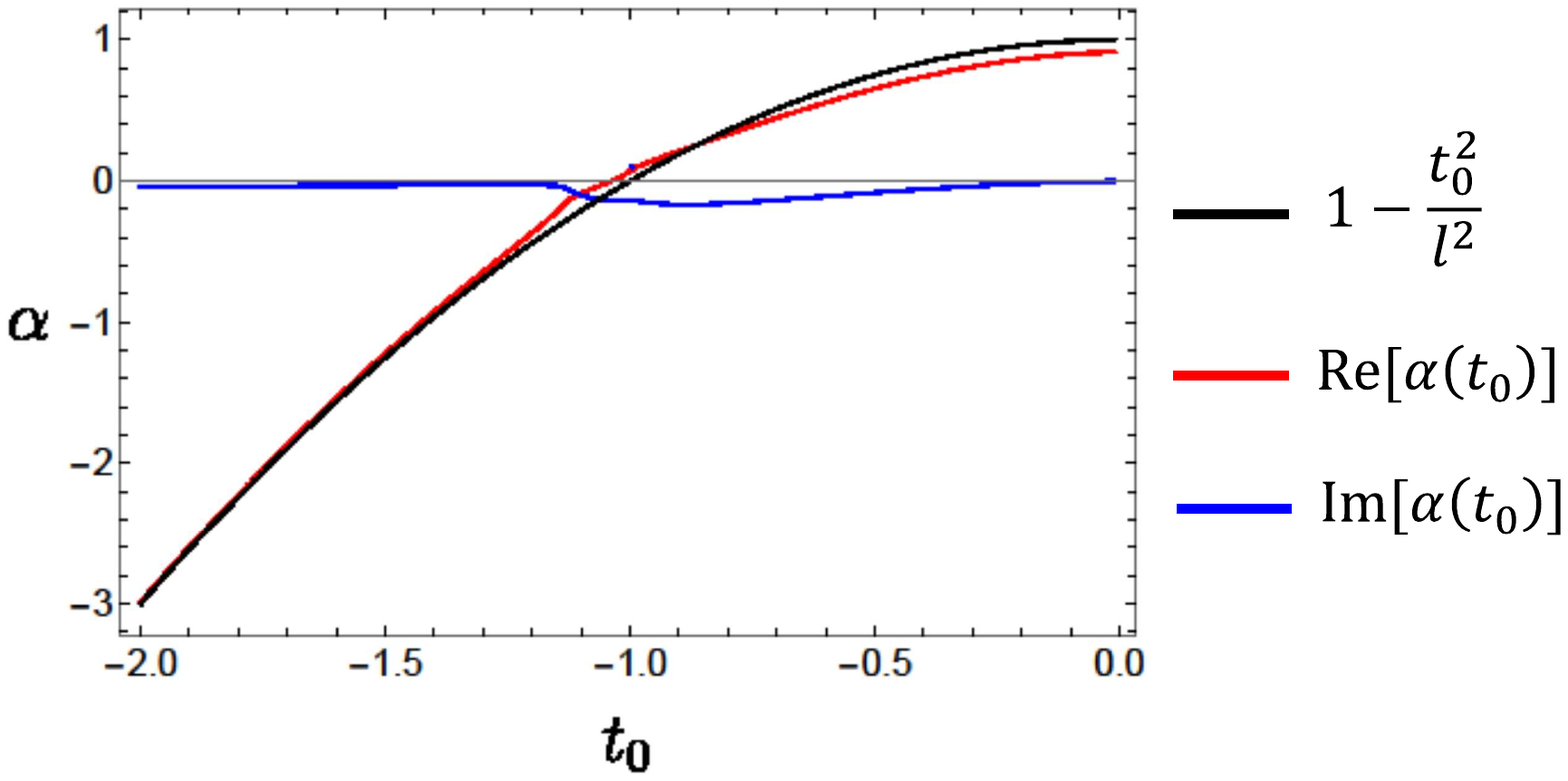}
      \caption{Plots for the $\alpha(t_0-i\ep')$. The parameters are chosen such that $l=1, \gamma=0.4,\ep=\ep'=0.01$. $1-\frac{t_0^2}{l^2}$ is the $\alpha$ for the vacuum AdS$_3$ case,\textit{i.e.}, $\gamma=0,\chi=1$.}
      \label{fig:lorentzian alpha}
  \end{figure}

\begin{figure}[htbp]
  \begin{minipage}[b]{0.53\linewidth}
    \centering
    \includegraphics[width=1.2\linewidth]{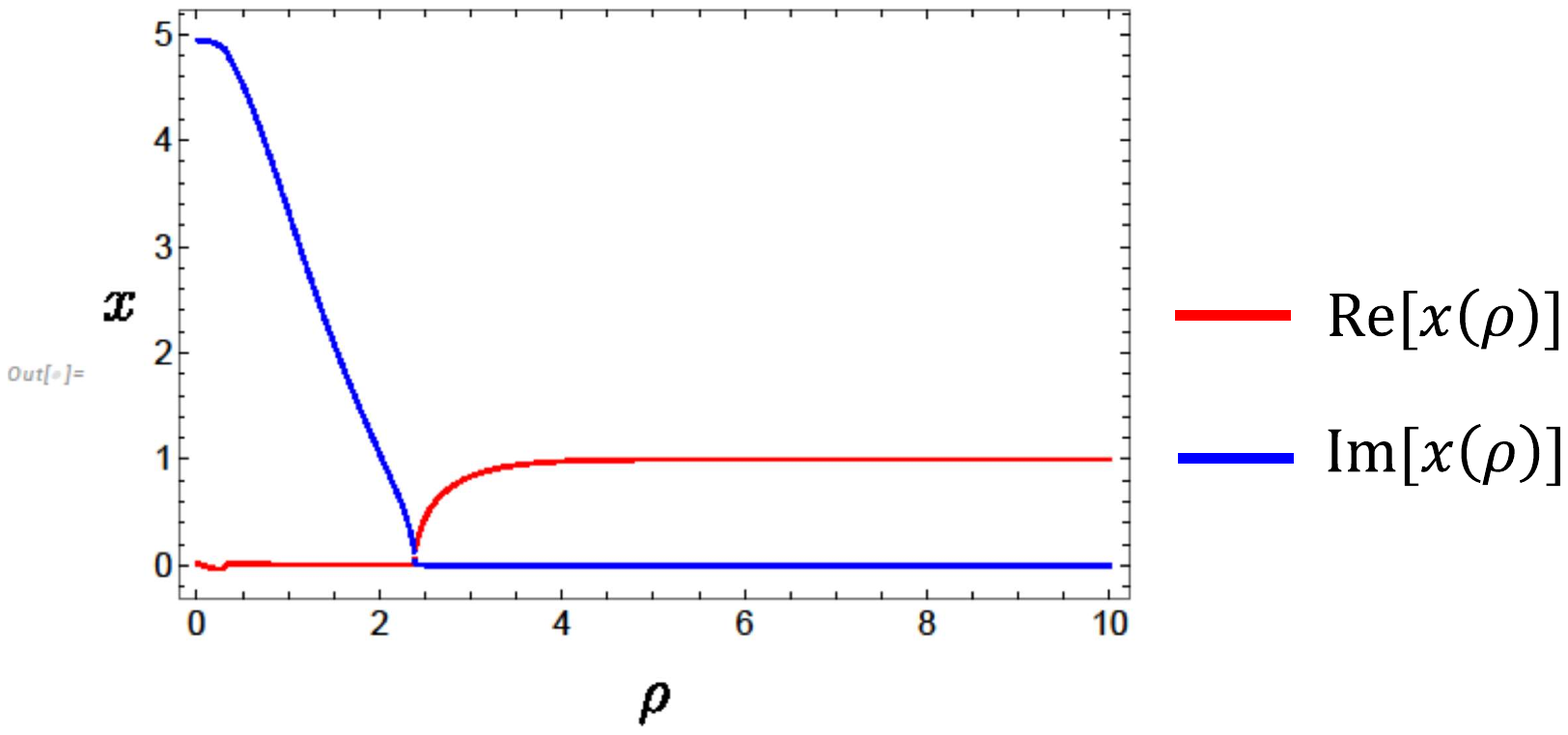}
  \end{minipage}
  \begin{minipage}[b]{0.53\linewidth}
    \centering
    \includegraphics[width=1.2\linewidth]{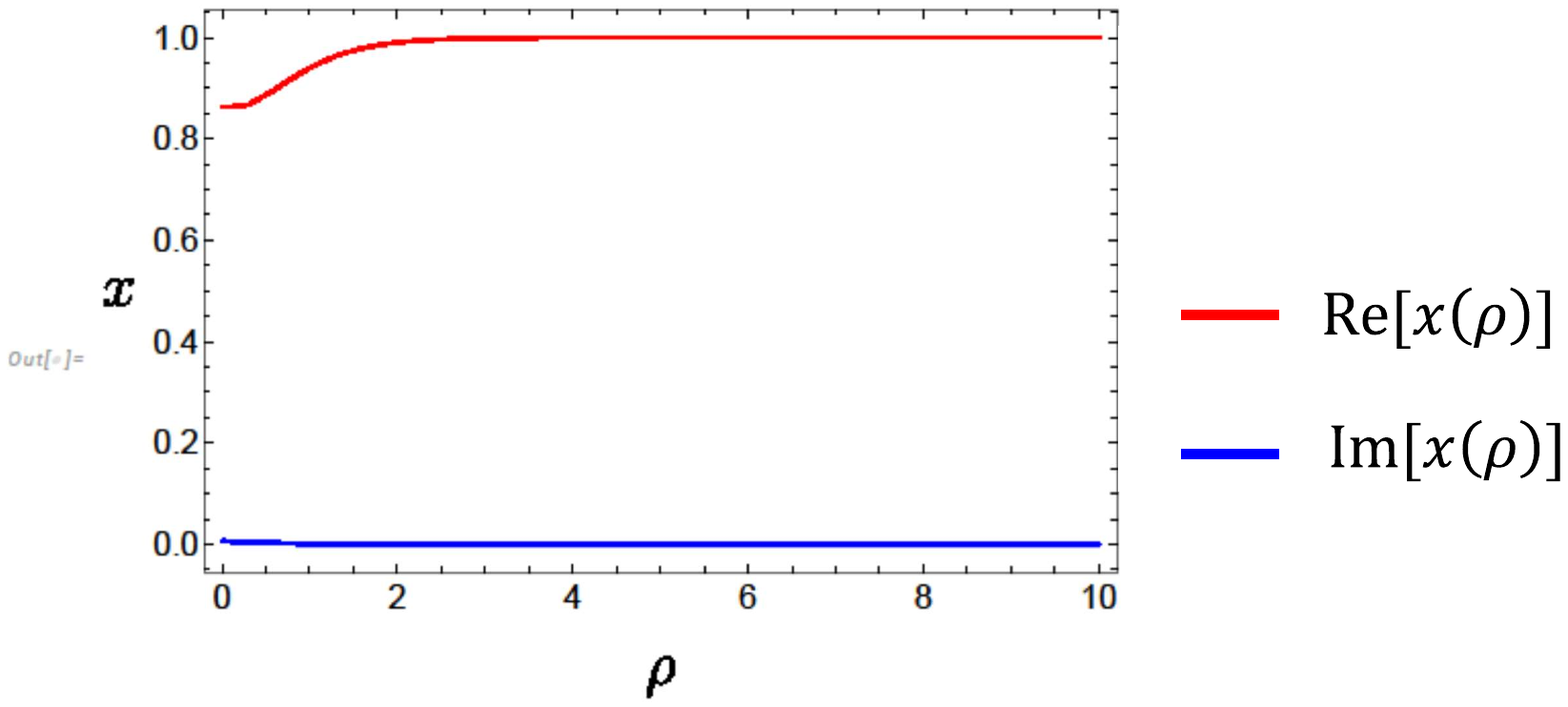}
  \end{minipage}
      \caption{Plots of $x(\rho)$. The left figure shows the $x(\rho)$ on $t_0=5$ and the right figure shows the one with on $t_0=0.5$. The parameters are chosen so that $\gamma=0.4, l=1, \ep=\ep'=0.01$. The $x(\rho)$ approaches to zero at $\rho=\rho_t$. }
      \label{fig:x(rho)}
  \end{figure}

  Let us compute the entanglement pseudo entropy holographically in this solution. Take subsystem $A$ as the interval $[x=-l,x=l]$ on the time slice at $t=t_0$. Here we will focus on the $t_0<0$. The $t_0>0$ case can be done in the same manner.
  In this case, the HRT surface starts from region $\grethree$. Let us denote the co-dimension two HRT surface as $\eta=\eta(\rho), x=x(\rho)$ and the corresponding Lagrangian as 
  \begin{equation}\label{eq:Lag three}
      \mL_{\grethree}= \sqrt{1-f(\rho)\frac{-\Dot{\eta}^2+\Dot{x}^2}{\eta^2}}.
  \end{equation}
  Similar to the model in the last subsection, this Lagrangian has two types of the symmetry:
  \begin{enumerate}
      \item AdS$_2$ scaling; 
      \begin{equation}
          \eta \to (1+\lambda)\eta,\quad  x\to (1+\lambda)x.
      \end{equation}
      The corresponding Noether charge is 
      \begin{equation}
          Q_{\mathrm{AdS}_2}=  -\frac{f(\rho)}{\mL_{\grethree}}\frac{-\Dot{\eta}\eta+\Dot{x}x}{\eta^2}.
      \end{equation}
      Because of the $\mathbb{Z}_2$ symmetry $x\to-x$, the HRT surface has a turning point at $x=0$, where $\frac{d\eta}{dx}=0$. Thus $Q_{\mathrm{AdS}_2}=0$. This leads to 
      \begin{equation}\label{eq:eta(x)}
          \eta^2 =x^2-A,
      \end{equation}
where $A$ is some constant. Similar to the Euclidean model considered above, so far we have not used the concrete form of $f(r)$ and thus it is natural to assume that $\eta(x)$ is independent of the value of $\gamma$. We assume that $A$ is also independent of $\gamma$, 
      \begin{equation}
          A=l^2-t_0^2.
      \end{equation}
      In other words, we set a boundary condition $\eta(l)=t_0$. 
      \item $x$ translation;
      \begin{equation}
          \eta \to \eta, \quad x\to x+c.
      \end{equation}
      This leads us to the corresponding Noether charge
      \begin{equation}\label{eq:QT}
          Q_T = -\frac{f(\rho)}{\mL_{\grethree}}\frac{\Dot{x}}{\eta^2}.
      \end{equation}
  \end{enumerate}
  By combining (\ref{eq:Lag three},\ref{eq:eta(x)},\ref{eq:QT}), we obtain
      \begin{align}\label{eq:x(rho)}
    \sqrt{A}\left(\frac{\Dot{x}}{x^2-A}\right)^2  &= \frac{\alpha}{f(\rho)(f(\rho)-\alpha)},\\
              \mL_{\grethree}^2 &= \frac{f(\rho)}{f(\rho)-\alpha},
      \end{align}
       where we introduce $\alpha=AQ_T^2$.
$x(\rho)$ can be determined from the first equation. Then we have
\begin{equation}
    x(\rho) = \sqrt{A}\tanh{\qty(-\sqrt{\frac{-k}{\chi}}\EllipticF{G(\rho)}{k}+B)},
\end{equation}
  where 
  \begin{equation}\label{eq:sol rhs}
  \begin{split}
      G(\rho) &= \pi-\Tilde{G}(\rho),\\
      \Tilde{G}(\rho)&=\Arcsin{\left( \sqrt{\frac{(1-\chi) (2 \alpha +\chi  \cosh{2 \rho }-1) \text{csch}^2\rho }{4\alpha  \chi }}\right)},\\
      k &= \frac{4\alpha\chi}{(1-\chi)(-2\alpha +\chi+1)}.
  \end{split}
  \end{equation}
  We can determine the constant $\alpha,B$ from the boundary condition for $x(\rho)$.
  \begin{itemize}
  \item Because of the $\mathbb{Z}_2$ symmetry, $\rho=\rho(x)$ is an even function of $x$. In this case, there is a turning point at $x=0$, where $\frac{d\rho}{dx}$=0. From (\ref{eq:x(rho)}), we find that the turning point is at $f(\rho)=0$ or $f(\rho)=AQ_T^2=\alpha$. Suppose the correct one is $f(\rho_t)=\alpha$, that is,
  \begin{equation}\label{eq:rho0}
      2\alpha+\chi\cosh{2\rho_t}-1=0
  \end{equation}
  holds. In fact, this choice is the correct one in the vacuum AdS$_3$ solution and $f(\rho)=0$ corresponds to $x=\pm \sqrt{A}$.
  It seems that we have two choices for arcsin in (\ref{eq:sol rhs}). When we take $\sin^{-1}(0)=0$ then we have $B=0$. When we take $\sin^{-1}(0)=\pi$, then
  \begin{equation}
  \begin{split}
      B= 2 \sqrt{\frac{-k}{\chi}} K(k). 
  \end{split}
  \end{equation}
  Especially in the vacuum AdS$_3$ limit $\chi \to 1$, $k\to +\infty$,
  this branch retrieves the correct solution. Indeed, from the asymptotic expansion of $K(k)$ \cite{eliptic}, we can see
  \begin{equation}
      B  \to \frac{1}{\sqrt{\chi}}\log\qty(-k)\to  \pm \infty.
  \end{equation}
      \item At the boundary $\rho\to \infty$, $x\to l$. From this condition we can get $\alpha$;
      \begin{equation}\label{eq:alpha}
          \begin{split}G(\rho\to\infty)=& \pi-\Arcsin{\sqrt{\frac{1-\chi}{2\alpha}}}\\
          l=&\sqrt{A}\tanh{\qty(\sqrt{\frac{-k}{\chi}}\EllipticF{\Tilde{G}(\rho_\infty)}{k})}.
          \end{split}
      \end{equation}
      We put the plots for the $\alpha(t_0)$ in figure \ref{fig:lorentzian alpha}. We see $\alpha$ can take complex values in the time period $t_0\gtrapprox -l$. $-i\ep'$ is prescribed so that the quantum state can be vacuum and we can obtain clean plots. 
  \end{itemize}
  Eventually, we get the geodesic 
  \begin{equation}
      x(\rho;t_0)= \sqrt{A}\tanh{\qty(\sqrt{\frac{-k}{\chi}}\EllipticF{\Tilde{G}(\rho)}{k})}.
  \end{equation}
  We have some comments on the turning points and geodesic. Firstly the turning point $\rho_t$ is a function of $t_0$ for general $\gamma<\gamma_c$. We naively expect that $\rho_t(t_0)$ is real positive until it reaches $\rho_c$. Roughly, this happens around $t_0 \sim -l$. In this time period, $\rho_t$ and also the whole geodesic lie in the parts of the region $\grethree$ where $f(\rho)>0$. See figure \ref{fig:Janus Penrose} for a sketch. After this time, the turning point is situated in the region $f(\rho)<0$. Later, it reaches region $\gretwo$, where $\rho_t$ is pure imaginary. This story is almost true for a small value of  $\gamma$. However, if we tune $\gamma$ to be larger where the correlation between the system and probe gets strong, $\rho_t$ takes general complex values after some time around $t_0\sim O(l)$. This comes from the fact that $\alpha$ takes complex values. Thus the geodesic lies in another region which analytically continues from region $\grethree$ rather than region $\gretwo$. The second comment is that, for a small enough time, the geodesic can be time-like since $f(\rho)<0$ happens in some parts of the geodesic, which leads to a complex valued holographic pseudo entropy. We put the plots of the profile $x(\rho) $ in figure \ref{fig:x(rho)}.
  
  \begin{figure}[H]
  \centering
    \begin{tabular}{cc}
        \begin{minipage}{.5\linewidth}
           \centering
      \includegraphics[width=1.08\linewidth]{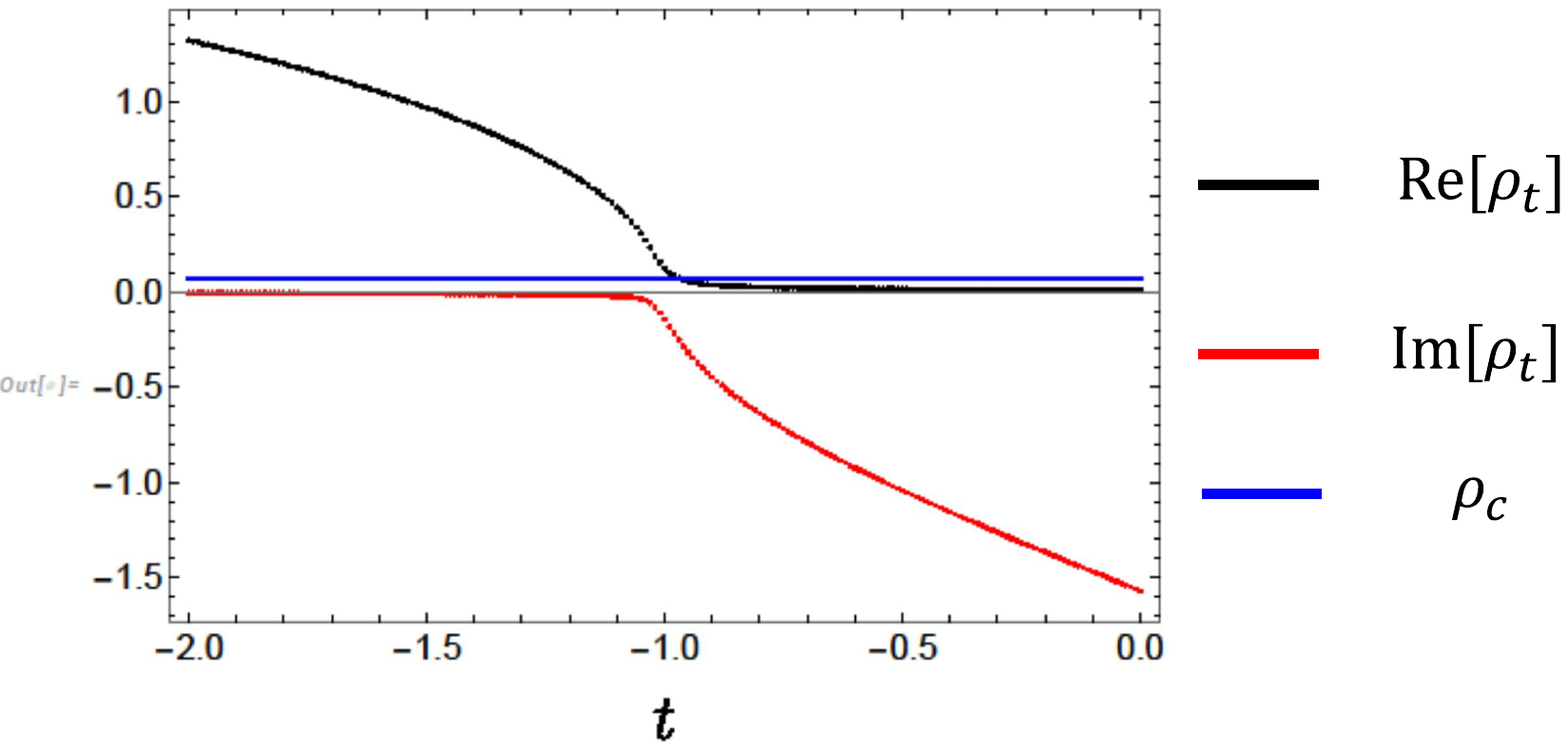}
        \end{minipage}
        \begin{minipage}{.49\linewidth}
              
      \includegraphics[width=1.08\linewidth]{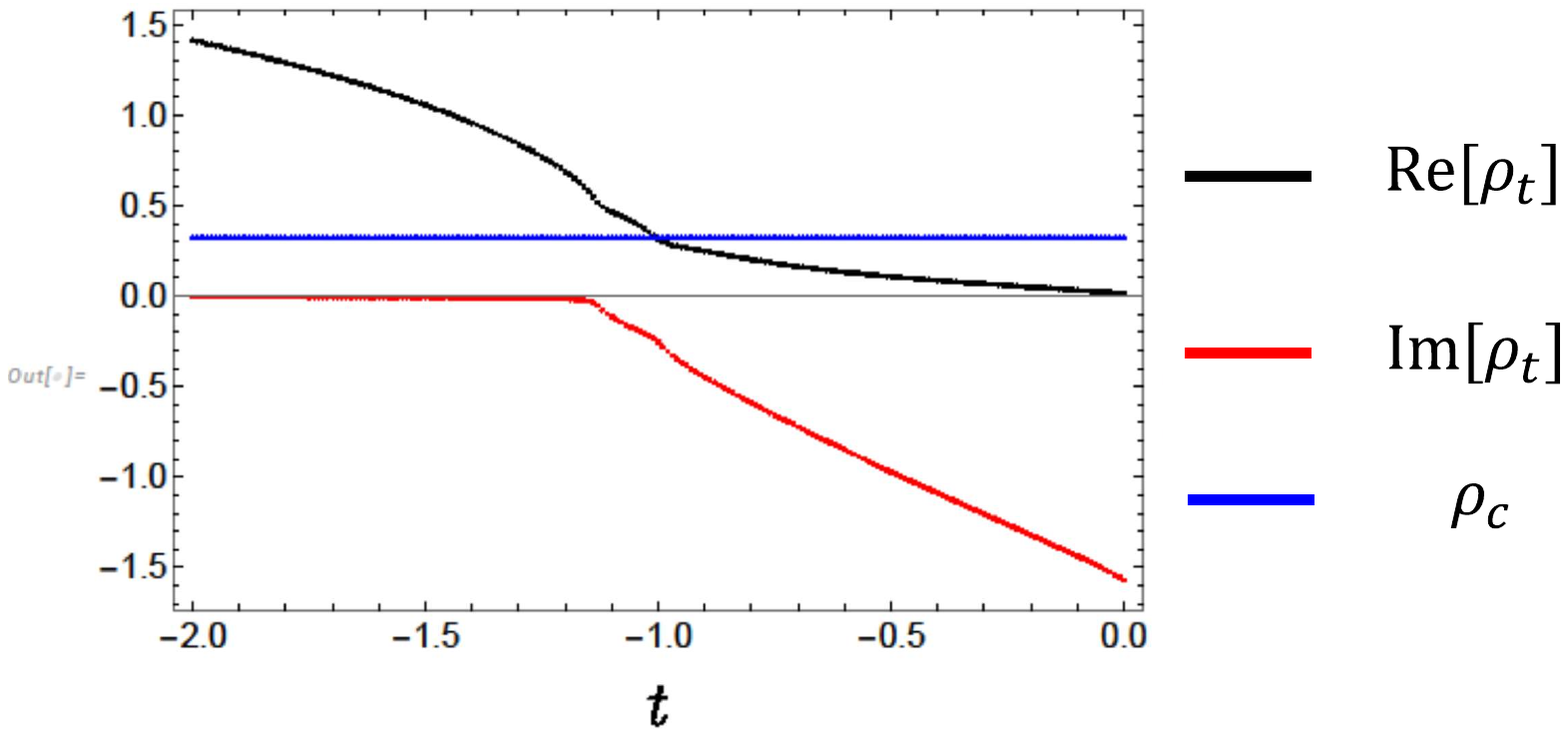}
           
        \end{minipage}
    \end{tabular}
    
      \caption{The plots of $\rho_t$. The Left figure shows when $\gamma=0.1$ and the right shows $\gamma=0.4$. The other parameters are set $\l=1,\ep=\ep'=0.01$. For the small $\gamma$, we see the transition of $\rho_t$ from real to pure imaginary around $t_0\sim -l$, however for large $\gamma$ it looks that $\rho_t$ takes a general complex number. }
      \label{fig:rhot}
\end{figure}
 
The geodesic length is computed as follows:
  \begin{equation}
  \begin{split}
      \mathrm{Area}[\Gamma_A] &= -2\int_C d\rho \sqrt{\frac{f(\rho)}{f(\rho)-f(\rho_t)}}\\ &=\qty[\sqrt{\frac{-k}{\alpha \chi}}\qty((1-\chi)\EllipticF{G(\rho)}{k}+(2\alpha+\chi-1)\EllipticPi{j}{G(\rho)}{k})]^{\rho_t}_{\rho_\infty},
  \end{split}
  \end{equation}
where 
  \begin{equation}
      j= \frac{2\alpha}{1-\chi}.
  \end{equation}
Here we introduce the cutoff surface $\rho=\rho_\infty$ by 
\begin{equation}
    \frac{\ep}{-t_0}= \frac{1}{\sqrt{1-f(\rho_\infty)}}.
\end{equation}
This will be a nice cutoff surface since the inside of the square root is always positive and it matches the one of the vacuum AdS case when $\gamma=0$. Note that this cutoff works well when $\abs{t_0}\gg \ep$. 
Since $G(\rho_t)=\pi$ and
\begin{equation}
    \begin{split}
        \EllipticF{\pi-\theta}{k}&= 2\EllipticK{k}-\EllipticF{\theta}{k},\\
        \EllipticPi{j}{\pi-\theta}{k} &= 2\EllipticPiC{j}{k}-\EllipticPi{j}{\theta}{k},
    \end{split}
\end{equation}
 we obtain
 \begin{equation}
     \mathrm{Area}\qty[\Gamma_A] = -\sqrt{\frac{-k}{\alpha \chi}}\qty((1-\chi)\EllipticF{\Tilde{G}(\rho_\infty)}{k}+(2\alpha+\chi-1)\EllipticPi{j}{\Tilde{G}(\rho_\infty)}{k}).
 \end{equation}

\begin{figure}[H]
\centering
  \begin{minipage}[b]{0.495\linewidth}
    \centering
    \includegraphics[width=1.1\linewidth]{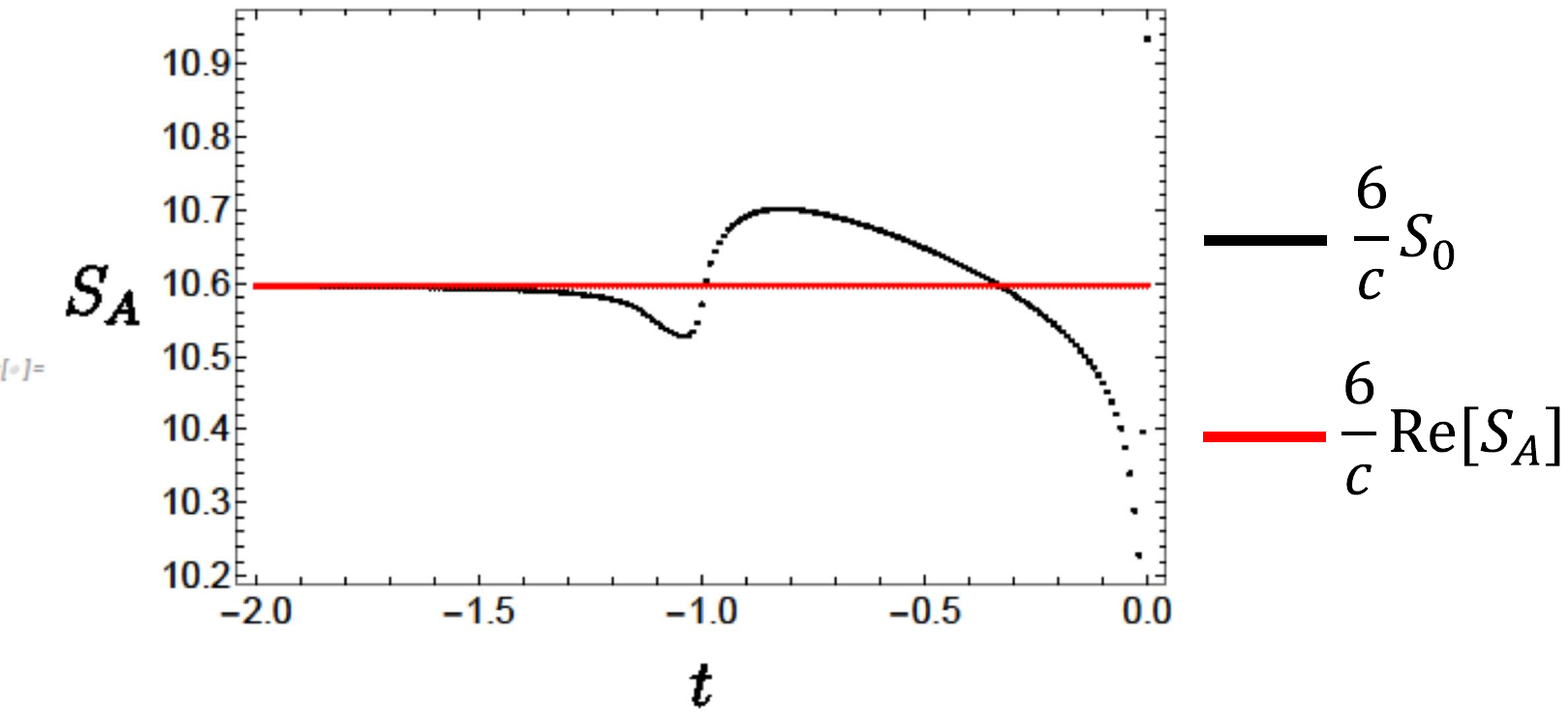}
  \end{minipage}
  \begin{minipage}[b]{0.495\linewidth}
    \centering
    \includegraphics[width=1.1\linewidth]{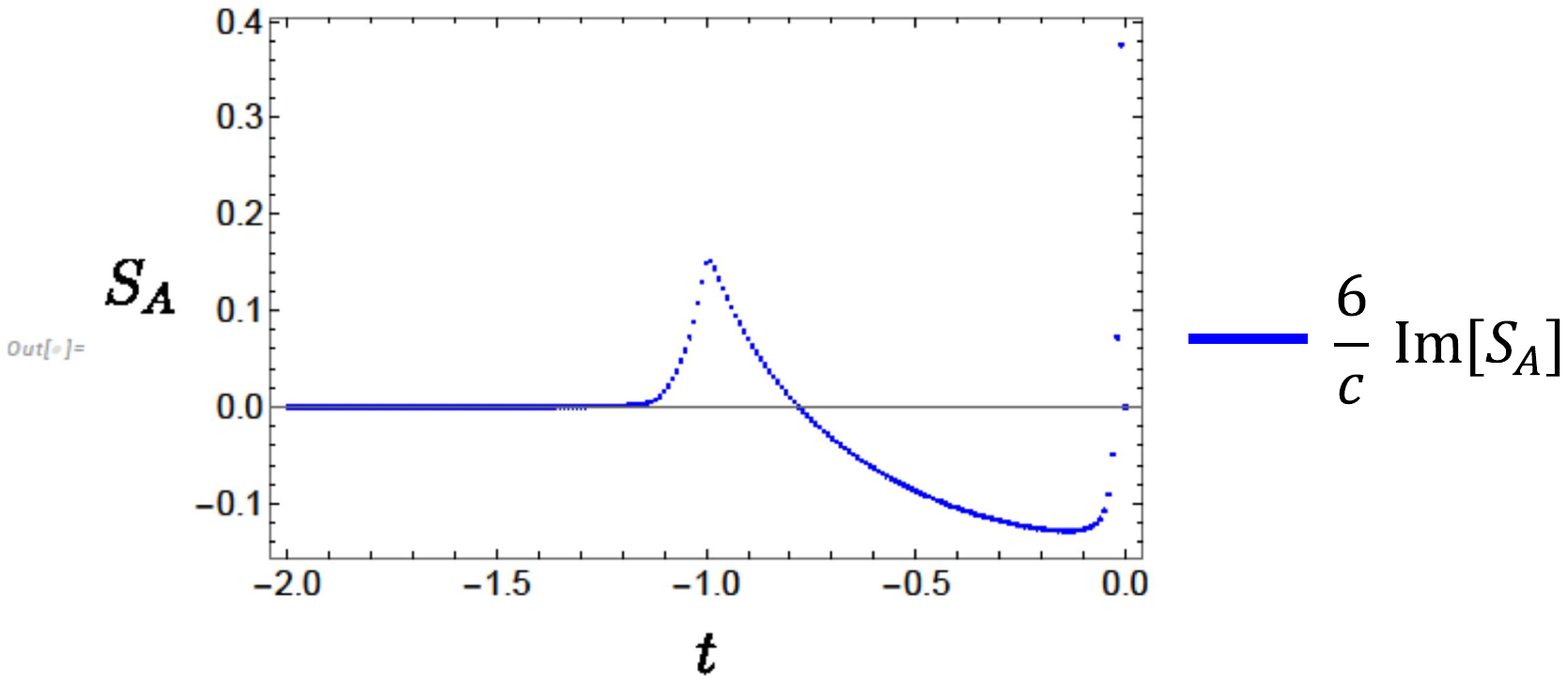}
  \end{minipage}
  \caption{The plot of the time evolution of the entanglement entropy. We put $\gamma=0.4,\ep=0.01,l=1$. $S_0$ denote the vacuum entanglement entropy without the defect $S_0= \frac{c}{3}\log{\frac{2l}{\ep}}$.}
    \label{fig:Janus/real gamma EE.}
\end{figure}

Similar to the Euclidean model in the last subsection, by using the formula (\ref{eq:Elliptic Pi formula}), we have the last expression 
\begin{equation}
    \begin{split}
        \mathrm{Area}\qty[\Gamma_A] &= \sqrt{\frac{-k}{\alpha \chi}}\qty(2 \alpha \EllipticF{\Tilde{G}(\rho_\infty)}{k}-(2\alpha+\chi-1)\EllipticPi{\frac{k}{j}}{\Tilde{G}(\rho_\infty)}{k})\\ &+ \log\qty(\frac{\sqrt{\frac{(j-k)(j-1)}{j}}\tan{\Tilde{G}(\rho_\infty)}+\sqrt{1-k \sin^2 \Tilde{G}(\rho_\infty)}}{-\sqrt{\frac{(j-k)(j-1)}{j}}\tan{\Tilde{G}(\rho_\infty)}+\sqrt{1-k \sin^2 \Tilde{G}(\rho_\infty)}}).
    \end{split}
\end{equation}
By using the condition (\ref{eq:alpha}), we can also write
\begin{equation}
\begin{split}
    \mathrm{Area}\qty[\Gamma_A] &= \sqrt{\frac{-k}{\alpha \chi}}\qty(2 \alpha \sqrt{\frac{\chi}{-k}}\arctanh{\qty(\frac{l}{\sqrt{A}})}-(2\alpha+\chi-1)\EllipticPi{\frac{k}{j}}{\Tilde{G}(\rho_\infty)}{k})\\ &+ \log\qty(\frac{\sqrt{\frac{(j-k)(j-1)}{j}}\tan{\Tilde{G}(\rho_\infty)}+\sqrt{1-k \sin^2 \Tilde{G}(\rho_\infty)}}{-\sqrt{\frac{(j-k)(j-1)}{j}}\tan{\Tilde{G}(\rho_\infty)}+\sqrt{1-k \sin^2 \Tilde{G}(\rho_\infty)}}).
\end{split}
\end{equation}
As shown in figure \ref{fig:Janus/real gamma EE.}, we see that the pseudo entropy behaves in completely the same manner as the analytic continuation of the pseudo entropy from the Euclidean case. This is because the transition matrix is equivalent in our setups.
 

\section{Conclusions and discussions}\label{sec:conclusions}

In this paper, we explored entanglement phase transitions in time evolution of pseudo entropy using holography. First we considered a model of AdS$_3$/BCFT$_2$ which describes the global quantum quench and added a scalar field localized on the end-of-the-world brane. In the Euclidean signature this is dual to a boundary conformal field theory (BCFT) on a strip where the boundary conditions on the two boundaries are different. In the gravity dual, the different boundary conditions are realized by requiring that the brane localized scalar field approaches different values at the two boundaries. Under the Lorentzian time evolution via its Wick rotation, the scalar field, which was originally real valued in the Euclidean solution, takes imaginary values. This effect suppresses the time evolution of area of the extremal surface, which is properly interpreted as the pseudo entropy instead of entanglement entropy. 
At a certain value of the imaginary scalar field, the growth of the entropy changes from the linear behavior to the logarithmic behavior. This exhibits the entanglement phase transition via the AdS/BCFT holography.
We analytically determined the coefficient of the time evolution in each phase as shown in (\ref{EPTb}).  In this analysis, we worked out the global extension of our AdS/BCFT solution using the Kruskal coordinate and found the profile of extremal surfaces in the extended geometry.

On the CFT side of this phase transition, we looked at the behavior of pseudo entropy of a transition matrix instead of entanglement entropy for a quantum state. However, since the gravity dual is given by a Lorentzian time evolving geometry, it might look dual to a certain quantum state in a CFT. The only difference from the standard gravity dual is that we have an imaginary valued scalar field localized on the brane, which is expected to give some dissipating effect. We noted that since the R\'enyi pseudo entropies computed from holography look real and non-negative, it is possible that it is pseudo Hermitian. If this is correct, it can be mapped to a regular Hermitian density matrix via a similarity transformation and our gravity solution can also be interpreted as a certain quantum state under a non-unitary time evolution instead of a transition matrix. It will be an intriguing future problem to explore this possibility. Besides, we also commented on the direct computation of the pseudo R\'enyi entropy from the CFT side. It would be an important future problem to understand how to compute the pseudo entanglement entropy, possibly nonperturbatively, from the CFT side. 

We also studied a class of AdS/BCFT solution with a gauge field localized on the end-of-the-world brane instead of a scalar field. In the AdS$_3$ case, we showed that the profile of the end-of-the-world brane is identical to the standard AdS/BCFT setup for pure gravity where the gauge flux gives a non-zero tension. This means that there is no critical point where the pseudo entropy grows logarithmically as opposed to the localized scalar case. By computing its free energy, we worked out its phase structure. We also analyzed solutions of the end-of-the-world brane with the gauge field in a higher dimensional Poincar\'e AdS. We found that the brane profile with the localized gauge flux in AdS$_{d+1}$ is the same as the one with the localized scalar field in AdS$_{d}$. One interesting future direction is to apply this new AdS/BCFT setup to model strongly correlated condensed matter systems with boundaries and study transport phenomena in them.
\par
Finally, we turned to a bulk version of such a solution where the scalar field is not localized on the brane but spreads throughout the entire bulk spacetime. We have discussed the transition matrices of two distinct vacuum states. We find that for both the Euclidean and Lorentzian models, the pseudo entropies decrease around the spacelike defect and take smaller values than the vacuum AdS case. It is intriguing to note that the Lorentzian solution describes a highly exotic spacetime. Especially, it has singularities and the dilation scalar field becomes complex valued in a certain region. As a future work, it would be interesting to demonstrate a measurement-induced phase transition with the bulk fields. More specifically, by considering the soft wall, the phase transition is expected to happen more mildly or in a crossover style. An important consideration for achieving entanglement phase transition in our bulk field model is the competition between the growth of entanglement due to a quantum quench and the reduction of entanglement through measurement. To realize entanglement phase transition by modifying our spacelike brane model, we should start with a quenched state which might be obtained from a black hole geometry.


\section*{Acknowledgements}

We are grateful to Rathindra Nath Das, Juan Maldacena, Shinsei Ryu and Mark Van Raamsdonk for useful discussions.
This work is supported by MEXT KAKENHI Grant-in-Aid for Transformative Research Areas (A) through the ``Extreme Universe'' collaboration: Grant Number 21H05187. This work is also supported by Inamori Research Institute for Science and by JSPS Grant-in-Aid for Scientific Research (A) No.~21H04469. TK is supported by Grant-in-Aid for JSPS Fellows No. 23KJ1315. YS is supported by Grant-in-Aid for JSPS Fellows No.23KJ1337. ZW is supported by the Society of Fellows at Harvard
University.

\appendix
\section{Formula for the Hyperbolic Coordinate and The Elliptic Integrals}
\label{app:tech_details}
\subsection{Hyperbolic coordinate and geodesic in the vacuum \texorpdfstring{AdS$_3$}{AdS3}}\label{appendix:Hyperbolic Coordinate in AdS}
The relation between the Poincar\'e coordinate $(z,t_E)$ and the Janus coordinate $(r,\xi)$ is
\begin{align}\label{eq:tf poincare amd ads2}
    \xi^2=\tau^2+z^2,\quad \frac{z}{\tau}=\frac{1}{\sinh{r}}.
\end{align}
We know the RT surface in the vacuum AdS$_3$ in the Poincar\'e coordinate
\begin{equation}
    z(x)= \sqrt{l^2-x^2},\quad \tau=\tau_0
\end{equation}
Then the RT surface we see is 
\begin{equation}
    x(r)=\sqrt{l^2-\frac{\tau_0^2}{\sinh{r}^2}},\quad \xi(r)=\tau_0 \coth{r}
\end{equation}
The Noether charges can be read off as
\begin{equation}
    Q_T = \frac{1}{l},\quad A= l^2 + \tau_0^2, \quad \alpha= \frac{A}{l}= 1+\frac{\tau_0^2}{l^2}
\end{equation}
To check consistency we should see the integral of the E.L. equation in a direct manner.
\begin{equation}
\begin{split}
    \int dr \sqrt{\frac{\alpha}{f(r)(f(r)-\alpha)}}&= \arctanh{g(r)}+B\\
    g(r) &= {\frac{\sqrt{\alpha}\sinh{r}}{\sqrt{1-\alpha+\sinh^2{r}}}}
\end{split}
\end{equation}
Thus the solution of the equation is
\begin{equation}
\begin{split}
    \arctanh{\frac{x(r)}{\sqrt{A}}}= \arctanh{g(r)}+B\\
    \frac{x(r)}{\sqrt{A}}= \frac{g(r)+B'}{1+g(r)B'},\quad B':= \tanh{B}
\end{split}
\end{equation}
Demand there is the turning point $x=0,r=r_t$ where $f(r_t)=\alpha$ we have $B'=+\infty$ and thus
\begin{equation}
    \frac{x(r)}{\sqrt{A}}= \frac{1}{g(r)}
\end{equation}
The geodesic length is evaluated by integrating the lagrangian. We should care about the sign of the Lagrangian. Correct one is
\begin{equation}
    L= 2 \mathrm{sgn}{(\tau_0)}\int_{r_\infty}^{r_t} dr \sqrt{\frac{f(r)}{f(r)-\alpha}}=  2 \mathrm{sgn}{(\tau_0)}\left[ \arctanh{\left(\frac{\sinh{r}}{\sqrt{-\alpha+1+\sinh^2{r}}}\right)}\right]_{r_\infty}^{r_t}
\end{equation}
By putting the cut-off as
\begin{equation}
    \frac{\ep}{\tau_0}= \frac{1}{\sinh{r_\infty}},\quad \cosh^2{r_t}=\alpha
\end{equation}
Then,
\begin{equation}
    \begin{split}
        \left[ \arctanh{\left(\frac{\sinh{r}}{\sqrt{-\alpha+1+\sinh^2{r}}}\right)}\right]_{r_\infty}^{r_t}&=\frac{1}{2}\log\left(-1\right)-\frac{1}{2}\log\left(\frac{1+\mathrm{sgn}(\tau_0)\left(1+\frac{\ep^2}{2\tau_0^2}\right)}{1+\mathrm{sgn}(\tau_0)\left(1-\frac{\ep^2}{2\tau_0^2}\right)}\right)\\
        & =\mathrm{sgn}(\tau_0)\log\frac{2l}{\ep}
    \end{split}
\end{equation}
Finally we see
\begin{equation}
    S_A = \frac{c}{3}\log{\frac{2l}{\ep}}
\end{equation}
  \subsection{Elliptic integrals}\label{appendix: elliptic integrals}
   In the computation in the Janus solution we see the expression with the elliptic integrals. In this section we summarize the definition and some formulas of them. See \cite{eliptic,elipticPi} for details.
   \subsubsection*{Definition}
   There are three kinds of the incomplete elliptic integrals. Incomplete elliptic integral of the first kind $F(z|m)$ is defined as 
   \begin{equation}
\EllipticF{z}{m}= \int_0^z \frac{1}{\sqrt{1-m \sin^2{t}}} dt.
   \end{equation}
   The complete elliptic integral of the first kind $K(m)$ is the specific version of the incomplete one
   \begin{equation}
       \EllipticK{m}=\EllipticF{\frac{\pi}{2}}{m}
   \end{equation}
   The definition of the incomplete elliptic function of the third kind is as follows
\begin{equation}
    \begin{split}
        \Pi\left(n;z|m\right)&= \int_0^z \frac{dt}{(1-n\sin^2{t})\sqrt{1-m \sin^2{t}}}\\
         \Pi\left(n|m\right) &=  \Pi\left(n;\frac{\pi}{2}|m\right)
    \end{split}
\end{equation}

\section{Another point of view for the spacelike brane}\label{app:int_Janus}
    In section \ref{sec:Lorentzian Model; Final State Projection}, we see a spacelike brane in the Lorentzian time evolution. 
    Here we want to give another viewpoint on the spacelike brane in the quantum field theories. To this end, we review the indirect measurement shortly based on \cite{ R300000001-I032210891-00,BRE02}.
    In general, it may be difficult to touch our interest system in direct ways. For such a case, it is useful to introduce a probe system and correlate it to our system to obtain some information. This indirect measurement can be done in the following schemes:
    \begin{enumerate}
        \item We prepare the quantum state which is a direct product of the system and probe Hilbert space.
        \begin{equation}
            \rho_{\mathrm{tot}} = \rho_S \otimes \rho_P
        \end{equation}

    \item Act a unitary operator $U$ to create correlations between the system and the probe.
    \begin{equation}
        \rho_{\mathrm{tot}} \to U \rho_S \otimes \rho_P U^\dag
    \end{equation}
    \item Doing the projection measurement $\{P_k\}_k$ for the probe system. A probability that the outcome of the measurement $k$ is 
    \begin{equation}
        p_k = \Tr_{SP}\qty[P_k U \rho_S \otimes \rho_P U^\dag P_k ].
    \end{equation}
    Also, the quantum state after the measurement with the outcome of the measurement $k$ is 
    \begin{equation}
        \rho_{\mathrm{tot}}^{(k)} = \frac{1}{p_k} P_k U \rho_S \otimes \rho_P U^\dag P_k
    \end{equation}
    \item Let us introduce the normalized orthogonal basis of the probe system $\{\ket{\Psi_m}\}$ and also denote the spectral decomposition of the initial state $\rho_P$ as $\rho_P = q_l \ket{\Phi_l}\bra{{\Phi_l}}$. Then by introducing the Kraus operator and the POVM as 
    \begin{equation}
    \begin{split}
        M_{ki} &= \sqrt{q_l} \bra{\Psi_m} P_k U \ket{\Phi_l},\quad i = (l,m)\\ 
        E_k &= \sum_i M^\dag_{ki}M_{ki}
    \end{split}  
    \end{equation}
     Then the quantum state of the system with the measurement outcome $k$ is 
     \begin{equation}
     \begin{split}
         \rho_S^{(k)} &= \frac{1}{p_k}\sum_i M_{ki}\rho_S M_{ki}^\dag \\
         p_k &= \Tr_S{\qty[E_k \rho_S]}
     \end{split}
     \end{equation}
     \item If we do not know the measurement results, the resulting state is 
     \begin{equation}
         \rho_S' = \sum_k p_k \rho_k = \sum_{ik}M_{ki}\rho_S M_{ki}^\dag
     \end{equation}
        \end{enumerate}
\underline{\textbf{Spacelike brane from tracing out the probe}}\par
 In total system point of view of the total system, we have the Schwinger Keldysh path integral formulation. 
 
  Suppose we are only interested in the system $S$ and trace out the probe system. Then we naturally expect that the system includes the spacelike brane if the total unitary $U$ is instantaneous. That is the effective action of the system will take the form of 
  \begin{equation}
      L_{S,\mathrm{eff}} = L_S + \delta(t_0) \gamma L_{\mathrm{sp. brane}}.
  \end{equation}
This will be done in the same manner as the Caldeira-Leggett model or more generally Feynman-Vernon influence action formulation\cite{CALDEIRA1983587,FEYNMAN2000547}.
  Here we can interpret $\gamma$ as a strength of the measurement. This can be explained as follows: the spacelike brane term is emergent from the trace out of the probe system and if the unitary $U$ is simply tensor product, then $\gamma=0$. In terms of measurement, if unitary $U$ is decoupled, we do not have any information about system $S$ from the probe and essentially we do not do any measurement in this case. 
  \par Here we introduce the instant spacelike brane, \textit{i.e.}, the thin brane localized on the $t=0$, we also consider the thick brane. It is known that there are classes of soliton solutions in quantum field theories and string theory, which are thick spacelike brane\cite{Gutperle:2002ai}. In these solutions, system field shows dissipative radiation as a tachyon condensation. It may be possible that we interpret this radiation as energy transfer between system and probe in the indirect measurement scheme.
  \par Note that it is clear that not all of the spacelike branes give an effective picture for the indirect measurement. For example, in our models in sec. \ref{sec:Lorentzian Model; Final State Projection}, the pseudo entropy clearly satisfies $S_A = S_{A^c}$, which holds since the geodesic is just determined by the entanglement surface. This means there is no decoherence and we can think of the path integral gives the two state amplitude.
  \begin{figure}
     \centering
   \begin{tikzpicture}
   \draw(1,0)--(1,1);
   \draw(0.5,1)--(1.5,1)--(1.5,2)--(0.5,2)--cycle;
   \draw(1,2)--(1,3);
   \draw(0.5,3)--(3.5,3)--(3.5,4)--(0.5,4)--cycle;
   \draw(1,4)--(1,5);
   \draw(0.5,5)--(1.5,5)--(1.5,6)--(0.5,6)--cycle;
   \draw(1,6)--(1,7);
   \draw(3,0)--(3,1);
   \draw(2.5,1)--(3.5,1)--(3.5,2)--(2.5,2)--cycle;
   \draw(3,2)--(3,3);
   \draw(3,4)--(3,5);
   \draw(2.5,5)--(3.5,5)--(3.5,6)--(2.5,6)--cycle;
   \draw(1,1.15)node[above]{$U_S$};
   \draw(1,5.15)node[above]{$U_S$};
   \draw(3,1.15)node[above]{$U_P$};
   \draw(3,5.15)node[above]{$P_k$};
   \draw(2,3.15)node[above]{$U$};
   \draw[>-Stealth] (4,3.5)--(5.9,3.5);
   \draw(5,3.7)node[above]{Trace out $P$};
   \draw[dashed,red](2.25,-0.2)--(2.25,6.5)--(3.75,6.5)--(3.75,-0.2)--cycle;
   \draw(3,-0.2)node[below]{\textcolor{red}{Probe}};
   \draw(7,0)--(7,1);
   \draw(6.5,1)--(7.5,1)--(7.5,2)--(6.5,2)--cycle;
   \draw(7,2)--(7,3);
   \draw(6.5,3)--(7.5,3)--(7.5,4)--(6.5,4)--cycle;
   \draw(7,4)--(7,5);
   \draw(6.5,5)--(7.5,5)--(7.5,6)--(6.5,6)--cycle;
   \draw(7,6)--(7,7);
   \draw(7,1.15)node[above]{$U_S$};
   \draw(7,5.15)node[above]{$U_S$};
   \draw(7,3.15)node[above]{$U_{{\mathrm{eff}},k}$};
   \end{tikzpicture}
     \caption{Sketch for the indirect measurement.If we trace out the probe system the system action modified to the effective action with a spacelike brane.}
     \label{fig:indirect measure}
 \end{figure}
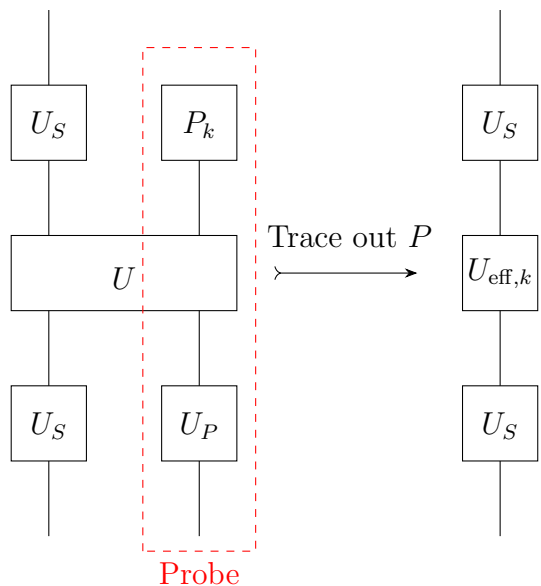

\section{CFT analysis of a defect marginal perturbation}\label{app:CFT_analysis}

\begin{figure}[H]
    \centering   \includegraphics[width=0.5\linewidth]{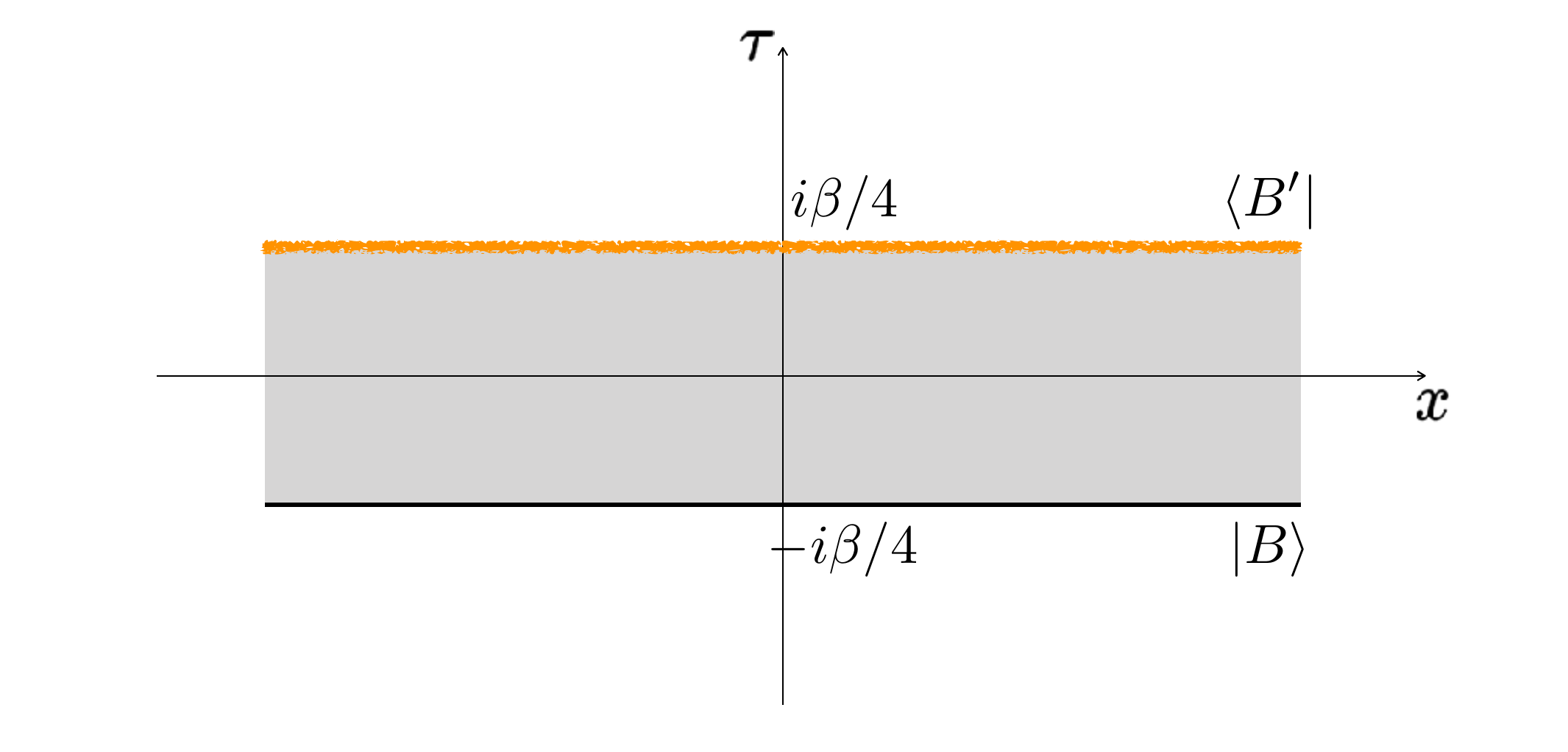}\includegraphics[width=0.5\linewidth]{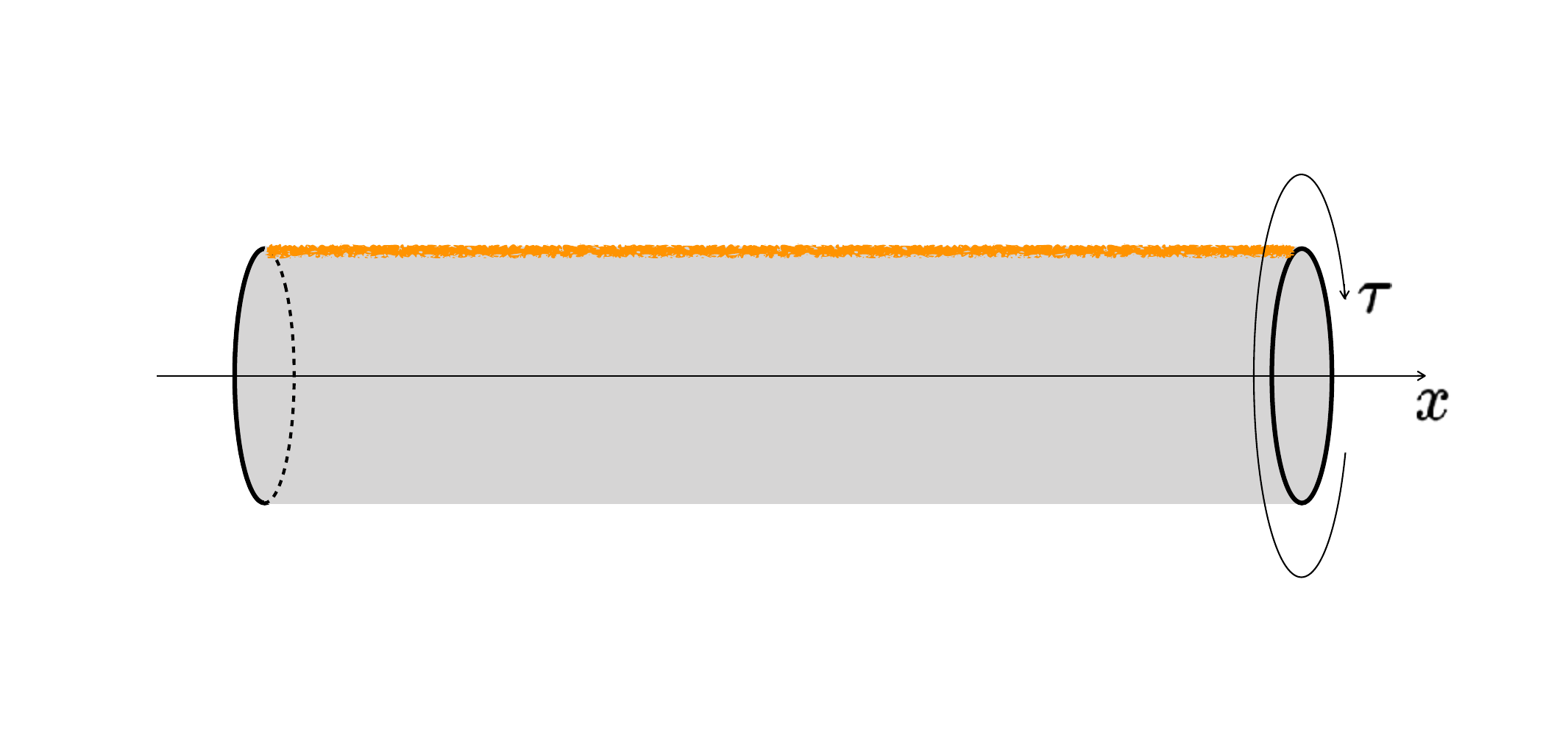}
    \caption{The left figure shows a path integral realization of the transition matrix \eqref{eq:TM_BCFT}, where the two boundaries states are taken to be $\ket{B}$ and $\ket{B'}$ respectively. $\ket{B'}$ is expected to be realized by performing a marginal perturbation (shown in orange) along $\tau = \beta/4$. However, there are some subtleties to perform marginal perturbations exactly along the boundary. The right figure shows an alternative path integral realizing the density matrix \eqref{eq:TMcylinder}. A marginal perturbation is performed along a one-dimensional defect at $\tau=\beta/4$.
    The lower half $-\beta/2<\tau\leq0$ prepares $\ket{\rm TFD}$ and the upper half $0<\tau\leq\beta/2$ prepares $\langle{\widetilde{\rm TFD}}|$. The two setups can be regarded as connected by a $\mathbb{Z}_2$ orbifold.}
    \label{fig:BCFTandTFD}
\end{figure}

In this section, we present a CFT analysis parallel to the holographic pseudo entropy studied in this paper. The transition matrix considered in this paper can be written as
\begin{align}\label{eq:TM_BCFT}
    \mathcal{T} = \frac{e^{-it H} e^{-\frac{\beta}{4}H}\ket{B}\bra{B'}e^{-\frac{\beta}{4}H}e^{it H}}{\bra{B'}e^{-\frac{\beta}{2}H}\ket{B}},
\end{align}
where $\ket{B'}$ and $\ket{B}$ are two different conformal boundary states connected by a marginal deformation. The subsystem considered in the current setup is $A=\{x|x>0\}$. The corresponding path integral setup is shown in the left of figure \ref{fig:BCFTandTFD}, which turns out to be a path integral over a strip with width $\beta/2$ along the imaginary time direction, where a marginal perturbation is performed on the one boundary located at $\tau = \beta/4$.


It would be great to study this transition matrix directly in BCFT. However, since it is a little bit subtle to perform the marginal perturbation exactly on the boundary in the path integral formalism, we consider an alternative version of the dynamics 
\begin{align}\label{eq:TMcylinder}
    \mathcal{T} = \frac{e^{-it (H_L+H_R)} \ket{\rm TFD}\langle{\widetilde{\rm TFD}}| e^{it (H_L+H_R)}}{\langle{\widetilde{\rm TFD}}|{\rm TFD}\rangle}.
\end{align}
Here, $|{\rm TFD}\rangle$ is the standard thermo-field double state 
\begin{align}
    |{\rm TFD}\rangle = \sum_i e^{-\beta E_i/2} |n_i\rangle_{L} |n_i\rangle_{R}, 
\end{align}
where $|n_i\rangle_{L(R)}$ is the energy eigenstate of $H_{L(R)}$ associated to energy $E_i$, and $L$ and $R$ label two copies of the original system. By contrast, $|\widetilde{\rm TFD}\rangle$ is realized by performing a marginal perturbation to the initial condition of the thermo field double state. More specifically, the corresponding setup is realized by considering a path integral on a Euclidean cylinder, where $S^1$ is parameterized by $\tau\in(-\beta/2, \beta/2]$, and $\mathbb{R}$ is parameterized by $x$. Additionally, the marginal perturbation is realized by adding $\lambda \int dx ~O(x+i\beta/4,x-i\beta/4)$ to the action of the path integral, where $\lambda$ is a dimensionless small parameter, and $O$ is a scalar primary with scaling dimension $\Delta = 1 $. In the following, we denote $O(x+i\beta/4,x-i\beta/4)$ as $O(x)$ for simplicity. See the right of figure \ref{fig:BCFTandTFD} for a sketch. A specific way to construct a BCFT on a strip is to take a $\mathbb{Z}_2$ orbifold of a standard CFT defined on a strip, by identifying $z = x+ i\tau$ and $z = x+ i(\beta/2-\tau)$. This justifies what we are doing here, i.e. considering the setup shown in the right of figure \ref{fig:BCFTandTFD} instead of that shown in the left. 

\begin{figure}[H]
    \centering   \includegraphics[width=0.7\linewidth]{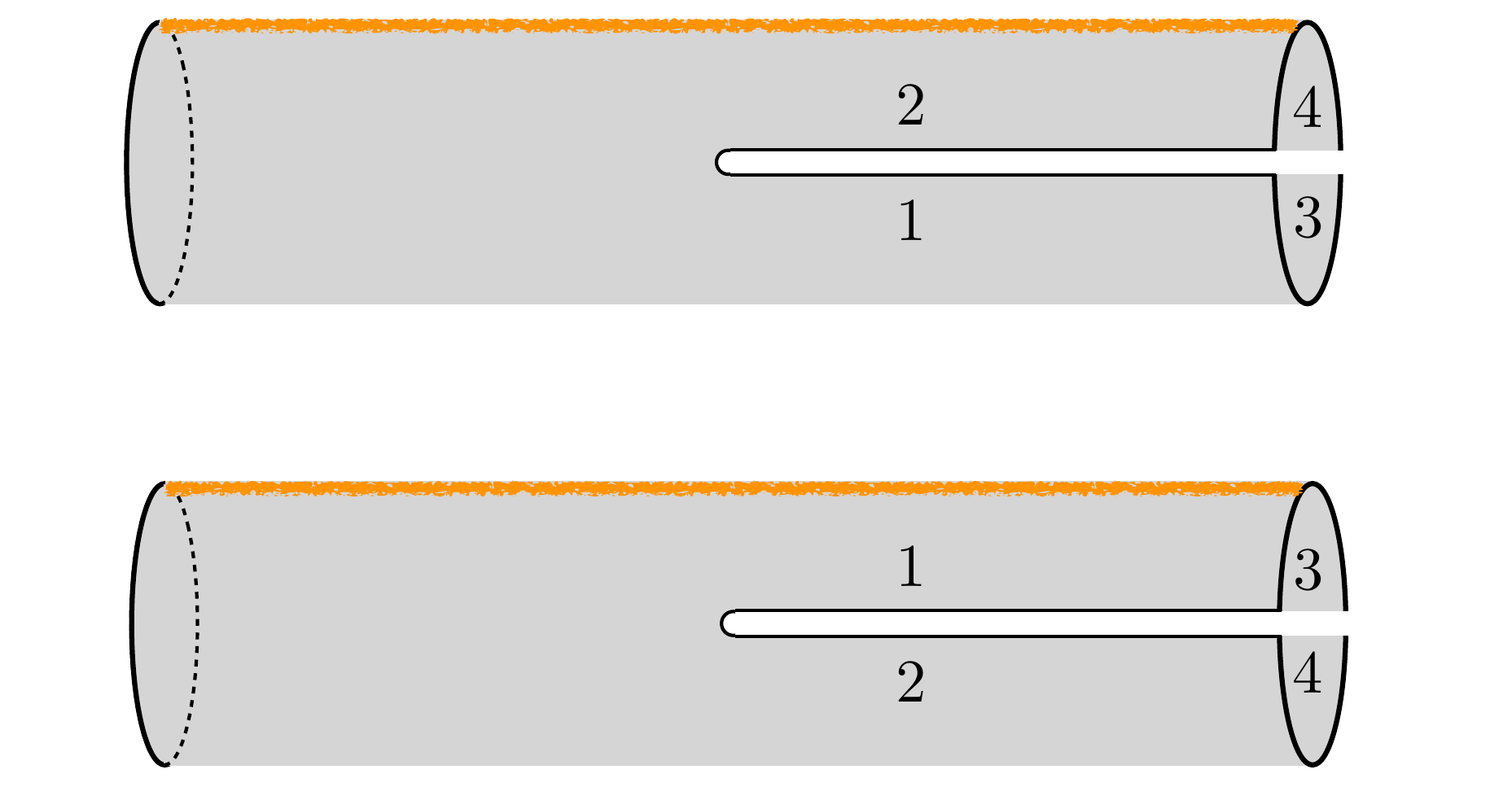}
    \caption{Replica manifold $\Sigma_2$ appearing in the computation of the 2nd R\'enyi entropy. Two copies of the original cylinder are cut along the $x>0$ half of imaginary time slices $\tau$ and $\beta/2-\tau$ and glued together by identifying the boundaries labeled by the same number. A marginal perturbation is performed along the orange lines.}
    \label{fig:replica_two_cylinder}
\end{figure}

\begin{figure}
    \centering   \includegraphics[width=0.7\linewidth]{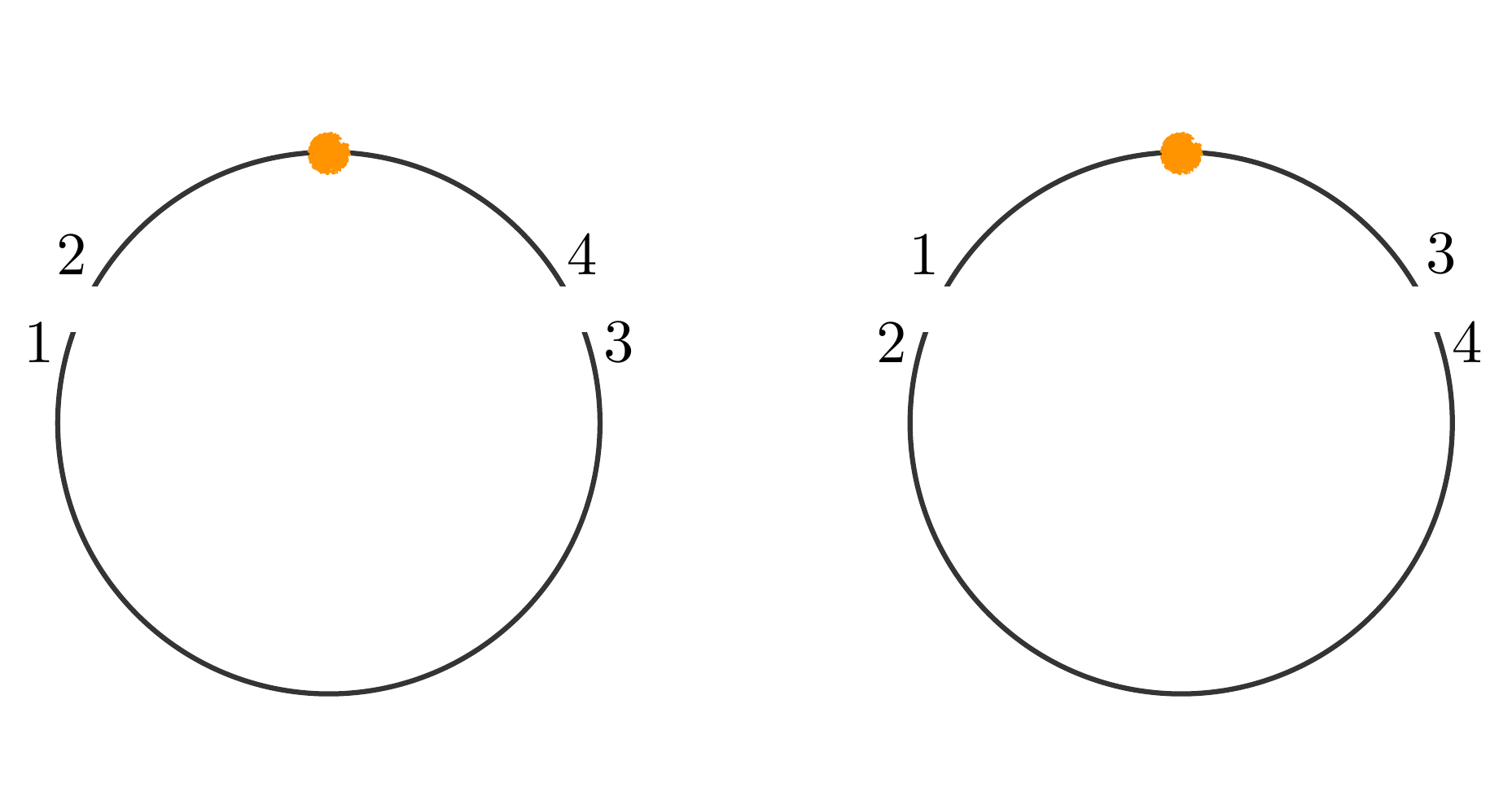}
    \caption{An $x={\rm const.}$ slice of $\Sigma_2$ at $x>0$. Points labeled by the same number are identified. A marginal perturbation is performed on the orange points. }
    \label{fig:replica_slices}
\end{figure}

Now we consider the pseudo R\'enyi entropy for the subsystem $A_L \cup A_R$, where $A_{L(R)}$ is the $x>0$ half of the left (right) system. 
Accordingly, we should put the two edges of the subsystem at $z=i\tau$ and $z=i(\beta/2 -\tau)$, perform the replica method, and then perform the analytic continuation $\tau\rightarrow it$. We use $\Sigma_n$ to denote the $n$-replicated manifold. A sketch of $\Sigma_2$ is shown in figure \ref{fig:replica_two_cylinder}, where boundaries labeled by the same number $1\leq k \leq 4$ are identified with each other. It is also convenient to think about the $x={\rm const.}$ slice at $x>0$, which is shown in figure \ref{fig:replica_slices}. 

Applying the replica method, the $n$-th pseudo R\'enyi entropy can be computed as 
\begin{align}
    S^{(n)}(\mathcal{T}_A) &= \frac{1}{1-n} \log \frac{\int_{\Sigma_n} D\phi \exp\left[\int dx \int d\tau \mathcal{L}_{CFT} + \lambda \sum_{i=1}^{n} \int dx_i ~O(x_i) \right]}{\left(\int_{\Sigma_1} D\phi \exp\left[\int dx \int d\tau \mathcal{L}_{CFT} + \lambda \int dx ~O(x) \right]\right)^{n}},
\end{align}
where $i$ labels the sheets of the replica manifold. Noting that the 1-point function of a primary operator vanishes, the leading nontrivial correction comes from the second order of $\lambda$ as 
\begin{align}\label{eq:PEperturbation}
    S^{(n)}(\mathcal{T}_A) \approx 
    &\frac{1}{1-n} \log \frac{\int_{\Sigma_n} D\phi \exp\left[\int dx \int d\tau \mathcal{L}_{CFT}\right]}{\left(\int_{\Sigma_1} D\phi \exp\left[\int dx \int d\tau \mathcal{L}_{CFT} \right]\right)^{n}} \nonumber\\
    &+ \frac{1}{1-n} \cdot \frac{\lambda^2}{2} \left\{\sum_{i=1}^n\sum_{j=1}^n \int dx_i dx_j \langle O(x_i)O(x_j)\rangle_{\Sigma_n} -n \int dx dx' \langle O(x)O(x')\rangle_{\Sigma_1} \right\}.
\end{align}
Let us treat the terms appearing here one by one. The first term gives back the standard global quench evolution as 
\begin{align}
    \frac{1}{1-n} \log \frac{\int_{\Sigma_n} D\phi \exp\left[\int dx \int d\tau \mathcal{L}_{CFT}\right]}{\left(\int_{\Sigma_1} D\phi \exp\left[\int dx \int d\tau \mathcal{L}_{CFT} \right]\right)^{n}}
    =   
    \frac{c}{6} \left(1+\frac{1}{n}\right) \log\left(\frac{\beta}{\pi\epsilon}\cosh\frac{2\pi t}{\beta}\right).
\end{align}
To evaluate the last term, note that the cylinder can be mapped to a plane via 
\begin{align}
    \xi = f(z) = e^{\frac{2\pi}{\beta}z}.
\end{align}
Accordingly, the two point function $\langle O(x)O(x')\rangle_{\Sigma_1}$ turns out to be 
\begin{align}\label{eq:2pointSigma1}
    \langle O(x)O(x')\rangle_{\Sigma_1}
    = \left(\frac{2\pi}{\beta}e^{\frac{2\pi}{\beta}x}\right) \left(\frac{2\pi}{\beta}e^{\frac{2\pi}{\beta}x'}\right) \frac{1}{\left(e^{\frac{2\pi}{\beta}x}-e^{\frac{2\pi}{\beta}x'}\right)^2} = \left(\frac{\pi}{\beta}\right)^2 \frac{1}{\sinh^2\left(\frac{\pi}{\beta}(x-x')\right)}.
\end{align}
Since the integral over $x$ and $x'$ diverges, we introduce both UV cutoff and IR cutoff as follows
\begin{align}
    \left(\int_{-\infty}^{x'-\epsilon} dx + \int_{x'+\epsilon}^{\infty} dx\right) \int dx'\langle O(x)O(x')\rangle_{\Sigma_1} = \frac{2\pi}{\beta} \left(\coth \frac{\pi\epsilon}{\beta}-1\right) \cdot L
\end{align}
where $L$ is the divergent volume of the time slice. 

\begin{figure}[H]
    \centering   \includegraphics[width=0.7\linewidth]{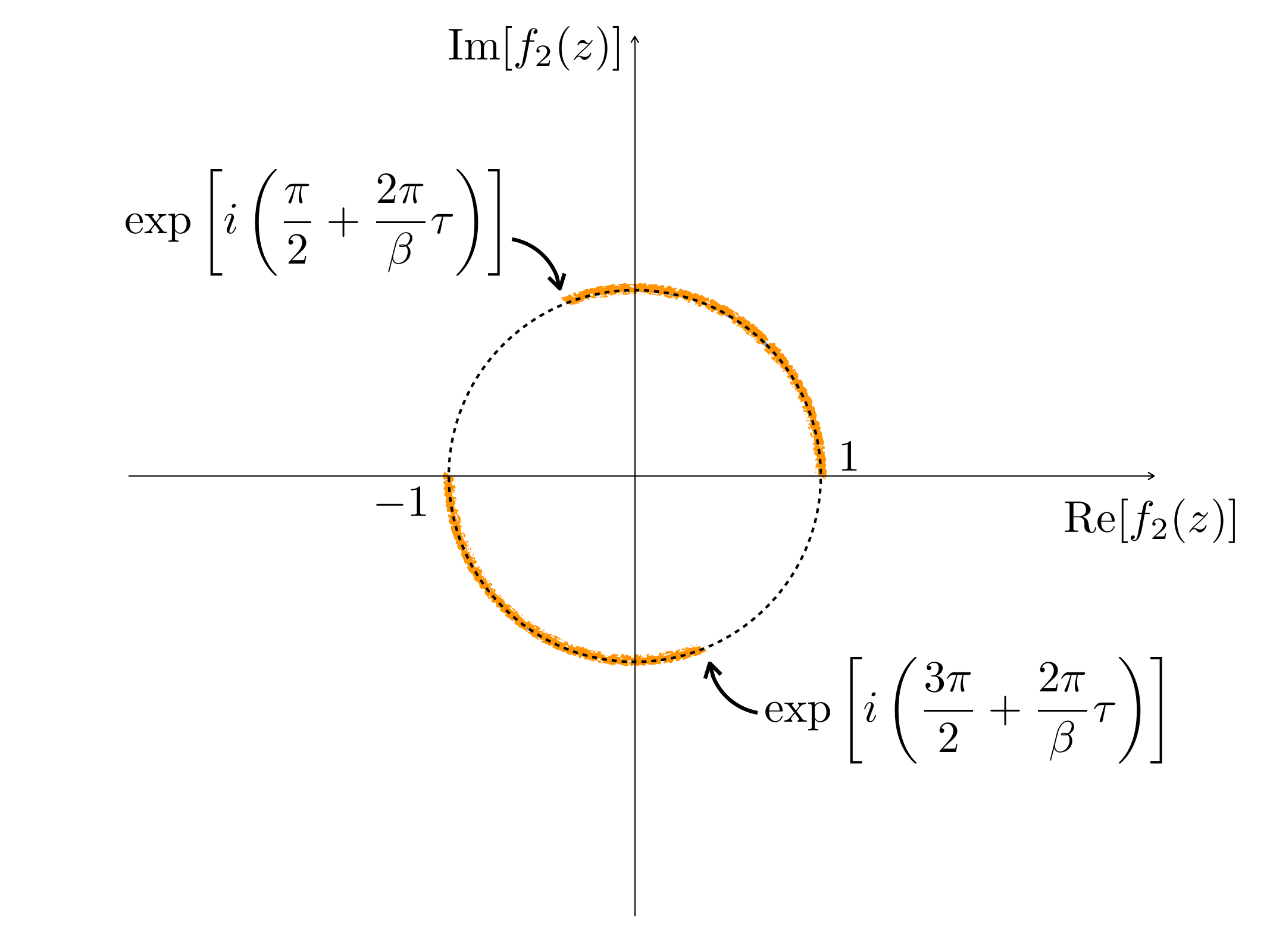}
    \caption{The lines $z=x+i\frac{\beta}{4}$ where the marginal perturbation is performed are shown in orange on the $f_2(z)$ plane for $n=2$. 
    The marginal perturbation lives on two segments of the unit circle. 
    The two segments correspond to the two different sheets of the replica manifold $\Sigma_2$ respectively. }
    \label{fig:unitcircle}
\end{figure}

Let us then move on to the second term, which is the most cumbersome. Noting that the two edges of the subsystem are located at $z=i\tau$ and $z=i(\beta/2 -\tau)$, we find that $\Sigma_n$ can be mapped to $\mathbb{R}^2$ via
\begin{align}
    f_n(z) = \left(\frac{e^{\frac{2\pi}{\beta}z}-e^{i\frac{2\pi}{\beta}\tau}}{e^{\frac{2\pi}{\beta}z}+e^{-i\frac{2\pi}{\beta}\tau}}\right)^{\frac{1}{n}}.
\end{align}
Note that, via this transformation, the line $z=x+i\frac{\beta}{4}$ is mapped to the segment of the unit circle shown below 
\begin{align}
    \frac{j-1}{n}2\pi \leq \theta \leq \frac{j-1}{n}2\pi + \frac{1}{n}\left(\pi + \frac{4\pi}{\beta}\tau\right), ~~ (1\leq j \leq n). 
\end{align}
where $\theta$ is the angular coordinate parameterizing the unit circle. 
A sketch of the $n=2$ case is shown in figure \ref{fig:unitcircle}.

\subsection{2nd pseudo \texorpdfstring{R\'enyi}{Renyi} entropy}
Let us consider the $n=2$ case as a specific example. The conformal map is 
\begin{align}
    f_2(z) = \left(\frac{e^{\frac{2\pi}{\beta}z}-e^{i\frac{2\pi}{\beta}\tau}}{e^{\frac{2\pi}{\beta}z}+e^{-i\frac{2\pi}{\beta}\tau}}\right)^{\frac{1}{2}},
\end{align}
and the first derivative is 
\begin{align}
    f_2'(z) = \frac{2\pi}{\beta} \cos\left(\frac{2\pi}{\beta}\tau\right) e^{\frac{2\pi}{\beta}z}
    \left(e^{\frac{2\pi}{\beta}z}-e^{i \frac{2\pi}{\beta}\tau}\right)^{-\frac{1}{2}}
    \left(e^{\frac{2\pi}{\beta}z}+e^{-i \frac{2\pi}{\beta}\tau}\right)^{-\frac{3}{2}}.
\end{align}
Accordingly, 
\begin{align}
    \left|f'_2(x+i\beta/4)\right|^2 = &\left[\frac{2\pi}{\beta}\cos\left(\frac{2\pi}{\beta}\tau\right) e^{\frac{2\pi}{\beta}x} \left(e^{\frac{4\pi}{\beta}x}-2 e^{\frac{2\pi}{\beta}x} \sin\frac{2\pi}{\beta}\tau +1 \right)^{-1}
 \right]^2 \nonumber\\
 =&  \left[\frac{\pi}{\beta}\cos\left(\frac{2\pi}{\beta}\tau\right) \left(\cosh{\frac{2\pi}{\beta}x} -\sin\frac{2\pi}{\beta}\tau \right)^{-1}
 \right]^2.
\end{align}
Therefore, the two point function on the replica manifold turns out to be 
\begin{align}
    \langle O(x_i)O(x_j)\rangle_{\Sigma_n} = \left|f'_2(x_i+i\beta/4)\right|\left|f'_2(x_j+i\beta/4)\right| \frac{1}{|f_2(x_i+i\beta/4)-f_2(x_j+i\beta/4)|^2}.
\end{align}
It would be cumbersome to compute this combination directly. However, note that there are only two situations: two points on the same sheet and two points on different sheets,
\begin{align}
    &\sum_{i=1}^2\sum_{j=1}^2 \int dx_i dx_j \langle O(x_i)O(x_j)\rangle_{\Sigma_n} \nonumber\\
    =& 
    2 \int dx dy \left( \langle O(x)O(y)\rangle_{\Sigma_n,{\rm same~sheet}} + \langle O(x)O(y)\rangle_{\Sigma_n,{\rm different~sheet}} \right).
\end{align}
Since different sheets correspond to different branches of $(\cdots)^{1/2}$, if we use $F_2(z)$ to denote the principal value of $f_2(z)$, then $\langle O(x)O(y)\rangle_{\Sigma_n,{\rm same~sheet}}$ and $\langle O(x)O(y)\rangle_{\Sigma_n,{\rm different~sheet}}$ can be written as 
\begin{align}
    &\langle O(x)O(y)\rangle_{\Sigma_n,{\rm same~sheet}} = \left|f'_2(x+i\beta/4)\right|\left|f'_2(y+i\beta/4)\right| \frac{1}{|F_2(x+i\beta/4)-F_2(y+i\beta/4)|^2}, \\
    &\langle O(x)O(y)\rangle_{\Sigma_n,{\rm different~sheet}} = \left|f'_2(x+i\beta/4)\right|\left|f'_2(y+i\beta/4)\right| \frac{1}{|F_2(x+i\beta/4)+F_2(y+i\beta/4)|^2},
\end{align}
respectively (refer figure \ref{fig:unitcircle}). 
It is convenient to combine them together and compute
\begin{align}
    &\frac{1}{|F_2(x+i\beta/4)-F_2(y+i\beta/4)|^2} + \frac{1}{|F_2(x+i\beta/4)+F_2(y+i\beta/4)|^2} \nonumber \\
    =&
   \frac{4  e^{\frac{2\pi}{\beta} x+\frac{2\pi}{\beta} y} (\cosh \frac{2\pi}{\beta} x-\sin   \frac{2\pi}{\beta}\tau) (\cosh \frac{2\pi}{\beta} y-\sin 
    \frac{2\pi}{\beta}\tau )}{\left(e^{\frac{2\pi}{\beta} x}-e^{\frac{2\pi}{\beta} y}\right)^2 \cos ^2 \frac{2\pi}{\beta} \tau } 
    \nonumber\\
    =&
   \frac{ (\cosh \frac{2\pi}{\beta} x-\sin   \frac{2\pi}{\beta}\tau) (\cosh \frac{2\pi}{\beta} y-\sin 
    \frac{2\pi}{\beta}\tau )}{\sinh^2 \frac{\pi}{\beta} (x-y) \cos ^2 \frac{2\pi}{\beta} \tau } 
\end{align}
Accordingly, 
\begin{align}\label{eq:2pointSigman}
    &\langle O(x)O(y)\rangle_{\Sigma_n,{\rm same~sheet}} + \langle O(x)O(y)\rangle_{\Sigma_n,{\rm different~sheet}} \nonumber \\
    =&
   \frac{ (\cosh \frac{2\pi}{\beta} x-\sin   \frac{2\pi}{\beta}\tau) (\cosh \frac{2\pi}{\beta} y-\sin 
    \frac{2\pi}{\beta}\tau )}{\sinh^2 \frac{\pi}{\beta} (x-y) \cos ^2 \frac{2\pi}{\beta} \tau } \nonumber\\
    &\times \frac{\pi}{\beta}\cos\left(\frac{2\pi}{\beta}\tau\right) \left(\cosh{\frac{2\pi}{\beta}x} -\sin\frac{2\pi}{\beta}\tau \right)^{-1} \times \frac{\pi}{\beta}\cos\left(\frac{2\pi}{\beta}\tau\right) \left(\cosh{\frac{2\pi}{\beta}y} -\sin\frac{2\pi}{\beta}\tau \right)^{-1}
 \nonumber\\
    =& \left(\frac{\pi}{\beta}\right)^2 \frac{1}{\sinh^2\left(\frac{\pi}{\beta}(x-y)\right)}.
\end{align}
Therefore, comparing \eqref{eq:PEperturbation}, \eqref{eq:2pointSigma1} and \eqref{eq:2pointSigman}, we find that the $\lambda^2$ correction to the 2nd pseudo R\'enyi entropy is exactly zero. 

\subsection{3nd pseudo \texorpdfstring{R\'enyi}{Renyi} entropy}
Let us then consider the $n=3$ case. Similarly, 
\begin{align}
    &\sum_{i=1}^3\sum_{j=1}^3 \int dx_i dx_j \langle O(x_i)O(x_j)\rangle_{\Sigma_n} \nonumber\\
    =& 
    3 \int dx_i dx_j \left( \langle O(x_i)O(x_j)\rangle_{\Sigma_n,{\rm same~sheet}} + \langle O(x_i)O(x_j)\rangle_{\Sigma_n,j = i+1} + \langle O(x_i)O(x_j)\rangle_{\Sigma_n,j = i+2} \right).
\end{align}
Note that 
\begin{align}
    \frac{1}{1+e^{i\varphi}}\frac{1}{1+e^{-i\varphi}} + \frac{1}{1+e^{i\left(\varphi+\frac{2\pi}{3}\right)}}\frac{1}{1+e^{-i\left(\varphi+\frac{2\pi}{3}\right)}} + \frac{1}{1+e^{i\left(\varphi+\frac{4\pi}{3}\right)}}\frac{1}{1+e^{-i\left(\varphi+\frac{4\pi}{3}\right)}} = \frac{9}{4} \left({\cos \frac{3}{2}\varphi}\right)^{-2}. 
\end{align}
Therefore, 
\begin{align}
    &\sum_{i=1}^3\sum_{j=1}^3 \int dx_i dx_j \langle O(x_i)O(x_j)\rangle_{\Sigma_n} \nonumber\\
    =& 
    3 \int dx dy \left|f'_3(x+i\beta/4)\right|\left|f'_3(y+i\beta/4)\right| \nonumber\\
    &\times \frac{9}{4} \left({\cosh \left[ \frac{3}{2} \times \frac{1}{3} \left(\log\frac{ie^{\frac{2\pi}{\beta}x}-e^{i\frac{2\pi}{\beta}\tau}}{ie^{\frac{2\pi}{\beta}x}+e^{-i\frac{2\pi}{\beta}\tau}} - \log\frac{ie^{\frac{2\pi}{\beta}y}-e^{i\frac{2\pi}{\beta}\tau}}{ie^{\frac{2\pi}{\beta}y}+e^{-i\frac{2\pi}{\beta}\tau}} \right)\right]}\right)^{-2} \nonumber\\
    =& 
    3 \int dx dy \left|f'_3(x+i\beta/4)\right|\left|f'_3(y+i\beta/4)\right| \nonumber\\
    &\times 9 \left( \left[ \frac{ie^{\frac{2\pi}{\beta}x}-e^{i\frac{2\pi}{\beta}\tau}}{ie^{\frac{2\pi}{\beta}x}+e^{-i\frac{2\pi}{\beta}\tau}} \frac{ie^{\frac{2\pi}{\beta}y}+e^{-i\frac{2\pi}{\beta}\tau}}{ie^{\frac{2\pi}{\beta}y}-e^{i\frac{2\pi}{\beta}\tau}} \right]^{\frac{1}{2}} + 
    \left[ \frac{ie^{\frac{2\pi}{\beta}y}-e^{i\frac{2\pi}{\beta}\tau}}{ie^{\frac{2\pi}{\beta}y}+e^{-i\frac{2\pi}{\beta}\tau}} \frac{ie^{\frac{2\pi}{\beta}x}+e^{-i\frac{2\pi}{\beta}\tau}}{ie^{\frac{2\pi}{\beta}x}-e^{i\frac{2\pi}{\beta}\tau}} \right]^{\frac{1}{2}} \right)^{-2} \nonumber\\
    =& 
    3 \int dx dy \left|f'_3(x+i\beta/4)\right|\left|f'_3(y+i\beta/4)\right| \nonumber\\
    &\times 9 \left( \frac{ie^{\frac{2\pi}{\beta}x}-e^{i\frac{2\pi}{\beta}\tau}}{ie^{\frac{2\pi}{\beta}x}+e^{-i\frac{2\pi}{\beta}\tau}} \frac{ie^{\frac{2\pi}{\beta}y}+e^{-i\frac{2\pi}{\beta}\tau}}{ie^{\frac{2\pi}{\beta}y}-e^{i\frac{2\pi}{\beta}\tau}} + 
    \frac{ie^{\frac{2\pi}{\beta}y}-e^{i\frac{2\pi}{\beta}\tau}}{ie^{\frac{2\pi}{\beta}y}+e^{-i\frac{2\pi}{\beta}\tau}} \frac{ie^{\frac{2\pi}{\beta}x}+e^{-i\frac{2\pi}{\beta}\tau}}{ie^{\frac{2\pi}{\beta}x}-e^{i\frac{2\pi}{\beta}\tau}} + 2 \right)^{-1} \\
    =& 
    3 \int dx dy \left(\frac{2\pi}{3\beta}\right)^2 \left(1+e^{\frac{4 i \pi  \tau }{\beta }}\right)^2
    \frac{ e^{\frac{2 \pi 
   x}{\beta }}}{\left(e^{\frac{2 \pi  (x+i \tau )}{\beta }}-i\right)
   \left(e^{\frac{2 \pi  x}{\beta }}+i e^{\frac{2 i \pi  \tau }{\beta }}\right)}
   \frac{  e^{\frac{2 \pi 
   y}{\beta }}}{\left(e^{\frac{2 \pi  (y+i \tau )}{\beta }}-i\right)
   \left(e^{\frac{2 \pi  y}{\beta }}+i e^{\frac{2 i \pi  \tau }{\beta }}\right)} \nonumber\\
    &\times 9 \left( \frac{\left(2 i e^{\frac{2 i \pi  \tau }{\beta }}-e^{\frac{2 \pi  (x+2 i \tau )}{\beta
   }}+e^{\frac{2 \pi  x}{\beta }}+2 i e^{\frac{2 \pi  (i \tau +x+y)}{\beta }}-e^{\frac{2
   \pi  (y+2 i \tau )}{\beta }}+e^{\frac{2 \pi  y}{\beta }}\right)^2}{\left(e^{\frac{2
   \pi  (x+i \tau )}{\beta }}-i\right) \left(e^{\frac{2 i \pi  \tau }{\beta }}-i
   e^{\frac{2 \pi  x}{\beta }}\right) \left(e^{\frac{2 \pi  (y+i \tau )}{\beta
   }}-i\right) \left(e^{\frac{2 i \pi  \tau }{\beta }}-i e^{\frac{2 \pi  y}{\beta
   }}\right)} \right)^{-1}  \nonumber\\
    =& 
    3 \int dx dy \left(\frac{2\pi}{\beta}\right)^2 
    \frac{ -\left(1+e^{\frac{4 i \pi  \tau }{\beta }}\right)^2 e^{\frac{2 \pi 
   x}{\beta } }e^{\frac{2 \pi 
   y}{\beta }}}{\left(2 i e^{\frac{2 i \pi  \tau }{\beta }}-e^{\frac{2 \pi  (x+2 i \tau )}{\beta
   }}+e^{\frac{2 \pi  x}{\beta }}+2 i e^{\frac{2 \pi  (i \tau +x+y)}{\beta }}-e^{\frac{2
   \pi  (y+2 i \tau )}{\beta }}+e^{\frac{2 \pi  y}{\beta }}\right)^2} \nonumber\\
    =& 
    3 \int dx dy \left(\frac{2\pi}{\beta}\right)^2 
    \frac{ -\left(1+e^{\frac{4 i \pi  \tau }{\beta }}\right)^2 e^{\frac{2 \pi 
   x}{\beta } }e^{\frac{2 \pi 
   y}{\beta }}}{\left(2 i e^{\frac{2 i \pi  \tau }{\beta }}-e^{\frac{2 \pi  (x+2 i \tau )}{\beta
   }}+e^{\frac{2 \pi  x}{\beta }}+2 i e^{\frac{2 \pi  (i \tau +x+y)}{\beta }}-e^{\frac{2
   \pi  (y+2 i \tau )}{\beta }}+e^{\frac{2 \pi  y}{\beta }}\right)^2}   \nonumber\\
    =& 
    3 \int dx dy \left(\frac{\pi}{\beta}\right)^2 
    \left(\frac{\cos \frac{2\pi}{\beta}\tau}{\cosh\frac{\pi}{\beta}(x+y) - \cosh\frac{\pi}{\beta}(x-y) \sin \frac{2\pi}{\beta}\tau}\right)^2 \nonumber\\ 
    =& - 6 \log \left( \sin \frac{2\pi}{\beta}\tau \right) \nonumber\\
    \approx& -\frac{12\pi}{\beta}t, 
\end{align}
where at the last line we write down the leading term at $t/\beta \gg 1$. Therefore, 
\begin{align}
    S^{(3)}(\mathcal{T}_A) = &\frac{2c}{9} \log\left(\frac{\beta}{\pi\epsilon}\cosh\frac{2\pi t}{\beta}\right) + \frac{3\lambda^2}{2} \left(\log\left(\sin i \frac{2\pi}{\beta} t \right)-\frac{\pi}{\beta}\left(\coth\frac{\pi\epsilon}{\beta}-1\right)L\right) + O(\lambda^3).\nonumber
\end{align}

\subsection{\texorpdfstring{R\'enyi}{Renyi} entropy for general \texorpdfstring{$n$}{n}}

Based on the results above, let us consider computing the Renyi entropy for general integer $n$. The map and its first derivative turn out to be 
\begin{align}
    f_n(z) = \left(\frac{e^{\frac{2\pi}{\beta}z}-e^{i\frac{2\pi}{\beta}\tau}}{e^{\frac{2\pi}{\beta}z}+e^{-i\frac{2\pi}{\beta}\tau}}\right)^{\frac{1}{n}},
\end{align}
and 
\begin{align}
    |f'_n(x+i\beta/4)|^2 = \left(\frac{2\pi}{n\beta}\right)^2 \frac{\left(1+e^{\frac{4 i \pi  \tau }{\beta }}\right)^2 e^{\frac{4 \pi 
   x}{\beta }}}{\left(e^{\frac{2 \pi  (x+i \tau )}{\beta }}-i\right)^2
   \left(e^{\frac{2 \pi  x}{\beta }}+i e^{\frac{2 i \pi  \tau }{\beta }}\right)^2},
\end{align}
respectively. Accordingly, 
\begin{align}
    &|f'_n(x+i\beta/4)| |f'_n(y+i\beta/4)| \nonumber\\
    =& \left(\frac{2\pi}{n\beta}\right)^2 \left(1+e^{\frac{4 i \pi  \tau }{\beta }}\right)^2
    \frac{ e^{\frac{2 \pi 
   x}{\beta }}}{\left(e^{\frac{2 \pi  (x+i \tau )}{\beta }}-i\right)
   \left(e^{\frac{2 \pi  x}{\beta }}+i e^{\frac{2 i \pi  \tau }{\beta }}\right)}
   \frac{  e^{\frac{2 \pi 
   y}{\beta }}}{\left(e^{\frac{2 \pi  (y+i \tau )}{\beta }}-i\right)
   \left(e^{\frac{2 \pi  y}{\beta }}+i e^{\frac{2 i \pi  \tau }{\beta }}\right)}. 
\end{align}
Note that 
\begin{align}
    &\sum_{j=1}^n \frac{1}{1+e^{i\left(\varphi+\frac{2\pi j}{n}\right)}}\frac{1}{1+e^{-i\left(\varphi+\frac{2\pi j}{n}\right)}} = \left(\frac{n}{2}\right)^2 \left({\sin \frac{n}{2}\varphi}\right)^{-2} = \frac{n^2}{2-e^{\frac{n}{2}\varphi}-e^{-\frac{n}{2}\varphi}}, ~~~~(n {\rm~even})  \\
    &\sum_{j=1}^n \frac{1}{1+e^{i\left(\varphi+\frac{2\pi j}{n}\right)}}\frac{1}{1+e^{-i\left(\varphi+\frac{2\pi j}{n}\right)}} = \left(\frac{n}{2}\right)^2 \left({\cos \frac{n}{2}\varphi}\right)^{-2} = \frac{n^2}{2+e^{\frac{n}{2}\varphi}+e^{-\frac{n}{2}\varphi}}.~~~~(n {\rm~odd})
\end{align}
Let us consider the two cases separately. 

\paragraph{When $n$ is odd}\par
\begin{align}
    &\sum_{i=1}^n\sum_{j=1}^n \int dx_i dx_j \langle O(x_i)O(x_j)\rangle_{\Sigma_n} \nonumber\\
    =& 
    n \int dx dy \left|f'_n(x+i\beta/4)\right|\left|f'_n(y+i\beta/4)\right| \nonumber\\
    &\times n^2 \left( 2+\frac{ie^{\frac{2\pi}{\beta}x}-e^{i\frac{2\pi}{\beta}\tau}}{ie^{\frac{2\pi}{\beta}x}+e^{-i\frac{2\pi}{\beta}\tau}} \frac{ie^{\frac{2\pi}{\beta}y}+e^{-i\frac{2\pi}{\beta}\tau}}{ie^{\frac{2\pi}{\beta}y}-e^{i\frac{2\pi}{\beta}\tau}} + 
    \frac{ie^{\frac{2\pi}{\beta}y}-e^{i\frac{2\pi}{\beta}\tau}}{ie^{\frac{2\pi}{\beta}y}+e^{-i\frac{2\pi}{\beta}\tau}} \frac{ie^{\frac{2\pi}{\beta}x}+e^{-i\frac{2\pi}{\beta}\tau}}{ie^{\frac{2\pi}{\beta}x}-e^{i\frac{2\pi}{\beta}\tau}}  \right)^{-1}  \nonumber\\
    =& 
    n \int dx dy \left(\frac{2\pi}{n\beta}\right)^2 \left(1+e^{\frac{4 i \pi  \tau }{\beta }}\right)^2
    \frac{ e^{\frac{2 \pi 
   x}{\beta }}}{\left(e^{\frac{2 \pi  (x+i \tau )}{\beta }}-i\right)
   \left(e^{\frac{2 \pi  x}{\beta }}+i e^{\frac{2 i \pi  \tau }{\beta }}\right)}
   \frac{  e^{\frac{2 \pi 
   y}{\beta }}}{\left(e^{\frac{2 \pi  (y+i \tau )}{\beta }}-i\right)
   \left(e^{\frac{2 \pi  y}{\beta }}+i e^{\frac{2 i \pi  \tau }{\beta }}\right)} \nonumber\\
    &\times n^2 \left( \frac{\left(2 i e^{\frac{2 i \pi  \tau }{\beta }}-e^{\frac{2 \pi  (x+2 i \tau )}{\beta
   }}+e^{\frac{2 \pi  x}{\beta }}+2 i e^{\frac{2 \pi  (i \tau +x+y)}{\beta }}-e^{\frac{2
   \pi  (y+2 i \tau )}{\beta }}+e^{\frac{2 \pi  y}{\beta }}\right)^2}{\left(e^{\frac{2
   \pi  (x+i \tau )}{\beta }}-i\right) \left(e^{\frac{2 i \pi  \tau }{\beta }}-i
   e^{\frac{2 \pi  x}{\beta }}\right) \left(e^{\frac{2 \pi  (y+i \tau )}{\beta
   }}-i\right) \left(e^{\frac{2 i \pi  \tau }{\beta }}-i e^{\frac{2 \pi  y}{\beta
   }}\right)} \right)^{-1}  \nonumber\\
    =& 
    n \int dx dy \left(\frac{2\pi}{\beta}\right)^2 
    \frac{ -\left(1+e^{\frac{4 i \pi  \tau }{\beta }}\right)^2 e^{\frac{2 \pi 
   x}{\beta } }e^{\frac{2 \pi 
   y}{\beta }}}{\left(2 i e^{\frac{2 i \pi  \tau }{\beta }}-e^{\frac{2 \pi  (x+2 i \tau )}{\beta
   }}+e^{\frac{2 \pi  x}{\beta }}+2 i e^{\frac{2 \pi  (i \tau +x+y)}{\beta }}-e^{\frac{2
   \pi  (y+2 i \tau )}{\beta }}+e^{\frac{2 \pi  y}{\beta }}\right)^2} \nonumber\\
    =& 
    n \int dx dy \left(\frac{2\pi}{\beta}\right)^2 
    \frac{ -\left(1+e^{\frac{4 i \pi  \tau }{\beta }}\right)^2 e^{\frac{2 \pi 
   x}{\beta } }e^{\frac{2 \pi 
   y}{\beta }}}{\left(2 i e^{\frac{2 i \pi  \tau }{\beta }}-e^{\frac{2 \pi  (x+2 i \tau )}{\beta
   }}+e^{\frac{2 \pi  x}{\beta }}+2 i e^{\frac{2 \pi  (i \tau +x+y)}{\beta }}-e^{\frac{2
   \pi  (y+2 i \tau )}{\beta }}+e^{\frac{2 \pi  y}{\beta }}\right)^2}   \nonumber\\
    =& 
    n \int dx dy \left(\frac{\pi}{\beta}\right)^2 
    \left(\frac{\cos \frac{2\pi}{\beta}\tau}{\cosh\frac{\pi}{\beta}(x+y) - \cosh\frac{\pi}{\beta}(x-y) \sin \frac{2\pi}{\beta}\tau}\right)^2 \nonumber\\ 
    =& - 2 n \log \left( \sin \frac{2\pi}{\beta}\tau \right) \nonumber\\
    \approx& -\frac{4n\pi}{\beta}t, 
\end{align}
where at the last line we write down the leading term at $t/\beta \gg 1$. Therefore, 
\begin{align}
    S^{(n)}(\mathcal{T}_A) = &\frac{c}{6}\left(1+\frac{1}{n}\right) \log\left(\frac{\beta}{\pi\epsilon}\cosh\frac{2\pi t}{\beta}\right) +  \frac{n\lambda^2}{n-1}\left(\log\left(\sin i \frac{2\pi}{\beta} t \right)-\frac{\pi}{\beta}\left(\coth\frac{\pi\epsilon}{\beta}-1\right)L\right) + O(\lambda^3),\nonumber
\end{align}
for odd $n$.

\paragraph{When $n$ is even}\par
\begin{align}
    &\sum_{i=1}^n\sum_{j=1}^n \int dx_i dx_j \langle O(x_i)O(x_j)\rangle_{\Sigma_n} \nonumber\\
    =& 
    n \int dx dy \left|f'_n(x+i\beta/4)\right|\left|f'_n(y+i\beta/4)\right| \nonumber\\
    &\times n^2 \left( 2-\frac{ie^{\frac{2\pi}{\beta}x}-e^{i\frac{2\pi}{\beta}\tau}}{ie^{\frac{2\pi}{\beta}x}+e^{-i\frac{2\pi}{\beta}\tau}} \frac{ie^{\frac{2\pi}{\beta}y}+e^{-i\frac{2\pi}{\beta}\tau}}{ie^{\frac{2\pi}{\beta}y}-e^{i\frac{2\pi}{\beta}\tau}} - 
    \frac{ie^{\frac{2\pi}{\beta}y}-e^{i\frac{2\pi}{\beta}\tau}}{ie^{\frac{2\pi}{\beta}y}+e^{-i\frac{2\pi}{\beta}\tau}} \frac{ie^{\frac{2\pi}{\beta}x}+e^{-i\frac{2\pi}{\beta}\tau}}{ie^{\frac{2\pi}{\beta}x}-e^{i\frac{2\pi}{\beta}\tau}}  \right)^{-1}  \nonumber\\
    =& 
    n \int dx dy \left(\frac{2\pi}{n\beta}\right)^2 \left(1+e^{\frac{4 i \pi  \tau }{\beta }}\right)^2
    \frac{ e^{\frac{2 \pi 
   x}{\beta }}}{\left(e^{\frac{2 \pi  (x+i \tau )}{\beta }}-i\right)
   \left(e^{\frac{2 \pi  x}{\beta }}+i e^{\frac{2 i \pi  \tau }{\beta }}\right)}
   \frac{  e^{\frac{2 \pi 
   y}{\beta }}}{\left(e^{\frac{2 \pi  (y+i \tau )}{\beta }}-i\right)
   \left(e^{\frac{2 \pi  y}{\beta }}+i e^{\frac{2 i \pi  \tau }{\beta }}\right)} \nonumber\\
    &\times n^2 \left( \frac{\left(1+e^{\frac{4 i \pi  \tau }{\beta }}\right)^2 \left(e^{\frac{2 \pi  x}{\beta }}-e^{\frac{2 \pi  y}{\beta }}\right)^2}{\left(e^{\frac{2 \pi  (x+i \tau )}{\beta }}-i\right)
   \left(e^{\frac{2 \pi  x}{\beta }}+i e^{\frac{2 i \pi  \tau }{\beta }}\right) \left(e^{\frac{2 \pi  (y+i \tau )}{\beta }}-i\right)
   \left(e^{\frac{2 \pi  y}{\beta }}+i e^{\frac{2 i \pi  \tau }{\beta }}\right)}  \right)^{-1}  \nonumber\\
    =& 
    n \int dx dy \left(\frac{\pi}{\beta}\right)^2 \frac{1}{\sinh^2\left(\frac{\pi}{\beta}(x-y)\right)}.
\end{align}
Therefore, comparing this with \eqref{eq:PEperturbation}, \eqref{eq:2pointSigma1}, we find that the $\lambda^2$ correction of the $n$-th pseudo R\'enyi entropy is zero when $n$ is even. 

\bibliographystyle{JHEP}
\bibliography{BMPT}


\end{document}